\documentclass[journal,twoside]{IEEEtran}

\normalsize

\usepackage[cmex10]{amsmath}
\usepackage{graphicx}
\usepackage{epstopdf}
\usepackage{multirow}
\usepackage{amssymb}
\usepackage{tensor}
\usepackage{boondox-cal}
\usepackage{ifthen,boxedminipage}
\usepackage{cite}
\usepackage{lipsum}
\usepackage{mathtools}
\usepackage{cuted}
\usepackage{varwidth}
\usepackage{tabularx}
\usepackage{rotating}
\usepackage{times,color}
\usepackage{hyperref} 

\DeclareMathOperator{\sign}{sign}
\DeclareMathOperator{\csch}{csch}
\definecolor{dgreen}{rgb}{0,0.48,0.3}

\begin{document}

\title{A graph representation based on fluid diffusion model for data analysis: theoretical aspects and enhanced community detection}

\author{Andrea~Marinoni, Christian Jutten, Mark Girolami
	\thanks{A. Marinoni is with Dept. of Physics and Technology, UiT the Arctic University of Norway, P.O. box 6050 Langnes, NO-9037, Tromsø, Norway, and Dept. of Engineering, University of Cambridge, Trumpington St., Cambridge CB2 1PZ, UK. E-mail: andrea.marinoni@uit.no.
	C. Jutten is with GIPSA-lab, University of Grenoble Alpes, 38000 Grenoble, France, and with the Institut Universitaire de France, 75005 Paris, France. E-mail: christian.jutten@gipsa-lab.grenoble-inp.fr. 
	M. Girolami is with Dept. of Engineering, University of Cambridge, Trumpington St., Cambridge CB2 1PZ, UK, and with The Alan Turing Institute, 96 Euston Rd, London NW1 2DB, UK. E-mail: mag92@cam.ac.uk. 
	}
}

\maketitle
\begin{abstract}

Representing data by means of graph structures
identifies one of the most valid approach to extract information
in several data analysis applications. This is especially true
when multimodal datasets are investigated, as records collected
by means of diverse sensing strategies are taken into account
and explored. Nevertheless, classic graph signal processing is
based on a model for information propagation that is configured
according to heat diffusion mechanism. This system provides
several constraints and assumptions on the data properties that
might be not valid for multimodal data analysis, especially when
large scale datasets collected from heterogeneous sources are
considered, so that the accuracy and robustness of the outcomes
might be severely jeopardized. In this paper, we introduce a
novel model for graph definition based on fluid diffusion. The
proposed approach improves the ability of graph-based data
analysis to take into account several issues of modern data
analysis in operational scenarios, so to provide a platform for
precise, versatile, and efficient understanding of the phenomena
underlying the records under exam, and to fully exploit the
potential provided by the diversity of the records in obtaining a
thorough characterization of the data and their significance. In
this work, we focus our attention to using this fluid diffusion
model to drive a community detection scheme, i.e., to divide multimodal datasets into many groups according to similarity among nodes in an unsupervised fashion.
Experimental results achieved by testing real multimodal datasets in diverse application scenarios show that our method is able to strongly outperform state-of-the-art schemes for community detection in multimodal data analysis.

\end{abstract}

\begin{IEEEkeywords}
Multimodal data analysis, fluid diffusion, information propagation, community detection, clustering.	
\end{IEEEkeywords}


\section{Introduction} 

The recent advancements in sensing technologies have allowed the collection of massive datasets that are able to describe almost every aspect of our everyday life, from traffic control to social media sentiment analysis, from biomedical studies to environmental monitoring. 
Nonetheless, modern data analysis still has to face a variety of challenges to face and solve so that effective information extraction can be performed, i.e., so that data analysis can be used to reliably drive decision-making processes and risk minimization procedures \cite{multimod1,multimod2,GEOMDL_limits_1,GEOMDL_limits_2,GEOMDL_limits_3,GEOMDL_limits_4, GEOMDL_limits_5, GEOMDL_limits_6,GEOMDL_limits_7,GEOMDL_limits_8}. 

In fact, to obtain a robust understanding of the system and problems at hand, large scale datasets have to be investigated. 
Thus, the collected records can be affected by multiple and diverse
sources of noise, leading to incomplete, non-uniform, and/or
unbalanced data to be analyzed. 
Moreover, data analysis solutions must be flexible so to properly address undesired effects such as misalignments, coregistration problems, and different record sizes, without having to develop bespoke methods for each set of records sources and each analysis goal. 
Finally, the efficiency and the explainability
of the data analysis must be addressed, so that near real time
applications and thorough interpretation of the data can be
carried out \cite{GEOMDL_limits_1,GEOMDL_limits_2,GEOMDL_limits_4,GEOMDL_limits_6,GEOMDL_limits_7}. 

In technical literature, several methods have been proposed to address these issues. 
For instance, architectures based on self-attention and reinforcement learning have been recently employed in several research fields. 
In particular, graph-based data analysis plays definitively a key role. 
In fact, it allows the investigation of complex datasets by means of a compact representation of mathematical manifolds, groups and varieties across different domains \cite{GEOMDL,fluid_graphsignalproc}. 
Graphs provide an abstract depiction of interactions among samples, so that they can be quickly adapted to multiple tasks and datasets.  
In this way, graph-based data analysis enables information extraction in several challenging situations (e.g., for temporally variable datasets acquired by diverse sources and sensing platforms), or complicated tasks (e.g., domain adaptation, transfer learning) \cite{multimod9,multimod1,GEOMDL_limits_1,GEOMDL_limits_3,GEOMDL_limits_4}. 
These properties are particularly interesting when semisupervised and unsupervised characterization of the records is required, e.g., clustering, outlier detection, ranking, label propagation \cite{COUILLET20,fluid_graphsignalproc}. 

Although the aforementioned recent data analysis strategies have been employed in different research fields with discrete success \cite{COUILLET18,COUILLET20,CommunityDetection_SC1,CommunityDetection_SC2,CommunityDetection_SC3,CommunityDetection_SC4,fluid_Communitydetection_Tang,fluid_graphsignalproc,COUILLET20,fluid_KGL}, they still show some limitations that can jeopardize their use in operational pipelines. 
In particular, these methods can hardly manage automatic learning in case of unreliable data, e.g., datasets gathered from corrupted observations or where records might be missing. 
To give an intuitive understanding of these points, let us consider a practical example. 

\begin{figure}[htb]
	\centering
	\includegraphics[width=1\columnwidth]{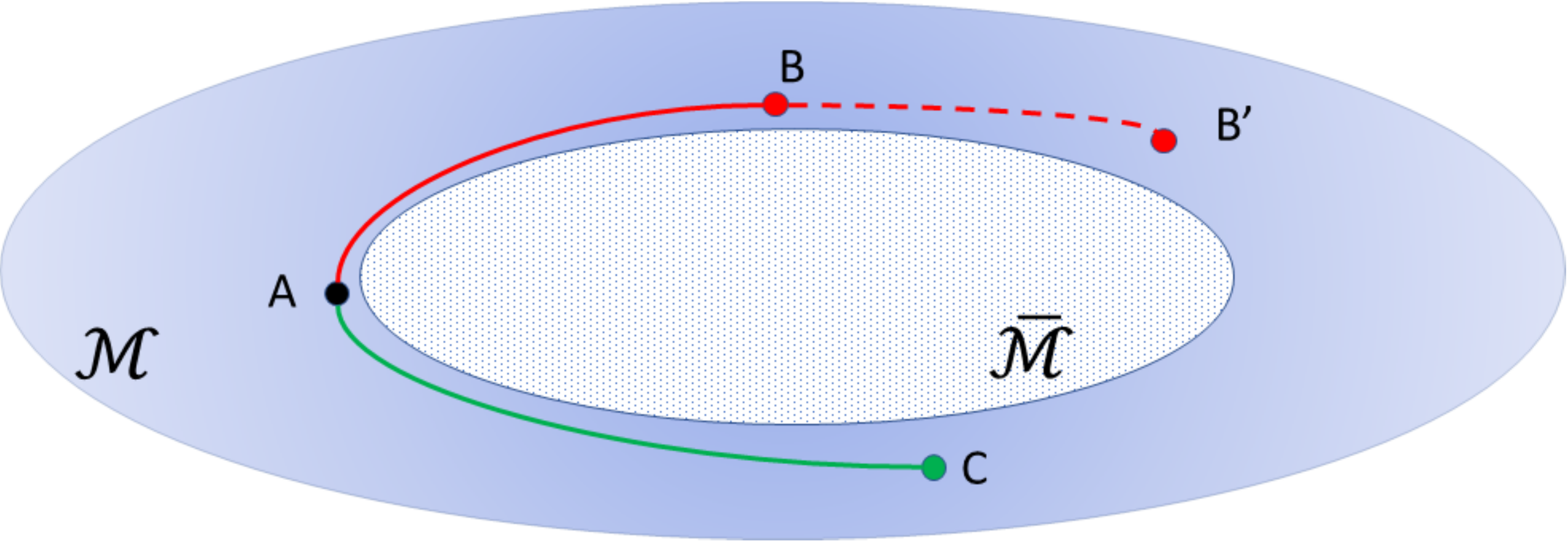} 
	\caption{Figurative example of the paradox induced by unreliable observations for semisupervised learning of samples on the manifold ${\cal M}$. }
	\label{fig_hike_sketch}
\end{figure}

Specifically, let us assume to have a dataset where each sample is identified by $n$ features. 
The whole dataset can be represented by a $n$-dimensional manifold
${\cal M}$ (figuratively depicted in Fig. \ref{fig_hike_sketch}, where the shaded area $\bar{\cal M}$ represents a region in the $n$-dimensional space where samples do not live, e.g., for unfeasible conditions with respect to the physics underlying the considered observed phenomena).  
Let us map on ${\cal M}$ three samples A, B, and C, and let us assume to have prior information on the semantics of samples B and C, e.g., B identifies an instance of the "red" class, whilst C belongs to the "green" class. 
At this point, let us assume to perform supervised learning on sample A. 
To this aim, we would have to compute a the distance (according to the characteristics of ${\cal M}$, e.g., by means of geometrical or statistical measures \cite{nielsen_metrics}) between A and B $d_{AB}$, and A and C $d_{AC}$, and then compare them. 
Since $d_{AB} < d_{AC}$, sample A would then be associated with class "red".
However, if a subset $m_B \leq n$ of features acquired for sample B were corrupted or unreliable (e.g., some detectors had experienced faults, or offsets in their measurements), the \textit{true} position of this sample on ${\cal M}$ would have been the one identified by the point B' in Fig. \ref{fig_hike_sketch}. 
Therefore, since $d_{AB'} > d_{AC}$, A should have been assigned to class "green", and we would have made a mistake in classifying it. 

This example shows how the reliability of the data is a key component to be taken into account to achieve solid understanding of the data structures, and to conduct robust information extraction. 
However, state-of-the-art methods for data analysis and automatic learning typically fail to address data reliability in their operations. 
Indeed, they eventually tend to compensate this issue by using highly nonlinear models to describe the data. This can affect the efficiency of the learning procedures, or eventually make the investigation prone to undesired effects such as overfitting or biasing \cite{GEOMDL_limits_1,GEOMDL_limits_2,GEOMDL_limits_3,GEOMDL_limits_4,GEOMDL_limits_5,GEOMDL_limits_6,GEOMDL_limits_7,GEOMDL_limits_8}. 

To address these issues, we focused our attention on graphs and in particular on graph representation. 
Our goal is to incorporate a quantification of the reliability of the data in the definition of the edge weights. 
In state-of-the-art graph-based data analysis, these quantities result from the assumption that information propagates through the graph according to heat diffusion mechanism. 
However, this model is not flexible enough to include data reliability in the description of the graph structure, therefore limiting and eventually jeopardizing the robustness of the analysis.  

To address these points, in this work we propose the consideration of a new model for information propagation across graphs.
Specifically, we introduce a \textit{fluid diffusion model} to shape the graph design, with special focus to the topology and connectivity of the data structure \cite{fluid_FokkerPlanck1,fluid_FokkerPlanck2}. 
In this way, the global and local interactions among samples and records are taken into account in terms of tensor representation, which can be expressed as permeability, diffusivity and flow velocity across the graph.   
This representation allows one to take into account the reliability of the data, so to achieve solid and robust automatic learning.  
By taking advantage of this novel data representation, we provide an efficient method for community detection that can be easily implemented in terms of spectral clustering approach.  
Then, we demonstrate that this approach enables an accurate understanding of the data structures, so that a strong enhancement with respect to the state-of-the-art methods can be observed.  

\color{black}

The main contributions of this paper can be then summarized as follows:
\begin{itemize}
	\item a new paradigm to model information propagation  - based on fluid diffusion - on graph structures which is able to grasp global and local scale interactions and patterns induced by multimodal datasets;
	\item the analysis of the proposed fluid diffusion system in terms of eigenvectors of the flow velocity matrix that can be employed to characterize the dependency among samples and the relevance of the features the considered dataset consists of. In this way, an effective understanding of the graph can be carried out in terms of eigenanalysis, enabling valid characterization of the data structures and topologies;
	\item an efficient method for non-overlapping community detection, taking advantage of the eigenanalysis of the flow velocity matrix used to describe the graph connections. 
\end{itemize}


The paper is organized as follows. 
Section \ref{sec_fluidgraph} introduces the theoretical aspects of classic graph representation of datasets based on the heat diffusion model. It continues with the motivation of the proposed novel graph representation based on fluid diffusion, and its main properties. 
Section \ref{sec_fluidCD} reports a thorough overview of the main works introduced in technical literature for the application task used as test case in this paper - community detection - as well as the proposed method for community detection based on the novel fluid graph representation of multimodal datasets. 
Section \ref{secresult} reports the performance results obtained over three multimodal datasets, as well as heuristic confirmation of the motivation and validity of the proposed fluid graph representation.  
Finally, Section \ref{secconcl} delivers our final remarks and some ideas on future research.

\section{Fluid graph representation}
\label{sec_fluidgraph}



In this Section, we introduce the main motivations for the novel graph representation  for multimodal data analysis we proposed in this work. 
First, the connection between classic graph representation and heat diffusion is summarized. 
Then, the main issues for classic graph representation are presented, leading to the motivation and the description of the graph representation based on fluid diffusion that we introduce in this paper. 

\subsection{Classic graph representation and heat diffusion}
\label{sec_meth_heat}

Graph representation of data manifolds is a valuable tool in extracting information from records, understanding their interactions,
and providing a thorough interpretation of their semantics. 
Indeed, graph-based signal processing has enabled exploiting data
structure and relational priors, improving data and computational
efficiency, and enhancing model interpretability in various domains \cite{fluid_graphsignalproc,GEOMDL}. 

The structure and meaning of the edges and nodes, as defined within graph representation, affects the accuracy and reliability of any information then derived from it  \cite{fluid_graphsignalproc,specclust3,specclust2,specclust1}. 
In fact, graph representation identifies a favorable trade-off between simplicity and explainability of the relationships between the samples in the dataset. 
The similarity and interactions among samples are represented by means of the weights of the edges of the graphs.
The edge weight is then typically computed as function of the proximity  of the corresponding
data points in the feature space induced by the records collected in the considered dataset. 
Hence, a connection between two samples in the dataset could be considered stronger as the proximity of their corresponding feature vectors increases \cite{fluid_diffusion3}. 

Characterizing the complex geometry of the data in the feature space is therefore crucial to obtain an accurate graph representation and therefore a reliable understanding of the interactions among samples. 
To this aim, combining the main properties of random walks and spectral analysis is a proven approach in finding relevant structures in complex manifolds, enabling the detection and classification of thematic clusters within the data \cite{fluid_diffusion2,fluid_diffusion3,COUILLET18}. 
Indeed,  
using the eigenfunctions of a Markov matrix defining a random walk on the data can help in achieving a new description of data sets, as well as in providing a thorough interpretation of the similarity modeled by the edge weights \cite{fluid_diffusion2,fluid_diffusion3}. 
To embed these samples in a Euclidean space, these quantities can be associated with transition mechanisms described in terms of diffusion processes. 
Further, processing the higher order moments of the Markov matrix this strategy aims to connect the spectral properties of the diffusion process to the geometrical characteristics of the dataset. 

Specifically, let $\textbf{X} = \{ \textbf{x}_i\}_{i=1, \ldots, N}$, $\textbf{x}_i \in \mathbb{R}^n$ be the considered dataset consisting of $N$ samples characterized by $n$ features. 
In general, it is possible to translate $\textbf{X}$ into a graph structure consisting of nodes and edges connecting them.  
Specifically, the $i$-th node identifies the sample $\textbf{x}_i$ in the dataset $\textbf{X}$. 
On the other hand, the weight of the edge connecting node $i$ to node $j$ is computed according to a function (or kernel) $\eta(\textbf{x}_i, \textbf{x}_j)$ of the features associated with the considered samples. 
In the classic derivation of graph structure, 
the goal of the metric $\eta$ is to capture the characteristics of the local geometry of the given dataset. 
It is then possible to construct a Markov matrix associated with $\textbf{X}$ that can describe the local geometry of the dataset by summarizing the node-to-node similarities. 
In other terms, the $(i,j)$ element of the Markov matrix is defined as probability
of transition in one time step from node $i$ to node $j$ in the graph. 
As such,  the $(i,j)$ element of the Markov matrix is also proportional to the edge weight $\eta(\textbf{x}_i, \textbf{x}_j)$. 
Moreover, it is possible to retrieve the transition probability in $t$ steps by elevating the Markov matrix to the power $t$
 \cite{fluid_diffusion2,fluid_diffusion3,fluid_diffusion1}. 

These properties of the Markov matrix are particularly interesting for the characterization of the graph structure and connections.  
Analyzing the behavior of the Markov matrix for long transitions, i.e., large power of the Markov matrix, can help in detecting and understanding the actual relationships among the samples in the dataset \cite{fluid_diffusion1,fluid_diffusion3}. 
To this aim, spectral theory plays a crucial role. 
In particular, the eigenanalysis of the aforesaid Markov matrix can help unveil hidden patterns among the samples, leading to a precise understanding of the interactions among samples. 
Moreover, a compact description of the random walk processes based on the eigenvectors and eigenvalues of the Markov matrix can be used to identify the information propagation mechanisms that can be drawn within the dataset according to the geometrical properties of the samples in the feature space \cite{fluid_diffusion2}. 

The metric $\eta$ is expected to provide a characterization of the local geometry of the dataset \cite{fluid_diffusion1}. 
On the other hand, the Markov matrix defines the direction of propagation according to the transition probabilities, which can lead to an exhaustive understanding of the overall properties of the dataset when long random walks induced by the Markov matrix are considered \cite{fluid_diffusion2}. 
This scenario can be investigated in terms of a stochastic dynamical system where the transitions summarized in the Markov matrix can be described as results of a differential equation. 
This can lead to a global characterization of the system when integrated on a long term scale \cite{fluid_diffusion1,fluid_diffusion3}. 
Hence, the graph is considered as a realization of a dynamical process at equilibrium \cite{fluid_diffusion1}. 

This analogy is particularly interesting, since it enables a robust description of the data interactions with respect to noise perturbation, as well as a multiscale analysis of the considered dataset \cite{fluid_diffusion3}. 
This investigation relies once again on the transition probability proportional to the weight of the edge
connecting the two nodes. 
In particular, the inference mechanism is based on the  transition probability density $p(\textbf{x}(t+\epsilon) = \textbf{x}_j|\textbf{x}(t)=\textbf{x}_i)$ of finding the system at location $\textbf{x}_j$ at time
$t+\epsilon$, given an initial location $\textbf{x}_i$ at time $t$, where $\textbf{x}_i$ identifies the point in the $n$-dimensional feature space corresponding to the $i$-th sample, using the notation previously introduced in this Section \cite{fluid_diffusion1,fluid_diffusion2,fluid_randomwalk}. 

In this way, the analysis of the relationships and interactions among samples can be less affected by the density of the data and the local geometry of the dataset \cite{fluid_diffusion3}. 
Nevertheless, it is also true that the characterization of the dataset in terms of dynamic system analysis requires that the Markov matrix and the metric used to quantify the edge weight in the graph representation would fulfill a number of conditions. 
Specifically: 
\begin{itemize}
	\item the transition probabilities must only depend on the current state and 
	not on the past ones (first-order Markov chain). 
	In this way, since the graph is connected, the
	Markov chain is irreducible, that is, every state can be reached from any other state \cite{fluid_diffusion2,fluid_diffusion1};
	\item the Markov matrix must be characterized by a unique stationary distribution to allow the existence of the eigenvalues of the Markov matrix \cite{fluid_diffusion2,fluid_diffusion3}; 
	\item the Markov chain must be ergodic since the state space of the Markov chain associated with the matrix of the node transitions is finite and the corresponding random walk is aperiodic 
	\cite{fluid_diffusion3};
	\item the kernel $\eta(\textbf{x}_i, \textbf{x}_j)$ used to quantify the edge weight must capture the relationships between pair of samples in $\textbf{X}$, so it is not surprising that it must be 
	non-negative. Moreover, the function $\eta$ must be a rotation invariant kernel \cite{rotinvkernel}, so that is possible to retrieve the manifold structure regardless of the distribution of the samples \cite{fluid_diffusion3}.
\end{itemize} 

When these conditions are satisfied, it is possible to prove that the solution of the aforementioned problem satisfies the forward Fokker-Planck equation associated with the \textit{heat diffusion process} which can be written for the density $p(\textbf{x}(t+\epsilon) = \textbf{x}_j|\textbf{x}(t)=\textbf{x}_i)$ as follows \cite{fluid_diffusion1}:

\begin{equation}
\frac{\partial p}{\partial t} = \nabla \cdot (\nabla p + p\nabla U(\textbf{x})),
\label{eq_FokkerPlanck_heat}
\end{equation}
 
\noindent where $\nabla = [\frac{\partial}{\partial x_i}]_{i=1, \ldots, n}$, and the state $\textbf{x}(t) \in \mathbb{R}^n$ (i.e., each sample in the dataset) is a realization of the dynamical system that can be written as follows:

\begin{equation}
\dot{\textbf{x}} = - \nabla U(\textbf{x}) + \sqrt{2}\dot{\textbf{w}},
\label{eq_dynsyst_heat}
\end{equation}

\noindent where $\dot{\textbf{x}}$ and $\dot{\textbf{w}}$ identify the derivatives with respect to $t$ of $\textbf{x}$ and $\textbf{w}$, respectively.  
Moreover, $U$ is the free energy at
$\textbf{x}$ (which can be also called the potential at $\textbf{x}$), and $\textbf{w}$(t) is
an $n$-dimensional Brownian motion process. 
From a practical point of view, in this scenario the considered dataset $\textbf{X} = \{ \textbf{x}_i\}_{i=1, \ldots, N}$, $\textbf{x}_i \in \mathbb{R}^n$ is assumed to be sampled from the aforesaid dynamical system in
equilibrium \cite{fluid_diffusion1,fluid_diffusion3}. 

In general, the solution of (\ref{eq_FokkerPlanck_heat}) 
can be written in terms of an eigenfunction expansion of the Fokker-Planck operator \cite{fluid_diffusion1,fluid_diffusion3}. 
In low dimensions, it is possible to calculate approximations to
these eigenfunctions via numerical solutions of the relevant partial differential equations. In
high dimensions, however, this approach is in general infeasible and one typically resorts to
simulations of trajectories of (\ref{eq_dynsyst_heat}). In this case, there is a need to
employ statistical methods to analyze the simulated trajectories, identify the slow variables,
the meta-stable states, the reaction pathways connecting them and the mean transition times
between them \cite{fluid_diffusion1,fluid_diffusion2,fluid_diffusion3}.

In particular, for the analysis of (\ref{eq_dynsyst_heat}), a key role is played by the normalized graph Laplacian matrix, i.e., the matrix which $(i,j)$ element is set to $\sum_{k=\{1,\ldots,N\}\setminus i} \eta(\textbf{x}_i,\textbf{x}_k)$ if $i=j$, and $- \eta(\textbf{x}_i,\textbf{x}_j)$ otherwise \cite{fluid_diffusion1,fluid_diffusion3,fluid_diffusion2,fluid_randomwalk}. 
In fact, 
it is possible to prove that the eigenvalues and eigenfunctions
of the normalized graph Laplacian operator asymptotically  correspond to the Fokker-Planck equation with 
a potential $2U(\textbf{x})$ \cite{fluid_diffusion1}. 
The crucial role of the Laplacian operator is further highlighted by considering that under special conditions the Fokker-Planck equation and consequently its solution can be strongly simplified. 
Specifically, assuming that the considered domain is regularly sampled, 
the solution $p(\textbf{x}(t+\epsilon) = \textbf{x}_j|\textbf{x}(t)=\textbf{x}_i)$ reduces to the value of the $(i,j)$ element of the 
Laplacian matrix associated with the Markov matrix 
 \cite{fluid_diffusion1,fluid_diffusion3,fluid_diffusion2,fluid_randomwalk}. 
This approach has then enabled the main contributions to graph-based data analysis, from normalized cut ratio derivation, to the development of spectral clustering techniques, to the most recent strategies for graph-based deep learning \cite{fluid_laplacian,specclust3,GEOMDL,fluid_graphsignalproc}. 

At this point, it is worth noting that this approach allows us to obtain a thorough description of the data interactions 
since affine data would result in high values of correlation among samples  
\cite{fluid_diffusion3}. 
It is also true that the correlation is quantified by taking into account the characterization of the local geometry of the dataset,
which in turn relies on the choice of the kernel metric,  $\eta$. 
Nevertheless, a given kernel will grasp specific properties and characteristics of the data set, so that the definition of the $\eta$ function should be based on the application scenarios and analysis tasks under exam, such as those in \cite{Luxburg,COUILLET20,CommunityDetection_SC1,CommunityDetection_SC2,CommunityDetection_SC3,CommunityDetection_SC4}. 
Therefore, the kernel choice is critical in achieving an accurate and reliable characterization of the data interactions \cite{fluid_diffusion1,fluid_diffusion3,COUILLET16}.
However, some kernel functions could be preferred for their versatility (e.g., Gaussian kernel). 
It is therefore difficult to understand (and eventually compare) \textit{a priori} the actual effectiveness and validity of the metrics used to quantify the edge weights in the graph representation derived according to the aforesaid approach \cite{fluid_diffusion1,COUILLET16}. 
The distribution of the higher order moments of the feature statistics might affect the validity of the different $\eta$ functions, thus making the search for an optimal kernel to derive the graph representation of the data very difficult to target \cite{COUILLET16,fluid_diffusion3}. 

On the other hand, it is possible to analyze the Laplacian matrix associated with the Markov matrix in terms of its eigenvalues to assess the ability of the considered graph representation to capture information about the data relationships \cite{COUILLET16}. 
Specifically, when the graph Laplacian matrix is able to retrieve information on the underlying phenomena and processes captured by the considered observations, its eigenvalues tend to isolate from each other, leading to a spiked model of the graph Laplacian \cite{Luxburg,COUILLET16,CommunityDetection_Fortunato}. 
Hence, understanding how the data covariance distribution could affect the eigenvalue spectrum of the Laplacian matrix would provide key information for the assessment of the kernel metric effectiveness. 
To this aim, describing the Laplacian matrix characteristics in terms of random matrix theory can help \cite{COUILLET16}. 
In particular, it is possible to prove that a Laplacian matrix with isolated eigenvalues corresponding to thematic clusters (e.g., classes, communities) within the data generated under the heat diffusion model can be written as a random matrix generated according to a spiked model with the same eigenvalues \cite{COUILLET16,randommatrix1, randommatrix2, randommatrix3}. 
This allows the analytic investigation of the characteristics of the graph Laplacian, with special focus to its covariance \cite{COUILLET16}. 

In fact, it is possible to identify several conditions that the mean and covariance matrices associated with each thematic cluster in the dataset must show 
so that the Laplacian matrix can have isolated eigenvalues \cite{COUILLET16,randommatrix3,fluid_diffusion2}. 
These conditions can be quantified by taking into account a few metrics derived from the heat diffusion model set-up and the inter-class covariance matrices \cite{COUILLET16}.
Specifically, let us assume that the samples in the dataset $\textbf{X}$ can be associated with $k$ classes. 
It is therefore possible to compute for each class a length-$n$ vector $\textbf{m}_l = [m_{l_i}]_{i=1, \ldots, n}$  ($l \in \{1, \ldots, k\}$), where $m_{l_i} \in \mathbb{R}$ identifies the average value that the $i$-th record assumes across the samples belonging to class $l$. 
Each class can be analogously
 characterized by a non-negative definite covariance matrix $\textbf{C}_l \in \mathbb{R}^{n \times n}$ computed across the samples associated with the $l$-th class. 
Moreover, let $N_l$ be the population of the $l$-th class within the dataset, i.e., the amount of samples belonging to the $l$-th class: hence, $\sum_{l=1}^{k} N_l = N$. 
It is thus possible to write $\textbf{m}_l'= \textbf{m}_l - \bar{\textbf{m}} = \textbf{m}_l - \sum_{j=1}^{k} \frac{N_j}{N} \textbf{m}_j$. 
Analogously, 
$\textbf{C}_l'= \textbf{C}_l - \bar{\textbf{C}} = \textbf{C}_l - \sum_{j=1}^{k} \frac{N_j}{N} \textbf{C}_j$.  Finally, we can define $\textbf{T} = \{T_{ij}\}_{(i,j) \in \{1, \ldots, k\}^2} \in \mathbb{R}^{k \times k}$, where $T_{ij}= \frac{1}{n} \text{Tr}[\textbf{C}_i'\textbf{C}_j']$, and $\textbf{t} = [t_l = \frac{1}{\sqrt{n}} \text{Tr}[\textbf{C}_l']]_{l=1, \ldots, k}$, $t_l \in \mathbb{R}$. 
 

With this in mind, it is possible to prove \cite{COUILLET16} that the Laplacian matrix would show isolated eigenvalues if 
the inter-class mean and covariance matrices must show as much energy (modeled by the $\textbf{m}$, $\textbf{T}$ and $\textbf{t}$ factors) as possible when the number of samples and/or features to be considered in the dataset increase, assuming that the first derivatives of the kernel function $\eta$ would not tend to 0 when $N \rightarrow +\infty$ and/or $n \rightarrow +\infty$ \cite{COUILLET16,randommatrix4,randommatrix1}. 
In other terms, one or more of the following conditions must hold: 

\begin{eqnarray}
||\textbf{m}_l'|| & \xrightarrow{N \rightarrow +\infty, n \rightarrow +\infty}&  +\infty \nonumber \\
t_l & \xrightarrow{N \rightarrow +\infty, n \rightarrow +\infty}&  +\infty \label{eq_isoleigen_cond}\\
T_{lj} & \xrightarrow{N \rightarrow +\infty, n \rightarrow +\infty}&  +\infty,\nonumber
\end{eqnarray}

\noindent for some $l,j \in \{1, \ldots, k\}$ \cite{COUILLET16}. 

These conditions are sufficient to guarantee that the graph Laplacian matrix derived under the heat diffusion model can show isolated eigenvalues. In other terms, it
 is able to summarize the main properties of the dataset, i.e., to lead to a thorough characterization of the interactions and relationships among the samples \cite{COUILLET16}. 
Nonetheless, when the data are sparse, these conditions might not be matched \cite{COUILLET20}. 
In this case, the graph Laplacian matrix should be regularized to ensure that the energy of the higher order statistics could be concentrated, thus avoiding the aforementioned vanishing phenomenon that could jeopardize the presence of isolated eigenvalues \cite{COUILLET18,COUILLET20,COUILLET16}. 
On the other hand, it is possible to show that this process can be valid only when the number of thematic clusters the considered records are meant to describe is very low (e.g., two) \cite{COUILLET20}.

\subsection{From heat to fluid: a new graph representation}
\label{sec_methmot}

\subsubsection{Motivations of a new graph representation}
\label{sec_meth_motnew}

As previously mentioned, investigating the graph structures induced by the datasets by exploiting the heat propagation analogy in terms of information inference has been proven to be effective and efficient for a wide range of applications and methodological research instances. 
Nonetheless, these architectures might fail in addressing several data analysis issues that can occur when dealing with multimodal records, especially in operational scenarios \cite{multimod1,capacity}. 

Specifically, we can summarize the major limitations of the classic heat diffusion model for graph investigation in the following points \cite{multimod1,GEOMDL,fluid_graphsignalproc}:
\begin{itemize}
	\item  \textbf{L1 - adaptivity}: The learning system would have to deal with records showing multiple resolutions (either in time, space, metrical units). Moreover, noise (i.e., any undesired effect) might affect attributes/features/classes in different ways across the whole dataset, as well as in intra- and inter-class relationships. Hence, a single data model (in terms of propagation mechanisms, label assignment, similarity computation) might not be adequate for obtaining accurate and solid characterization of the records;
	\item \textbf{L2 - sparsity/missing data}: Not all the attributes of each sample might be relevant (by corruption, or by linear correlation). Using all the records to compute the similarity among samples might lead to dramatic degradation and/or bias of the analysis.  
	Further, the complexity of the data to be investigated might make classic impainting/interpolation techniques inadequate, thus jeopardizing the validity of the outcomes;
	\item \textbf{L3 - data mismatch/unbalance}: 
	the distribution of the thematic clusters in the dataset might be strongly unbalanced, and/or the training set might not contain samples associated with all the classes actually present in the dataset. 
	Thus, relying on a uniform statistical distribution as the source of the samples to be investigated might be an assumption too hard to match.
\end{itemize}

These issues would result in strong limitations of the data analysis schemes used to characterize the records.
They would in fact limit the full exploitation of the available training set, 
either in terms of information extraction or context-aware inference. 
Moreover, the aforementioned points would reverberate in terms of degradation of confidence and precision of the analysis, as well as restriction of the ability to fully explain and interpret the records under exam \cite{capacity,multimod2,multimod1,multimod7}.  

With this in mind, the graph representation based on the heat diffusion model might sound intuitively inadequate to deal with all these limitations induced by modern data analysis. 
Nevertheless, several approximate solutions have been proposed in technical literature, in order to mitigate the effect of these conditions whilst maintaining the data analysis steps compliant to the main assumptions presented in Section \ref{sec_meth_heat} \cite{GEOMDL,fluid_graphDL1,fluid_graphDL2,fluid_graphsignalproc,fluid_structeqmodel1}. 
Thus, it is useful to provide a practical example to show how the properties in the previous Section that motivate the use of heat diffusion model are not matched when multimodal datasets are considered. 
In particular, we can focus on the conditions in (\ref{eq_isoleigen_cond}), as they must be fulfilled for the classic graph representation to be adequate for information extraction from the considered datasets. 

To this aim, we report in 
Appendix \ref{app_motiv} 
an analysis we conducted on  a multimodal dataset that is considered as a benchmark in the remote sensing community \cite{Trentodata}. 
Investigating this dataset from a theoretical, methodological, and experimental perspective supports the need for a novel graph representation so that the major limitations of the classic heat diffusion model could be addressed. 
In particular, we have shown how the necessary conditions for the heat diffusion model to reliably characterize the data interactions (i.e., the properties in (\ref{eq_isoleigen_cond})) might not hold. 

For these reasons, we propose using a fluid diffusion model to derive a new graph representation, that could be more flexible and versatile to address the modern data analysis needs and limitations. 
Our findings are reported in the following subsection. 

\subsubsection{Proposed approach}
\label{sec_fluid_repr}
We need to define the graph topology and the diffusion model to be applied in order to take into account the data analysis needs mentioned in the previous subsection. 
%
To this aim, the definition of the process underlying the diffusion mechanisms across the graph should reflect a higher flexibility of the system, so to address the relevance of the features and the modeling of a confidence score for the propagation structure \cite{fluid_FokkerPlanck5,fluid_FokkerPlanck6}. 
Hence, the system in (\ref{eq_dynsyst_heat}) should be replaced by a more complex stochastic differential model, such as follows:

\begin{equation}
\dot{\textbf{x}} = \textbf{a}(\textbf{x}) + \textbf{B}(\textbf{x})\dot{\textbf{w}},
\label{eq_dynsyst_fluid}
\end{equation}

\noindent where $\textbf{w}(t)$ is a $N$-dimensional Wiener process, 
$\textbf{a}(\textbf{x})$ is a length-$n$ vector, whilst 
$\textbf{B}(\textbf{x})$ identifies a $n \times N$ matrix  \cite{fluid_FokkerPlanck3,fluid_FokkerPlanck5,fluid_FokkerPlanck6,fluid_FokkerPlanck4}.  

In a fluid diffusion system model, \textbf{a} regulates the \textit{flow rate}, i.e., the velocity by which the diffusion can take place from one node to another in the system \cite{fluid_FokkerPlanck5,fluid_FokkerPlanck2}. 
In general, it depends on the characteristics of the $\textbf{x}$ state, as well as on the local \textit{conductivity} properties of the fluid diffusion at local scale, and on the \textit{diffusivity} properties of the model at global scale \cite{fluid_FokkerPlanck2,fluid_FokkerPlanck3}. 
In particular, the $\textbf{B}(\textbf{x})$ matrix summarizes the rate by which the diffusion can take place across the features of the considered system \cite{fluid_FokkerPlanck5,fluid_FokkerPlanck2}.

In more detail, the \textit{conductivity} properties (typically summarized by a $N \times N \times n$ tensor $\cal K$) model the ease with which the fluid diffusion can take place from one node to another in the system \cite{fluid_FokkerPlanck2,fluid_FokkerPlanck1,fluid_FokkerPlanck3}. 
In our analogy, $\cal K$ would model the relevance of each single feature in computing the weight of each edge in the graph, hence quantifying the information diffusion at \textit{local} scale in the dataset. 
On the other hand, the aforementioned \textit{diffusivity} is expected to model the ability of each feature to permit diffusion across the nodes in the diffusion system. 
As such, it represents a \textit{global} quantity, that is summarized by the matrix $\tilde{\textbf{B}}(\textbf{x}) = \frac{1}{2} \textbf{B}(\textbf{x})\textbf{B}^T(\textbf{x})$. 
In our analogy, $\tilde{\textbf{B}}(\textbf{x})$ quantifies a contextual weight,
aiming to model the ability of each edge in the graph
to convey the information, and is hence linked with a notion of
confidence that can be associated with local portions of
the whole manifold \cite{fluid_FokkerPlanck4,fluid_FokkerPlanck5,fluid_FokkerPlanck2}. 

With this in mind, the exchange of information across nodes (modeled by $\textbf{a}$) would depend on the convection process ruled by the conductivity tensor ${\cal K}$ and the diffusion rate, computed as a derivative over the features of the diffusivity matrix $\tilde{\textbf{B}}$ \cite{fluid_FokkerPlanck8}. 
Thus, the \textbf{a} vector is typically written as follows: 

\begin{equation}
\textbf{a}(\textbf{x}) = \textbf{v}(\textbf{x}, {\cal K}) + \nabla \tilde{\textbf{B}}(\textbf{x}),   
\label{eq_flowrate}
\end{equation}

\noindent where $\nabla = [\frac{\partial}{\partial x_i}]_{i=1, \ldots, n}$ as in (\ref{eq_dynsyst_heat}), and $\textbf{v}$ is the \textit{fluid transport velocity}, i.e., a function of the conductivity and the state \textbf{x}, ultimately modelling a dynamic weight based on ${\cal K}$ for the different components of \textbf{x} \cite{fluid_FokkerPlanck1,fluid_FokkerPlanck2}.

Analogously to the case reported in Section \ref{sec_meth_heat}, we are interested in deriving the expression of the transition probability density for the system in (\ref{eq_dynsyst_fluid}). 
To achieve this goal, we can investigate the properties of this stochastic differential equation  by taking advantage of the results provided by the It\^{o}'s lemma \cite{fluid_FokkerPlanck7} to the diffusion process in the form of (\ref{eq_dynsyst_fluid}). 
This approach first introduces an arbitrary twice-differentiable scalar function $g(\textbf{x})$ where $\textbf{x}$ is defined as in (\ref{eq_dynsyst_fluid}). 
Then, it considers the expansion in Taylor series of $g(\textbf{x})$. 
Specifically, considering that $\textbf{w}$ is by definition a Wiener process, the Taylor expansion of $g(\textbf{x})$ can be truncated at the second order. 
As such, it is possible to write as follows \cite{fluid_FokkerPlanck7}: 
%

\begin{eqnarray}
dg(\textbf{x}) = &  \{ \textbf{a}(\textbf{x}) \cdot \nabla g(\textbf{x}) \\
+ &  \frac{1}{2} \sum_{i,j=1}^{n} \tilde{B}_{ij} \partial x_i \partial x_j g(\textbf{x}) \}  dt \nonumber \\
+ &  \nabla g(\textbf{x}) \cdot \textbf{B}(\textbf{x}) \textbf{w}, \nonumber
\label{eq_Ito1_fluid}
\end{eqnarray}

\noindent where $\tilde{B}_{ij} = \frac{1}{2}[\textbf{B}(\textbf{x}) \textbf{B}^T(\textbf{x})]_{ij}$. 
At this point, recalling that expected value of an It\^{o} integral is zero \cite{fluid_FokkerPlanck7} and that $\textbf{w}$ identifies a Wiener process with $\frac{d}{dt}\mathbb{E}[\textbf{w}] = 0$, it is possible to write as follows: 

\begin{equation}
\frac{d}{dt}\mathbb{E}[g(\textbf{x})] = \mathbb{E}\left[ \textbf{a}(\textbf{x}) \cdot \nabla g(\textbf{x}) +  \sum_{i,j=1}^{n} \tilde{B}_{ij} \partial x_i \partial x_j g(\textbf{x}) \right]. 
\label{eq_Ito2_fluid}
\end{equation}
 
\noindent 
The term $\frac{d}{dt}\mathbb{E}[g(\textbf{x})]$ can be computed taking into account the conditional probability density of a particle starting at $(\textbf{x}_0,t_0)$, i.e., $p(\textbf{x},t) = p(\textbf{x},t| \textbf{x}_0,t_0)$ \cite{fluid_FokkerPlanck7,fluid_FokkerPlanck5}.  
Specifically, we can write (\ref{eq_Ito2_fluid}) as follows:

\begin{eqnarray}
\frac{d}{dt}\mathbb{E}[g(\textbf{x})] = &  \frac{d}{dt} \int g(\textbf{x}) p(\textbf{x},t) d\textbf{x} = \int g(\textbf{x}) \frac{\partial p(\textbf{x},t)}{\partial t} d\textbf{x} \nonumber \\
= &  \int \textbf{a}(\textbf{x}) \cdot \nabla g(\textbf{x}) p(\textbf{x},t) d\textbf{x} \\ 
+ & \sum_{i,j=1}^{n} \int \tilde{B}_{ij} \partial x_i \partial x_j g(\textbf{x})p(\textbf{x},t) d\textbf{x}.  \nonumber
\label{eq_Ito3_fluid}
\end{eqnarray}

\noindent 
Integrating by parts, this equation can be rewritten as follows \cite{fluid_FokkerPlanck7}: 

\begin{equation}
\int \left[ \frac{\partial p(\textbf{x},t)}{\partial t} + \nabla \cdot \left[\textbf{a}(\textbf{x})p(\textbf{x},t) \right] - \beta'(\textbf{x},t) \right] g(\textbf{x}) d\textbf{x} = 0,
\label{eq_Ito4_fluid}
\end{equation}

\noindent where $\beta'(\textbf{x},t) =  \sum_{i,j=1}^{n} \partial x_i \partial x_j (\tilde{B}_{ij} p(\textbf{x},t))$. 
Then, since the $g$ function is arbitrary by construction, the aforesaid equation is satisfied when the term inside the square brackets is null. 
With this in mind and expanding the $\beta'$ term, it is possible to write as follows (the details of the algebraic steps are detailed in 
Appendix \ref{app_deriv1}): 

\begin{eqnarray}
\frac{\partial p(\textbf{x},t)}{\partial t}  + &  \nabla \cdot \left[ \textbf{a}(\textbf{x})p(\textbf{x},t) \right] - \left[ \nabla \cdot (\nabla \tilde{\textbf{B}}(\textbf{x})) \right] p(\textbf{x},t) \nonumber \\
= & \left[ \nabla \tilde{\textbf{B}}(\textbf{x}) \right] \cdot \nabla p(\textbf{x},t) - \nabla \cdot \left[ \tilde{\textbf{B}}(\textbf{x}) \nabla p(\textbf{x},t)\right]
\label{eq_Ito5_fluid}
\end{eqnarray}

%
%
At this point, 
using the representation in (\ref{eq_flowrate}),  and considering the linearity of the divergence operator and its product rule \cite{fluid_FokkerPlanck7}, it is now possible to write the diffusion equation for this system as follows:

\begin{eqnarray}
\frac{\partial p(\textbf{x},t)}{\partial t} = & -&  \nabla \cdot \left \{ \left[\textbf{a}(\textbf{x}) - \nabla \tilde{\textbf{B}}(\textbf{x})\right]p(\textbf{x},t) \right \} \nonumber \\
&+& \nabla \cdot \tilde{\textbf{B}}(\textbf{x})\nabla p(\textbf{x},t).
\label{eq_FokkerPlanck_fluid}
\end{eqnarray}

It is possible to recognize in this expression the Fokker-Planck equation for \textit{fluid diffusion in porous media} 
 \cite{fluid_FokkerPlanck1,fluid_FokkerPlanck2}. 
%
Analogously to the heat diffusion analysis, the solution of (\ref{eq_FokkerPlanck_fluid}) can be retrieved by eigenanalysis of the Fokker-Planck operator.
Furthermore, the asymptotic analysis of the trajectories of the system in (\ref{eq_dynsyst_fluid}) can help in obtaining a thorough characterization of its solution \cite{fluid_FokkerPlanck1}. 

It is indeed possible to analyze the geometry of the dataset by investigating the data by means of an approach based on Markov chain scheme \cite{fluid_diffusion1}. 
In fact, it is possible to characterize the transition probability density  
$p(\textbf{x}(t+\epsilon) = \textbf{x}_j|\textbf{x}(t)=\textbf{x}_i)$ by using a time domain random walk approach \cite{Fluid_TDRW1,Fluid_TDRW2}. 
This scheme achieves the characterization of the whole system by exploiting the adjacency of the nodes. 
Specifically, it aims to solve a Green function problem derived from (\ref{eq_dynsyst_fluid}) by imposing initial conditions and absorbing boundary conditions to the diffusion system centered on the node $i$ \cite{Fluid_TDRW1}. 

This strategy aims to determine the transition probability density $p(\textbf{x}(t+\epsilon) = \textbf{x}_j|\textbf{x}(t)=\textbf{x}_i)$ by means of the first arrival time density $\phi_{ij}$ at the boundary between nodes $i$ and $j$, that would denote the joint probability of the transition to occur from node $i$ to node $j$ \cite{Fluid_TDRW1}. 
Specifically, it is possible to write this transition probability density as follows: 

\begin{equation}
p(\textbf{x}(t+\epsilon) = \textbf{x}_j|\textbf{x}(t)=\textbf{x}_i) = p_{ij} = \int_0^{+\infty} \phi_{ij} dt
\label{eq_fluid_trans_prob_ij}
\end{equation}

The details of the time domain random walk strategy are detailed in Appendix \ref{app_deriv2}. 
By projecting the solution of the Green function problem in a Laplace space, we can write $p_{ij}$ as follows \cite{Fluid_TDRW1,Fluid_TDRW2}:

\begin{equation}
p_{ij} = \frac{|v_{+}^\dagger| \exp\left[v_{+}^\dagger \right] \csch \left[|v_{+}^\dagger|\right]}{\sum_{u \in \{+,-\}} |v_{u}^\dagger| \exp\left[\tilde{u}\cdot v_{u}^\dagger \right] \csch \left[|v_{u}^\dagger|\right]},
\label{eq_fluid_trans_prob_pij}
\end{equation}

\noindent where $\csch[z] = 1/\sinh[z] = 2/(\exp[z] - \exp[-z])$. Moreover, 
$\tilde{u}$ is set to 1 when $u=+$, whilst $\tilde{u}=-1$ if $u$ is -. Finally, $v_{\pm}^\dagger = v_{\pm}/2\tilde{B}_\pm$, being: 

\begin{eqnarray}
v_+ & = & v_{ij}, \\
v_- & = & \sum_{m \in {\cal N}(i) \setminus j} \frac{v_{im}}{|{\cal N}(i)|- 1}, \nonumber \\
\tilde{B}_+ & = & \tilde{B}_{ij}, \nonumber \\
\tilde{B}_- & = & \sum_{m \in {\cal N}(i) \setminus j} \frac{\tilde{B}_{im}}{|{\cal N}(i)|- 1}, \nonumber
\label{eq_v_B_pm}
\end{eqnarray}

%

\noindent where ${\cal N}(i)$ identifies the neighborhood of node $i$, i.e., the set of nodes adjacent to node $i$.  
Finally, $\tilde{B}_{ij} \in [0,1]$ is the $(i,j)$-th element of the matrix $\tilde{\textbf{B}}$, whereas $v_{ij}$ is the transport velocity between node $i$ and node $j$ as in (\ref{eq_flowrate}). 
This quantity is typically computed as $v_{ij} = - || {\cal K}_{ij:}^T \odot (\textbf{x}_i - \textbf{x}_j) ||_2 \in [0,1]$, where ${\cal K}_{ij:}$ is the length-$n$ row vector collecting the third dimension elements of the conductivity tensor $\cal K$ on the $(i,j)$ coordinates and $\odot$ is the Hadamard product  \cite{fluid_FokkerPlanck1,fluid_FokkerPlanck2,fluid_FokkerPlanck3,fluid_FokkerPlanck4, fluid_FokkerPlanck5, fluid_FokkerPlanck6, Fluid_TDRW1, Fluid_TDRW2}.

Hence, it is possible to define a Markov matrix $\textbf{Q} = \{Q_{ij}\}_{(i,j) \in \{1, \ldots, N\}^2}$ that summarizes the edge weights of the new graph representation according to the transition probabilities in (\ref{eq_fluid_trans_prob_pij}), i.e., $Q_{ij}= p_{ij}/\sum_{i} p_{ij}$ \cite{fluid_diffusion1,fluid_FokkerPlanck6,fluid_diffusion3}.  
The matrix $\textbf{Q}$ can be explored and used to address several tasks in multimodal data analysis and to improve the information extraction from the considered datasets. 
In particular, the properties of the tensor ${\cal K}$ and of the quantities in (\ref{eq_fluid_trans_prob_pij}) and (\ref{eq_v_B_pm}) make the proposed graph structure able to address the main issues that affect modern data analysis, especially with respect to the reliability of the data to be processed by automatic learning techniques. 
In fact, the proposed graph structure is:
\begin{itemize}
	\item  \textit{adaptive} (\textbf{L1}): the distance between two samples $i$ and $j$ depends only on the features that are relevant to determine the similarity between these samples. This information is stored in the ${\cal K}_{ij:}$ vector. Also, if contextual information is available (e.g., semantic knowledge provided by a training set on the categories to which samples $i$ and $j$ are respectively assigned to), the similarity can be made more robust by adjusting the term $\tilde{B}_{ij}$;
	\item \textit{able to address sparse and missing data} (\textbf{L2}): to avoid inducing bias or polarize the generalization ability, the proposed approach allows to set the elements of ${\cal K}$ and $\tilde{\textbf{B}}$ to 0. In this way, the similarity between samples are computed taking into account only the significant features in the dataset, hence avoiding to add spurious terms to the computation of the distance between samples. 
	\item \textit{able to address data imbalance} (\textbf{L3}): the stratification in the data population with respect to their semantic categories is addressed by means of the $\tilde{B}_{ij}$ terms. In this way, we can incorporate the semantic information on the samples directly in the graph construction, hence reducing the impact of data imbalance. Moreover, the adaptive nature of ${\cal K}$ allows to incorporate information about the data distribution, so to avoid to rely on a uniform statistical distribution as the source of the samples to be investigated.
\end{itemize}

These points are proven by the experimental results provided in the next Sections, and are further supported by the definition of the eigenvalues and eigenvectors associated with the $\textbf{Q}$ matrix. 
\color{black}
The eigenanalysis of the 
Laplacian matrix associated with $\textbf{Q}$ (whose $(i,j)$ element can be defined as $\sum_{k=\{1,\ldots,N\}\setminus i}Q_{ik}$ if $i=j$, and as $- Q_{ij}$ otherwise) can be directly connected to the solution of the system in (\ref{eq_FokkerPlanck_fluid}). 
In fact, in general the solution of the fluid diffusion equation can be written in terms of the eigenfunction expansion, i.e., $p(\textbf{x},t)$ can be expressed as follows:

\begin{equation}
p(\textbf{x},t) = \sum_{i=0}^{+ \infty} \omega_i \exp[-\lambda_i t] \varphi_i(\textbf{x}), 
\label{eq_FokkerPlanck_solution} 
\end{equation}

\noindent where $\lambda_i$ are the sorted eigenvalues of the fluid Fokker-Planck operator (with $\lambda_0 =0$), 
$\varphi_i$ are
their corresponding eigenfunctions, and the coefficients $\omega_i$ depend on the initial conditions \cite{fluid_diffusion1,fluid_FokkerPlanck6,fluid_diffusion3}. 

It is worth noting that numerical approximations of these eigenfunctions can be computed  
when the considered data are characterized by a low amount of records (e.g., three). 
On the other hand, when high dimensional data are considered - such as the multimodal data we are considering in this work - it is not possible to use numerical solutions to solve this equation. 
The only valid approach would be to simulate the trajectories of the stochastic differential model in (\ref{eq_dynsyst_fluid}), which implies the use of statistical methods to analyze the simulated results and explore the validity of low and high frequency trends, as well as the mean transition times among them \cite{fluid_FokkerPlanck5,fluid_FokkerPlanck6,fluid_diffusion1}.

It has been proven that the solution of (\ref{eq_FokkerPlanck_solution}) can be described by reduced set of $\kappa$ eigenfunctions, which can carry significant information on the density and geometry of the data under exam \cite{fluid_FokkerPlanck1,fluid_diffusion1,fluid_FokkerPlanck6}. 
To obtain a reliable characterization of the relevant eigenvectors and eigenvalues, 
it is useful to explore the asymptotic behavior of the diffusion process in the probability space. 
This analysis shows that the eigenvalues of the matrix $\textbf{Q}$  asymptotically 
correspond to the relevant $\kappa$ eigenfunctions that can be used to achieve a solid understanding of the solution in (\ref{eq_FokkerPlanck_solution}), as previously mentioned \cite{fluid_FokkerPlanck2,fluid_FokkerPlanck1,fluid_FokkerPlanck5,fluid_FokkerPlanck6,fluid_diffusion1,fluid_diffusion3}. 
This result is extremely interesting, and it summarizes the key-role that the matrix $\textbf{Q}$ can have in providing a  thorough and reliable understanding of the properties underlying the graph topology induced by the considered datasets, as well as the information propagation mechanisms. 

In Appendix \ref{app_Qmat}, 
we provide an example to visualize how the proposed definition of the \textbf{Q} matrix could improve the characterization of data interactions with respect to the classic graph representation based on heat diffusion mechanism.

In this work, we focus our attention on the investigation of $\textbf{Q}$ in order to learn the structure of the data under exam, and to enable an effective functional analysis of the records, with special focus to multimodal data analysis. 
The next Section summarizes the main steps of the proposed method for fluid community detection.

\section{Fluid community detection}
\label{sec_fluidCD}

\subsection{Background and related works} 
\label{sec_methback}

Several 
criteria at global and local scale can be used to identify communities in graphs. 
Since in this work we focus our attention towards the detection of communities that are separated, we report in this Section an overview of non-overlapping community detection methods. 
Moreover, we summarize the main categories community detection algorithms can be grouped in, according to the strategy they employ \cite{CommunityDetection_Fortunato}. 

In particular, it is possible to categorize these methods in seven main groups: 1) graph partitioning; 2) hierarchical clustering; 3) partitional clustering; 4) spectral clustering; 5) dynamic community detection; 6) statistical inference-based community detection; and 7) hybrid methods. 
We report in \ref{app_review} 
a brief overview of these algorithms.  
The methods 
are typically designed to address problems that could be described in a monovariate data analysis system \cite{CommunityDetection_Fortunato,Fluid_multimodML, Luxburg}. 
In particular, these architectures are developed at theoretical level to address community detection problems when a single source of information is used to generate the data to be analyzed. 
Nevertheless, multimodal community detection is typically addressed by extending these approaches to records acquired by multiple modalities \cite{multimod1,multimod7,multimod2,Fluid_multimodML}. 
In particular, the similarity between samples (either in terms of edge betweenness, modularity, Euclidean distance, geodesic metric) is computed along all the features available \cite{GEOMDL,graphclust_schaub,fluid_graphsignalproc}. 
In other terms, the aforesaid methods are applied to datasets where the records acquired by the diverse modalities are stacked and vectorized, so that each sample could be considered as a point in an extended multidimensional feature space. 

This approach is very popular within the scientific community, because of its high degree of implementability. 
However, this does not always reflect in good performance in terms of community detection, especially in operational scenarios \cite{multimod1,fluid_graphDL2,fluid_graphsignalproc,capacity,Fluid_multimodML}. 
Hence, directly applying these architectures to multimodal data analysis might lead to strong limitations of the multimodal community detection performance.  
In other terms, 
successful community detection is achieved only when datasets characterized by low diversity, low sparsity, high reliability, and low variability can be found across the considered records (see for instance \cite{CommunityDetection_Bertozzi1, CommunityDetection_Bertozzi2, CommunityDetection_Bertozzi3, CommunityDetection_Bertozzi4}). 
As these characteristics are hard to be found in multimodal datasets, especially when addressing operational scenarios \cite{multimod1,Fluid_multimodML}.  
Hence, this makes this research avenue an open field for a successful development of multimodal data analysis methods \cite{Fluid_multimodML,CommunityDetection_DL}. 

Nevertheless, it is also true in recent years methods designed to address multimodal data analysis have been proposed. 
One first approach relies on the design of $n$-partite graphs, where different kind of nodes are used to represent the diverse modalities under exam \cite{CommunityDetection_multi1, CommunityDetection_multi2,CommunityDetection_multi3}. 
On these graph structures, partitional clustering techniques are applied to retrieve the community structures hidden within the data. 
These methods are typically showing pretty high computational complexity, such that it is proven that they might achieve successful performance when the trade-off between diversity and thematic clusters to be identified is good (i.e., when few modalities with numerous communities are considered, or when multiple modalities and a small amount of communities are taken into account). 
This is a major factor that must be considered when employing these techniques \cite{Fluid_multimodML,CommunityDetection_Fortunato,CommunityDetection_DL}. 

Similar results can be registered when the records acquired by diverse modalities are separately processed, to be then fused at a later stage \cite{CommunityDetection_align1,CommunityDetection_align2,CommunityDetection_align3}. 
In this case, methods retrieved from graph partitioning, partitional clustering, and/or hybrid approaches are used to analyze the different sources of information. 
Then, the obtained information is integrated according to optimization criteria designed to maximize the alignment between modalities and therefore identify communities are are more homogeneous across the diverse sources of information \cite{CommunityDetection_align2}. 
Although this approach can be performed with pretty low latency (especially when high performance computing platforms are available), it can hardly be generalized for multimodal datasets where diverse records, characterized by various statistical distributions, high variability and sparsity ae considered, limiting the range of applications that could actually benefit of this strategy. 

Recently, methods relying on the multiview data analysis approach have been introduced for community detection \cite{CommunityDetection_view1,CommunityDetection_view2}. 
In this case, the different $n$ modalities are assumed to generate $n$ graphs that are then investigated by means of spectral clustering techniques. 
Then, an optimization process based on normalized cut approach is performed, in order to identify the most informative clusters across the whole dataset. 
This approach typically shows low computational complexity. 
However, it is also true that it requires the modalities to be as less as diverse as possible, so that the joint cut across the graphs can be accurately performed \cite{CommunityDetection_view1}. 
Moreover, the multiple graphs are expected to show homogeneous characteristics imposed by the communities underneath. This is a pretty strong requirement, since in multimodal data analysis not all the features are typically reliable, significant and/or informative at the same level \cite{CommunityDetection_view2}. 
As such, using this scheme to integrate heterogeneous sources of information at large scale might be cumbersome.  


\subsection{Proposed approach}
\label{secmeth}

Taking into account the definition of the $\textbf{Q}$ matrix  as a result of the fluid diffusion model (as introduced in Section \ref{sec_fluid_repr}), there are several properties that can be particularly interesting to address the community detection task in an accurate and efficient way. 
In this work, we take advantage of the aforementioned characteristics of the eigenanalysis of the flow velocity matrix to identify the most informative clusters within the considered multimodal dataset.
As such, the approach we propose in this paper could fall within the spectral clustering category of community detection mentioned in Section \ref{sec_methback}. 

This choice helps us in achieving accurate and reliable understanding of the data interactions in closed form and with rigorous convergence, while guaranteeing a simple implementation and high efficiency of the unsupervised community detection approach. 
It is also worth noting that the focus of this paper is on unsupervised analysis of the functional relationships among samples, i.e., no contextual information (either in shape of side information, or \textit{a priori} knowledge, nor semantic knowledge) could be taken into account to achieve a fully data drive investigation. 
As such, the diffusivity term $\tilde{\textbf{B}}$ in (\ref{eq_flowrate}) can be then set to the identity matrix throughout the following Sections.
   
\begin{figure}[htb]
	\centering
	\includegraphics[width=1\columnwidth]{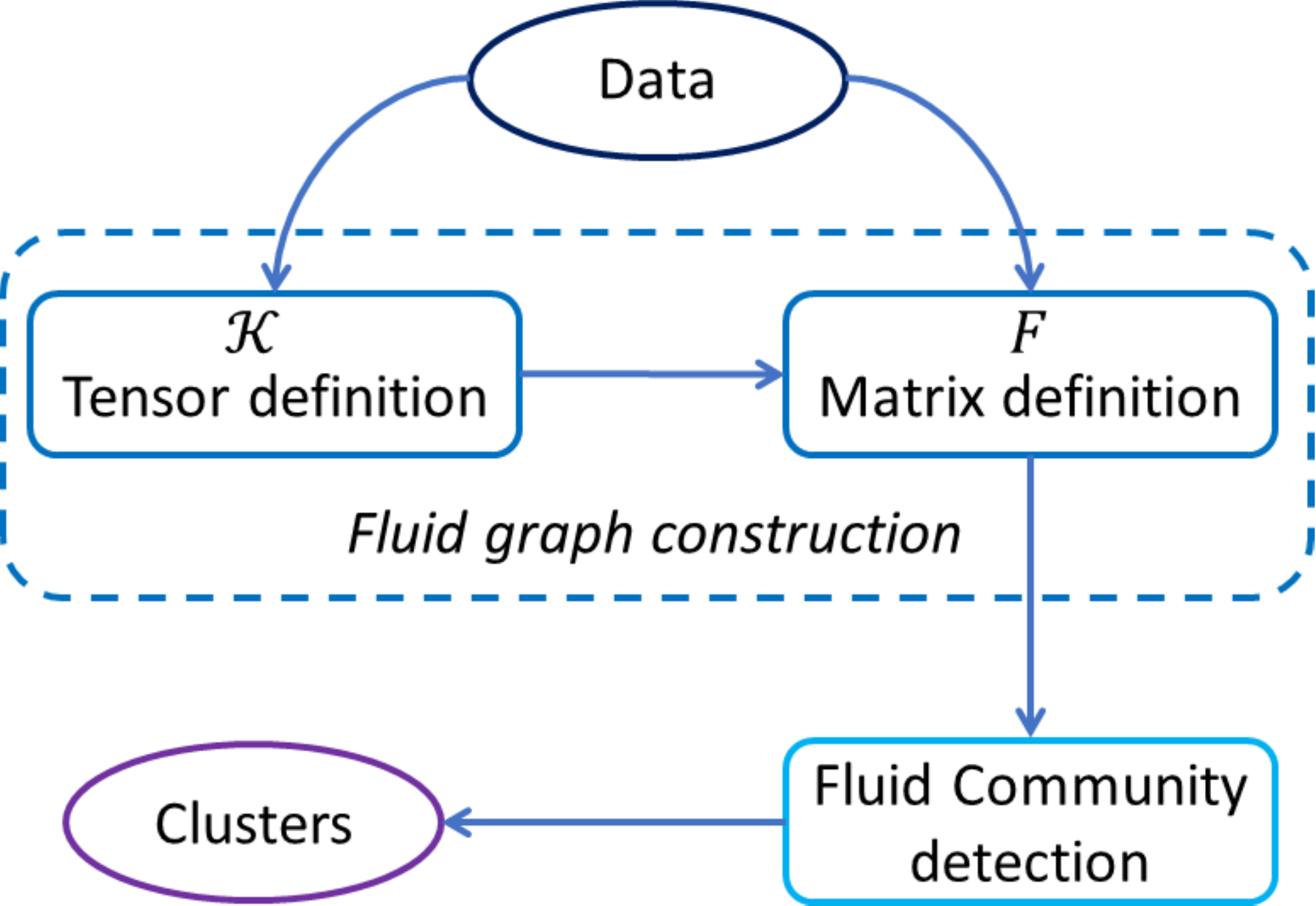} 
	\caption{Flowchart of the main steps of the proposed unsupervised approach for community detection based on fluid graph representation.}
	\label{fig_flowchart}
\end{figure}

The main steps of the proposed approach are summarized in Figure \ref{fig_flowchart}, and detailed in the following Sections. 

\subsubsection{Definition of the \textbf{Q} matrix}
\label{sec_meth_ADR}
In order to provide a thorough investigation of the complex relationships hidden within the records in multimodal datasets according to the fluid graph representation previously introduced, we first need to define the permeability tensor ${\cal K}$ in (\ref{eq_flowrate}). 
To this aim, it would be instrumental to investigate the significance of the features associated with each sample in the considered dataset. 
This goal can be achieved by using several strategies for feature selection (e.g., \cite{CommunityDetection_FS4,CommunityDetection_FS1, CommunityDetection_FS2,CommunityDetection_FS3}). 

In this work, we propose to address this task by exploring the relevance of the features at global (i.e., across all modalities) and local (i.e., across samples for each feature) scale. 
Following the successful approach proposed in \cite{ensembleADR}, we quantify the multiscale significance of the features by using information theory-based metrics. 
Specifically, we consider to measure the degree of redundancy and intercorrelation between features across the whole dataset by employing mutual information \cite{mackaybook,MUTINF6}. 
This choice allows us to assess the redundancy and dependence among features we could record across the dataset. 
In fact, mutual information quantifies the shared information between two random variables \cite{MUTINF6}. 
This is especially relevant when complex datasets, which lead to fully connected graph representations, are taken into account \cite{ELdata_pruning5,mackaybook}. 

In other terms, let us consider a dataset $\textbf{X}$ that consists of $N$ samples and $n$ features, i.e., $\textbf{X} = \{\textbf{x}_i^T\}_{i=1, \ldots, N}$, $\textbf{x}_i  = [x_{ij}]_{j=1, \ldots, n}$, that induces a graph ${\cal G} =({\cal V}, {\cal E})$ where ${\cal V}$ and  ${\cal E}$ identify the node and edge sets, respectively. 
Moreover, the $i$-th node $v_i$ is associated with the $i$-th sample $\textbf{x}_i$.
Then, we can write the mutual information between two features $n_1$ and $n_2$ (where $n_j \in \{1, \ldots, n\}$) as follows: 

\begin{eqnarray}
w_{n_1 n_2}^{\text{MI}} =& \sum_{i = 1}^N\sum_{j = 1}^N p(x_{in_1},x_{jn_2})\log \frac{p(x_{in_1},x_{jn_2})}{p(x_{in_1})p(x_{jn_2})}, \label{eq_GKMI2}	
\end{eqnarray}
 
\noindent where $p(y,z)$ is the joint density function of $y$ and $z$, and $p(z)$ is marginal of $z$. 
It is worth to note that, according to this definition, large values of $w_{n_1 n_2}^{\text{MI}}$
imply redundancy in information. On the other hand, low values of $w_{n_1 n_2}^{\text{MI}}$
imply synergy (novelty) \cite{mackaybook}. 

On the other hand, it is important to evaluate
the significance of the local properties of the features for each sample. 
In this way, we can take into account the local characteristics of each feature, so to address 
 the variability of the statistical properties of the data across the complete dataset. 
In other terms, we should identify a metric for which, if the features $n_1$ and $n_2$ are very similar across the $m$-th sample, the value will be large. 
In this case, it would be possible to assume that using just one of these features would be enough to obtain a robust and precise understanding of the given sample.
Conversely, a small value of this metric would mean that the two features are independent from each other such that they should be both taken into account to characterize the sample \cite{Luxburg,specclust1}. 

The distance metric based on Gaussian kernel would shows all these properties. 
Thus, we propose to quantify the difference between two features $n_1$ and $n_2$ for the $m$-th sample as follows: 

\begin{eqnarray}
w_{m_{n_1 n_2}}^{\text{GK}} & =&  \exp \left[-\frac{|| \underline{x}_{m n_1}- \underline{x}_{m n_2} ||^2}{2\sigma^2} \right], \label{eq_GKMI1}
\end{eqnarray}

\noindent where the value of $\sigma$ controls the width of the Gaussian kernel \cite{Luxburg,COUILLET16}. 

At this point, we can build for each sample $m$ a graph $\mathbb{G}_m=({\cal V},{\cal E}_m^{\text{GK}}, {\cal E}^{\text{MI}})$. 
The $n$-th vertex $v_n$ in ${\cal V}$ identifies the $n$-th feature of the $m$-th sample. 
Two vertices $v_{n_1}$ and $v_{n_2}$ are connected by two kinds of edges, ${\cal E}_{m_{n_1 n_2}}^{\text{GK}}$ and ${\cal E}_{{n_1 n_2}}^{\text{MI}}$, whose weights are computed according to (\ref{eq_GKMI1}) and (\ref{eq_GKMI2}), respectively. 
This structure can be used as a platform to perform adaptive feature selection across the dataset according to the guidelines of spectral clustering approach \cite{Luxburg,specclust1,specclust2,specclust3,COUILLET18,COUILLET20}. 
In particular, the aforesaid weights can be arranged in matrix form, so to generate two adjacency matrices associated with $\mathbb{G}_m$, i.e., $\textbf{W}_m^{\text{GK}} = \{w_{m_{ij}}^{\text{GK}}\}_{(i,j)\in \{1, \ldots, n\}^2}$ and $\textbf{W}^{\text{MI}} = \{w_{ij}^{\text{MI}}\}_{(i,j)\in \{1, \ldots, n\}^2}$. 
For the first matrix, it is possible to define a degree matrix $\textbf{D}_m^{\text{GK}} = 
\text{diag}(\textbf{d}_m^{\text{GK}})$, $\textbf{d}_m^{\text{GK}} = [d_{m_j}^{\text{GK}} = \sum_{l=1}^{n} w_{m_{jl}}^{\text{GK}}]_{j=1, \ldots, n}$. 
Hence, it is possible to define a normalized Laplacian matrix associated with $\textbf{W}_m^{\text{GK}}$ as $\bar{\textbf{L}}_m^{\text{GK}} = \textbf{I} -  \textbf{D}_m^{\text{GK}^{-1/2}}
\textbf{W}_m^{\text{GK}}
\textbf{D}_m^{\text{GK}^{-1/2}}$.  
Analogously, we can define $\textbf{D}^{\text{MI}} = \text{diag}(\textbf{d}^{\text{MI}})$, $\textbf{d}^{\text{MI}} = [d_j^{\text{MI}} = \sum_{l=1}^{n} w_{jl}^{\text{MI}}]_{j=1, \ldots, n}$, as the degree matrix associated with $\textbf{W}^{\text{MI}}$, and 
$\bar{\textbf{L}}^{\text{MI}} = \textbf{I} -  \textbf{D}^{\text{MI}^{-1/2}}
\textbf{W}^{\text{MI}}
\textbf{D}^{\text{MI}^{-1/2}}$ as its associated normalized Laplacian matrix. 
 
With this in mind, identifying the relevant features for each sample in the dataset can be described as partitioning the graph $\mathbb{G}_m$
such that the vertices of the same subgraph have strong connections via both links, while the vertices from different subgraphs have one or two weak connections. 
In spectral clustering, this problem can be written as follows: 

\begin{eqnarray}
\begin{cases}
\min_{\textbf{H}} & \text{Tr}( \textbf{H}^T \bar{\textbf{L}}_m^{\text{GK}}  \textbf{H}) 
\\ 
\min_{\textbf{H}}  & \text{Tr}( \textbf{H}^T \bar{\textbf{L}}^{\text{MI}}  \textbf{H}) 
\end{cases}
\label{eq_GKMI3}
\end{eqnarray}

\noindent where $\textbf{H}$ represents the matrix of the indicator vectors, and $\textbf{H}\textbf{H}^T = \textbf{I}$ \cite{Luxburg,specclust1,specclust2}. 
The solution of (\ref{eq_GKMI3}) is given by the common
eigenspace of the two normalized Laplacian matrices. 
Hence, this problem translates in identifying the set of joint eigenvectors $\textbf{V}_m = [\textbf{v}_{m1}, \dots, \textbf{v}_{mn}]$ that solves the following \cite{Ablin2018,Pham}: 

\begin{eqnarray} 
\min_{\textbf{V}_m}
\log  \frac{\left|\text{diag}(\textbf{V}_{m}^T\bar{\textbf{L}}_m^{\text{GK}} \textbf{V}_{m})\right|}{\left|\textbf{V}_{m}^T\bar{\textbf{L}}_m^{\text{GK}} \textbf{V}_{m}\right|}  
+ \log \frac{\left|\text{diag}(\textbf{V}_{m}^T\bar{\textbf{L}}^{\text{MI}} \textbf{V}_{m})\right|}{\left|\textbf{V}_{m}^T\bar{\textbf{L}}^{\text{MI}} \textbf{V}_{m}\right|},
\label{eqn:criterion}
\end{eqnarray} 

At this point,   $\textbf{H}$ in (\ref{eq_GKMI3}) contains the eigenvectors corresponding to the  $K_m$ lowest and non-null eigenvalues. 
It is worth recalling that the cardinality $K_m$ identifies the number of relevant features in the $m$-th sample according to the spectral clustering guidelines \cite{Luxburg,COUILLET18,COUILLET20,specclust1,specclust2,specclust3}. 
In fact, the number of relevant features equals the number of informative eigenvalues that 
can be defined as the local minima of the eigenvalues' difference curve \cite{Luxburg,COUILLET20,specclust1}. 
To this aim, the kneedle algorithm can be employed to select the optimal $K_m$ \cite{kneedle}. 
It is therefore crucial that the difference among the eigenvalues is well pronounced, so that the separation between eigenvalues associated with relevant and non-informative features can be easily carried out. 
Once the $K_m$ eigenvalues have been identified, it is possible to select the set of relevant features for the $m$-th sample as the centroids of the associated clusters.  

This information will finally be used to define the elements of the ${\cal K}$ tensor in (\ref{eq_flowrate}).
Specifically, if the $l$-th feature has been selected as relevant for both sample $m_1$ and $m_2$, then ${\cal K}_{m_1 m_2 l} =1$; otherwise, ${\cal K}_{m_1 m_2 l} =0$. 
It is worth noting that this simple set-up could be made more sophisticated by allowing the values of ${\cal K}$ to live in $\mathbb{R}$. 
Future works will be dedicated to investigate the impact of this choice in the effective use of the proposed fluid diffusion model in multimodal data analysis. 
Analogously, the definition of the distance operator to be used to determine the $v$ values in (\ref{eq_v_B_pm}) can be subject for deep investigation in the near future. 
Nevertheless, in this work we can assume without losing generality (and considering the observations on the continuity of the information propagation drawn in Section \ref{sec_methmot}) that the each $v_{ij}$ in (\ref{eq_v_B_pm}) will be based on the norm-2 between nodes \cite{Fluid_TDRW1,Fluid_TDRW2}. 

At this point, we can compute the \textbf{Q} matrix that has been introduced in Section \ref{sec_fluid_repr}. 
The next steps of the proposed community detection strategy consist in the eigenanalysis of the \textbf{Q} matrix. The next paragraphs summarize the main steps of the approach we present in this work. 

\subsubsection{Community detection based on fluid Laplacian matrix}
\label{sec_meth_fluidLapl}
As previously mentioned, the \textbf{Q} matrix is the core of the community detection algorithm based on fluid diffusion that we introduce in this work. 
We can indeed build a new matrix $\textbf{F}= \textbf{D}-\textbf{Q}$, where \textbf{D} is a $N \times N$ diagonal matrix such that $D_{ii} = \sum_{j=\{1,\ldots,N\}\setminus i} Q_{ij}$. 
As such, \textbf{F} is a matrix where all the
diagonal elements are positive, and the other elements
are negative. 
Therefore, \textbf{F} is invertible. 
Let us further analyze the properties of \textbf{F}. 
Specifically, let us consider a generic vector \textbf{z}, and let us derive the analytical solution of the $\textbf{z}^T \textbf{F} \textbf{z}$ function. 
It is possible to prove that the following holds \cite{CommunityDetection_SC2,COUILLET20}: 

\begin{eqnarray}
\textbf{z}^T\textbf{F}\textbf{z} & = & \textbf{z}^T \textbf{D} \textbf{z} - \textbf{z}^T \textbf{Q} \textbf{z} \nonumber \\ 
& = &  \sum_{i=1}^{N} D_{ii} z_i^2 - \sum_{i,j=1}^{N} Q_{ij}z_i z_j \\
& = &  \frac{1}{2} \left( \sum_{i=1}^{N} D_{ii} z_i^2 - 2\sum_{i,j=1}^{N} Q_{ij}z_i z_j + \sum_{j=1}^{N} D_{jj} z_j^2 \right) \nonumber \\
& = & \frac{1}{2} \sum_{i,j=1}^{N} Q_{ij} (z_i - z_j)^2. \nonumber
\label{eq_fluidLapl_deriv}
\end{eqnarray}

\noindent Therefore, the matrix \textbf{F} can be considered as a Laplacian matrix, and will take the name of \textit{fluid Laplacian matrix}. 
Moreover, this system can be used to construct a graph that could be partitioned in communities. 
In particular, in order to find a partition of the graph such
that the edges between different communities have lower
weight and the edges within the same community have a
higher weight, we can apply the Normalized Cut algorithm \cite{normalizedCut}. 
In other terms, the graph induced by \textbf{Q} can be partitioned in $K_F$ connected components $C_\kappa$, $\kappa=1, \ldots, K_F$ (where $\bigcup_{\kappa=1}^{K_F} C_\kappa = {\cal V}$, and $C_{\kappa_1} \cap C_{\kappa_2} = \emptyset$ $\forall \kappa_1, \kappa_2 \in \{1, \ldots,K_F\}^2$, $\kappa_1 \neq \kappa_2$) by minimizing over $\{C_\kappa\}_{\kappa = 1, \ldots, K_F}$ the NormalizedCut function $NC_{K_F}$, which can be written as follows:

\begin{eqnarray}
NC_{K_F} &=& \frac{1}{2} \sum_{\kappa=1}^{K_F} \frac{\zeta(C_\kappa,\bar{C}_\kappa)}{\tilde{C}_\kappa}, 
\label{eq_specclust2}
\end{eqnarray} 

\noindent where $\bar{C}_\kappa$ represents the complement of the $\kappa$-th partition over the vertex set ${\cal V}$, $\tilde{C}_\kappa$ is a measure of the width of the $\kappa$-th partition (typically expressed in volume $\mathsf{vol}(C_\kappa) = \sum_{v_i \in C_\kappa} \sum_{j=1}^{|{\cal V}|} Q_{ij}$), and $\zeta(C_\kappa,\bar{C}_\kappa) = \sum_{v_i \in C_\kappa, v_j \in \bar{C}_\kappa} Q_{ij}$.  \cite{normalizedCut}. 

The minimization of $NC_{K_F}$ leads to have large  weights for the edges connecting the nodes within $C_\kappa$, while the edges connecting nodes within $C_\kappa$ with the nodes in its complement $\bar{C}_\kappa$ will show small weights. 
Furthermore, this operation can be described in terms of the eigenvectors of the normalized fluid Laplacian matrix $\bar{\textbf{F}} = \textbf{D}^{-1/2} \textbf{F} \textbf{D}^{-1/2}$. 
In other terms, the $NC_{K_F}$ optimization can be written as follows:

\begin{eqnarray}
\min_{\textbf{J}} \text{Tr}( \textbf{J}^T \bar{\textbf{F}} \textbf{J}) & \text{subject to} &  \textbf{J}^T\textbf{J}=\textbf{I},
\label{eq_specclust4}
\end{eqnarray}

\noindent where \textbf{J} is the matrix of the first smallest $K_F$ eigenvectors of $\bar{\textbf{F}}$. 
Hence, in order to solve this problem, it is possible to employ the kneedle algorithm \cite{kneedle} to select the best value of $K_F$, and then run a traditional $K_F$-means algorithm over \textbf{J} (considering the rows of \textbf{J} as nodes) in order to identify the $K_F$ communities \cite{COUILLET18,Luxburg,CommunityDetection_SC2}. 
In this way, we can guarantee a high degree of implementability and efficiency of the system, whilst achieving a thorough unsupervised characterization of the multimodal data under exam. 
The following Section provides tests to validate this set-up with respect to state-of-the-art methods.

\section{Experimental results and discussion}
\label{secresult}

We conducted several experiments to show how the proposed fluid graph construction is beneficial to address the limits of modern data analysis that we mentioned in Section \ref{sec_meth_motnew}. 
To this aim, we tested the scheme we introduced in Section \ref{secmeth} on four diverse datasets form four research fields (remote sensing, brain-computer interface, photovoltaic energy, and off-shore wind farms). 
These datasets show high diversity in terms of heterogeneity of sensing platforms and acquisition strategies (e.g., multimodal and unimodal datasets; static and dynamic records). 
Thus, the considered datasets display various characteristics in terms of noise distributions across the samples, as well as of semantics of the data (i.e., inter- and intra-class relationships). 

This high degree of diversity make these datasets a great fit to analyze the robustness of the proposed unsupervised approach for community detection in case of missing data, and unbalanced data distribution. 
We tested these points by simulating the occurrence of missing data (by setting subsets of features to a null value) and of unbalanced datasets (by changing the density of samples belonging to each thematic cluster in the dataset). 
We therefore organized this Section to address the limits that we previously listed, and assessed the outcomes we obtained by comparing with state-of-the-art methods. 

Several multimodal datasets representative of different research fields have been used to validate the novel graph representation based on fluid diffusion model and to test the performance of the proposed community detection method. 
In this Section, we first summarize the main characteristics of the datasets we have taken into account. 
Then, 
we report the performance of the proposed community detection framework based on fluid diffusion model.

\subsection{Datasets}
\label{sec_exp_res}
We tested the proposed approach on four very diverse datasets, focusing on four different research fields: remote sensing, brain-computer interface, photovoltaic energy, and off-shore wind farm monitoring. 

\subsubsection{Multimodal remote sensing (RS)}
\label{sec_exp_RS}
First, 
we considered a multimodal 
dataset consisting of LiDAR and hyperspectral records acquired over the University of Houston campus and the neighboring urban area, and was distributed for the 2013 IEEE GRSS Data Fusion Contest~\cite{data_houston}. 
Specifically:
\begin{itemize}
	\item the size of the dataset is $N=$1905$\times$349 pixels, with spatial resolution equal to 2.5m;
	\item the final dataset consists of $n$=151 features. In fact, the hyperspectral dataset includes 144 spectral bands ranging from 0.38 to 1.05 $\mu{m}$, whilst the LiDAR records includes one band and 6 textural features;
	\item the available ground truth labels consists of $K_C=15$ classes. 
\end{itemize}

\subsubsection{Multimodal brain-computer interface (BCI)}
\label{sec_exp_BCI}
The second dataset we considered was collected by means of brain-computer interface \cite{BrainComputer1}. 
Specifically, the records were collected by means of 
 60-channel electroencephalography
(EEG),  7-channel electromyography
(EMG) and 4-channel electro-oculography (EOG) on $K_C=11$ intuitive upper extremity
movements from 25 participants. 
A 3-sessions experiment was carried out, and 3300 trials per participant were collected. 
According to the notation we have used in Section \ref{secmeth}, the final dataset consists of $n=71 (=60+4+7)$ multimodal features for a total sum of $N=$82500 samples across all the participants.

\subsubsection{Multimodal photovoltaic energy (PV)}
\label{sec_exp_PV}
The final dataset was acquired in order to monitor the photovoltaic energy produced between July 21 and Aug. 17, 2018 at the University of Queensland, Australia \cite{data_PhotoVoltaic}. 
The records that have been collected by weather ground stations can be listed as follows:
\begin{itemize}
	\item instantaneous and average wind speed [km/h] and direction [deg];
	\item temperature [deg];
	\item relative humidity [\%];
	\item mean surface level pressure [hPa];
	\item accumulated rain [mm];
	\item rain intensity [mm/h];
	\item accumulated hail [hits/cm2];
	\item hail intensity [hits/cm2hr];
	\item solar mean [W/m2].
\end{itemize}
This summed to 1440 samples acquired for each day, summing up to a dataset of $N=$1440 $\times$ 28 records. 
For each sample, the photovoltaic energy [W/h] is recorded: $K_C=10$ classes uniformly drawn based on this parameter are considered. 
The considered data analysis task consists of assigning a class to all the samples by investigating the $n=12$ heterogeneous features. 

\subsubsection{Off-shore wind farm (OSWF) monitoring (Hywind)}
\label{sec_exp_HYWIND}
The final dataset we considered was a collection of GPS measurements of surge and sway motions conducted by Equinor ASA to monitor the Hywind Scotland wind turbines \cite{EquinorHywind}. 
In particular, we considered the records that have been acquired by a motion sensor mounted on a floating wind mill (HS4) off the coast of Scotland. 
The sensor captures longitude and latitude misplacements per second for 
11 intervals of 30 minutes: these misplacements are calculated averaging 10 measurements per second.  
The dataset has been arranged so that one sample would identify a sequence of misplacements measured for 5 minutes. 
Therefore, the Hywind dataset we considered consists of $N=$198 samples of $n=100$ \textit{unimodal} records, i.e., each sample is a temporal sequence of $n=100$ measured misplacements. 
Moreover, $K_C=4$ classes across these samples are considered.


\subsection{Results}
\label{sec_eval}

In order to provide a thorough investigation of the actual impact of the proposed approach, we conducted several experiments at different levels. 
Specifically, we tested the proposed architecture in terms of design choices, validity of the proposed model, parameter sensitivity, and community detection performance. 

\subsubsection{Validity of the fluid graph model}
First, we investigated whether the proposed adaptive dimensionality reduction method for the definition of the ${\cal K}$ tensor in (\ref{eq_fluid_trans_prob_pij}) that is used then to determine the fluid Laplacian matrix in Section \ref{sec_meth_fluidLapl}. 
In particular, we tested the ability of the method in Section \ref{sec_meth_ADR} to reliably identify relevant features across complex datasets, so that the construction of the ${\cal K}$ tensor could be carried out. 
We reported these results in Appendix \ref{app_K_ablation}, where we show that the proposed strategy is able to track better than other comparable methods
 the data structure in complex systems, hence leading to a more precise characterization of the relevance of the features in the considered datasets. 

Let us focus our attention on the actual procedure for community detection based on the fluid diffusion model that we propose in this work. 
In particular, it is worth to recall that the main steps for the community detection technique reported in Section \ref{sec_meth_fluidLapl} are fundamentally based on the eigenanalysis of the fluid Laplacian matrix \textbf{F}. 
Thus, the method in Section \ref{sec_meth_fluidLapl} could be considered as an instance of the spectral clustering approach for community detection, according to the characteristics summarized in Appendix \ref{sec_back_spec}. 
As such, the ability to discriminate the lowest eigenvalues from the overall eigenvalues set, so that the identification of the communities in the dataset can be accurately carried out \cite{Luxburg}. 
Therefore, it is important to analyze the eigenspectrum of the computed eigenvalues, so to retrieve a solid understanding of the actual characterization ability the considered spectral clustering-based architecture might have. 
Hence, in order to obtain a first assessment of the actual impact of the proposed community detection method, we assessed the improvement provided by the use of the fluid Laplacian matrix for spectral clustering. 

Specifically, we computed the eigenspectrum resulting from the analysis of the datasets in Section \ref{sec_exp_res} by means of the scheme proposed in Section \ref{sec_meth_fluidLapl} and several spectral clustering methods introduced in technical literature, i.e., using \textit{unnormalized} and \textit{normalized} Laplacian matrix \cite{Luxburg}, \textit{graph distance}-based spectral clustering \cite{CommunityDetection_SC2}, \textit{covariate}-assisted spectral clustering \cite{CommunityDetection_SC1}, spectral clustering using \textit{probability} matrix \cite{CommunityDetection_SC3}, \textit{self tuning} spectral clustering \cite{CommunityDetection_SC4}, and \textit{regularized} Laplacian matrix \cite{COUILLET20}. 
It is worth noting that all these methods are relying on the graph representation based on heat diffusion. 
In particular, the parameter $\zeta_p$ for the regularized Laplacian matrix was set to $\sqrt{c\Phi}$, according to the guidelines in \cite{COUILLET20}. 


In this respect, when considering a dataset composed by $K_C$ communities, the gap between the $K_C$-th and the $K_C+1$-th eigenvalues plays a crucial role to the achievement of an effective clustering of the datasets, according to the theoretical aspects of spectral clustering (briefly reported in Appendix \ref{sec_back_spec}). 
In particular, it is possible to achieve a more accurate community detection for large eigengaps. 
Thus, to quantify the difference between the aforesaid approaches 
, we computed the difference $|\lambda_{K_C+1} - \lambda_{K_C}|$ (where $\lambda_i$ identifies the $i$-th eigenvalue) for all the methods in these figures. We reported the eigengaps  we obtained for the considered datasets in Table \ref{tab1}, where $K_C$ is set to 15, 11, and 10 for the datasets in Section \ref{sec_exp_RS}, \ref{sec_exp_BCI}, and \ref{sec_exp_PV}, respectively. 

At this point, the improvement provided by using the fluid Laplacian matrix as in Section \ref{sec_meth_fluidLapl} with respect to state-of-the-art methods appears dramatic. 
Indeed, the results in Table \ref{tab1} emphasize the ability of the fluid diffusion model in addressing the complex interactions among samples that can occur at global and local scale in multimodal datasets. 
This impact of this result is further highlighted by Table \ref{tab2}, where the gap between the $K_C+1$-th and $K_C+2$-th eigenvalue is displayed. 
In fact, it is possible to appreciate how this difference is sensibly smaller than their corresponding values in Table \ref{tab1}. 

\color{blue}
\begin{table}[!th]
	\tiny
	\caption{Eigengap $|\lambda_{K_C+1} - \lambda_{K_C}|$ for the proposed fluid graph construction and state-of-the-art spectral clustering methods.}
	\label{tab1}
	\centering
	\begin{tabular}{|c|c|c|c|c|}
		\hline
		\bfseries Method & \bfseries RS & \bfseries BCI & \bfseries PV & \textbf{Hywind}\\
		& (multimodal & (multimodal & (multimodal & (unimodal\\
		& remote sensing) & brain-computer & photovoltaic & temporal \\
		&  & interface) & energy) & OSWF monitoring) \\
		\hline
		\textbf{Fluid} & 45 & 48 & 54 & 36 \\
		\hline
		\textbf{Unnormalized} &  $2 \times 10^{-14}$ & $1.4 \times 10^{-14}$ & $3 \times 10^{-15}$ & $4.1 \times 10^{-16}$ \\
		\hline
		\textbf{Normalized} & $6 \times 10^{-15}$  & $3 \times 10^{-15}$  & $ 2\times 10^{-15}$ & $2.2 \times 10^{-16}$ \\
		\hline
		\textbf{Graph Distance} & $6 \times 10^{-14}$  & $3 \times 10^{-14}$ &  $4 \times 10^{-14}$ & $3.7 \times 10^{-16}$ \\
		\hline
		\textbf{Covariate} &  $7 \times 10^{-14}$ & $3 \times 10^{-14}$  & $2.9 \times 10^{-14}$ & $8 \times 10^{-16}$ \\
		\hline
		\textbf{Probability} & $8 \times 10^{-15}$  & $2 \times 10^{-15}$ &  $3 \times 10^{-14}$ & $6.9 \times 10^{-16}$\\
		\hline
		\textbf{Self tuning} & $4 \times 10^{-15}$  & $1.4 \times 10^{-15}$ &  $2.3 \times 10^{-14}$ & $9.2 \times 10^{-16}$ \\
		\hline
		\textbf{Regularized} & $3 \times 10^{-15}$  & $1 \times 10^{-15}$ &  $1 \times 10^{-14}$& $7.4 \times 10^{-16}$ \\
		\hline
	\end{tabular}
\end{table}
\color{black}
\color{blue}
\begin{table}[!th]
	\tiny
	\caption{Eigengap $|\lambda_{K_C+2} - \lambda_{K_C+1}|$ for the proposed fluid graph construction and state-of-the-art spectral clustering methods.}
	\label{tab2}
	\centering
	\begin{tabular}{|c|c|c|c|c|}
		\hline
		\bfseries Method & \bfseries RS & \bfseries BCI & \bfseries PV & \textbf{Hywind} \\
		& (multimodal & (multimodal & (multimodal & (unimodal\\
		& remote sensing) & brain-computer & photovoltaic & temporal\\
		&  & interface) & energy) &OSWF monitoring) \\
		\hline
		\textbf{Fluid} & 0.2 & 1.2 & 0.3 & 0.17\\
		\hline
		\textbf{Unnormalized} &  $3.2 \times 10^{-14}$ & $1.8 \times 10^{-14}$ & $2.2 \times 10^{-15}$ &  $1.2 \times 10^{-16}$\\
		\hline
		\textbf{Normalized} & $6.4 \times 10^{-15}$  & $3.9 \times 10^{-15}$  & $ 2.3\times 10^{-15}$  &  $2.3 \times 10^{-16}$\\
		\hline
		\textbf{Graph Distance} & $5.5 \times 10^{-14}$  & $2.8 \times 10^{-14}$ &  $3.8 \times 10^{-14}$ &  $3.3 \times 10^{-16}$\\
		\hline
		\textbf{Covariate} &  $6.8 \times 10^{-14}$ & $2.9 \times 10^{-14}$  & $2.77 \times 10^{-14}$  &  $2.4 \times 10^{-16}$\\
		\hline
		\textbf{Probability} & $7.7 \times 10^{-15}$  & $2.1 \times 10^{-15}$ &  $2.8 \times 10^{-14}$ &  $3.1 \times 10^{-16}$\\
		\hline
		\textbf{Self tuning} & $3.8 \times 10^{-15}$  & $1.5 \times 10^{-15}$ &  $2.4 \times 10^{-14}$ &  $1.7 \times 10^{-16}$\\
		\hline
		\textbf{Regularized} & $2.8 \times 10^{-15}$  & $1.26 \times 10^{-15}$ &  $1.03 \times 10^{-14}$ &  $4 \times 10^{-16}$\\
		\hline
	\end{tabular}
\end{table}
\color{black}

\subsubsection{Community detection performance}
With this in mind, we can further explore the capacity of the proposed method by assessing its actual ability of detecting communities within the considered datasets. 
In particular, we can assess how the proposed strategy can address the limits of modern data analysis we summarized on Section \ref{sec_meth_motnew}. 
In order to obtain a quantitative assessment in this sense, we compute cluster assignments on the graph and evaluate the partitions that are delivered. 
We evaluate the performance of the methods by computing three indices: modified purity ($mP$), modified adjusted Rand index ($mARI$), and modified normalized mutual information ($mNMI$) \cite{fluid_metrics,fluid_metrics2}. 
These metrics are able to quantify the ability of the methods to understand the graph properties, and 
to provide a proper set-up to extract information at semantic and functional level from the considered dataset. 
Moreover, in order to obtain a more reliable evaluation of the graph learning performance, these metrics take into account graph topology whilst their "traditional" counterparts do not \cite{fluid_metrics,fluid_metrics2}. 
We detailed the aforesaid metrics in Appendix \ref{app_metrics}. 

Thus, when assessing the functional information retrieval performance of the different graph learning frameworks, we considered the metrics as defined according to these set-ups, i.e., three values for each metric, for a total of nine metrics for performance comparison. 
In particular, we used these metrics to compare the strategy we introduced in this work with the following state-of-the-art methods: 
\begin{itemize}
	\item clustering via hypergraph modularity (\textit{CNM}) \cite{CommunityDetection_GP7};
	\item hierarchical community detection (\textit{HCD}) \cite{CommunityDetection_HC5};
	\item community detection based on distance dynamics (\textit{Attractor}) \cite{CommunityDetection_HC6};
	\item joint criterion for community detection (\textit{JCDC}) \cite{CommunityDetection_HC7};
	\item a standard \textit{k}-means algorithm, where $k=K_C$;
	\item variational Bayes community detection (\textit{VB}) \cite{CommunityDetection_SI5};
	\item node importance-based label propagation (\textit{NI-LPA}) \cite{CommunityDetection_LP3};
	\item fluid label propagation (\textit{FLP}) \cite{CommunityDetection_LP4}
	\item weighted stochastic block model (\textit{WSBM}) \cite{CommunityDetection_Hyb4};
	\item multiview spectral clustering (\textit{Multiview}) \cite{CommunityDetection_view1};
	\item covariate-assisted spectral clustering (\textit{CASC}) \cite{CommunityDetection_SC1};
	\item regularized spectral clustering (\textit{RegularizedSC}) \cite{COUILLET20};
	\item deep multimodal clustering (\textit{MMClustering}) \cite{CommunityDetection_multi3};
	\item heterogeneous graph embedding model leveraging on metapath-based embedding (\textit{MP2V}) \cite{MP2V}; 
	\item Attributed
	social network embedding (\textit{ASNE}) \cite{ASNE}; 
	\item semantic-associated heterogeneous networks (\textit{SHNE}) \cite{SHNE};
	\item inductive representation
	learning (\textit{GraphSAGE}) \cite{GraphSAGE}; 
	\item graph attention network (\textit{GAT}) \cite{GAT};
	\item heterogeneous graph neural network (\textit{HetGNN}) \cite{HetGNN}.
\end{itemize}

\noindent All these methods rely on graph representation based on heat diffusion. It is worth to recall that the method introduced in \cite{CommunityDetection_LP4} does not show any overlap whatsoever with the strategy for community detection we introduce in this work. The authors in \cite{CommunityDetection_LP4} do not discuss fluid diffusion indeed, nor introduce any novel graph representation of datasets.  

We organized the discussion on the experimental results with respect to the limits of modern data analysis \textbf{L1-L3} mentioned in Section \ref{sec_meth_motnew}. 

\begin{figure}[htb]
	\centering
	\includegraphics[width=1\columnwidth]{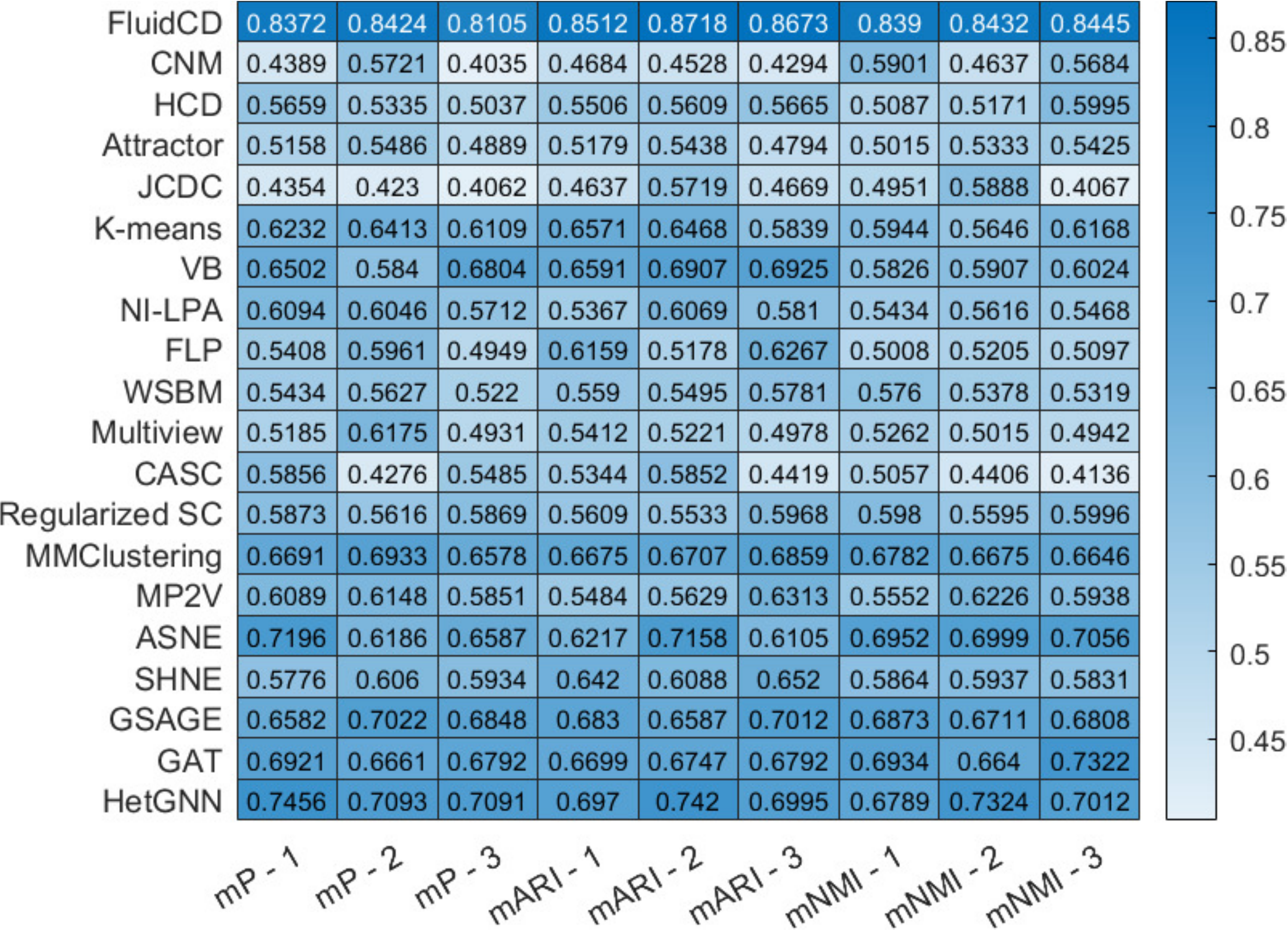} 
	\caption{Heatmap of the community detection results achieved when analyzing the dataset in Section \ref{sec_exp_RS} (multimodal remote sensing), for each method considered. The results in terms of modified purity index (mP), modified adjusted Rand index (mARI), and modified normalized mutual information (mNMI) for different configurations discussed in Section \ref{sec_exp_res} (degree measure - 1; embeddedness measure - 2; weighted embeddedness measure - 3) are reported.}
	\label{fig_res_heat_RS}
\end{figure}

\begin{figure}[htb]
	\centering
	\includegraphics[width=1\columnwidth]{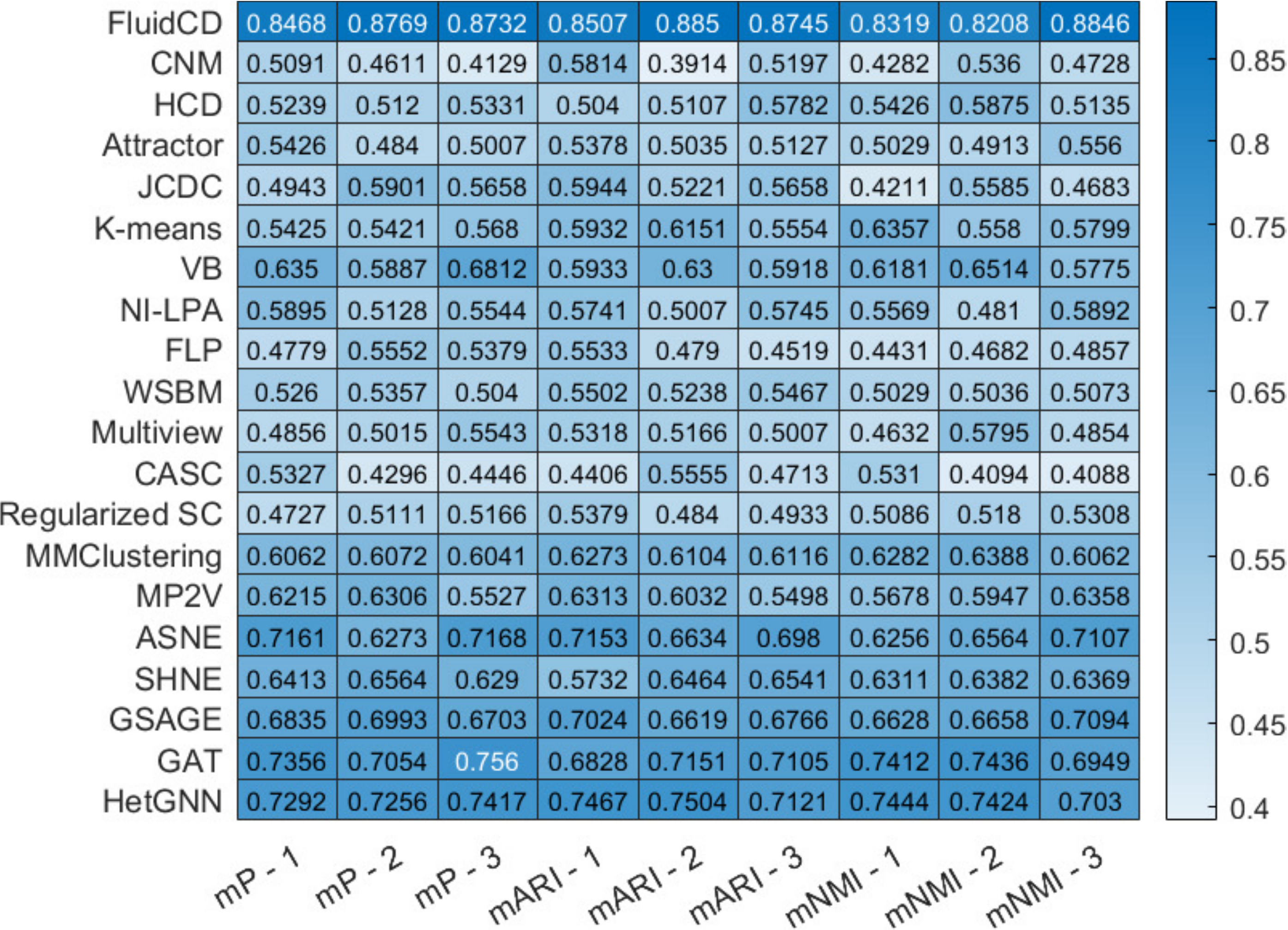} 
	\caption{Heatmap of the community detection results achieved when analyzing the dataset in Section \ref{sec_exp_BCI} (multimodal brain-computer interface), for each method considered. The same notation as in Fig. \ref{fig_res_heat_RS} is used here.}
	\label{fig_res_heat_BCI}
\end{figure}

\begin{figure}[htb]
	\centering
	\includegraphics[width=1\columnwidth]{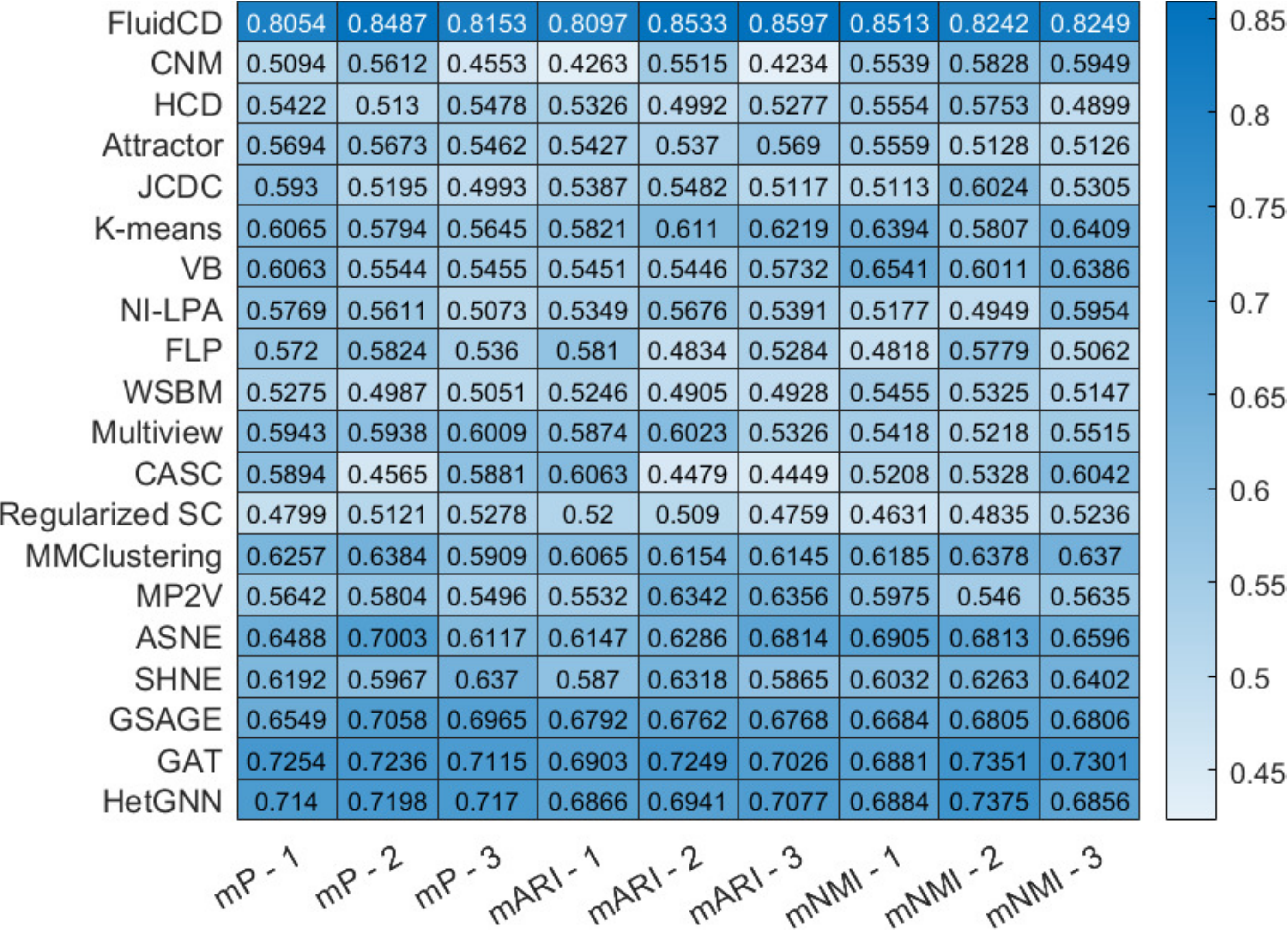} 
	\caption{Heatmap of the community detection results achieved when analyzing the dataset in Section \ref{sec_exp_PV} (multimodal photovoltaic energy), for each method considered. The same notation as in Fig. \ref{fig_res_heat_RS} is used here.}
	\label{fig_res_heat_PV}
\end{figure}

\begin{figure}[htb]
	\centering
	\includegraphics[width=1\columnwidth]{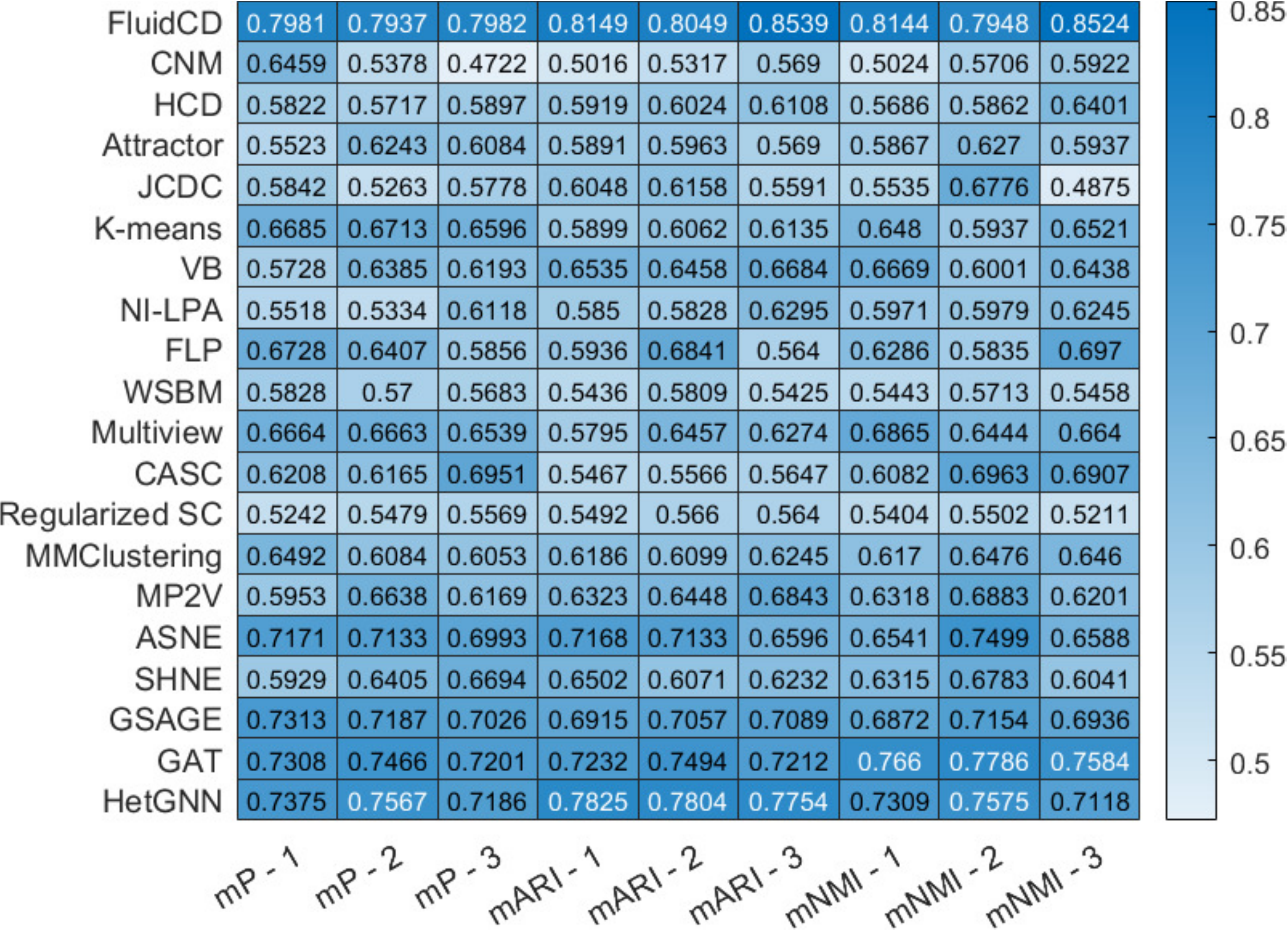} 
	\caption{Heatmap of the community detection results achieved when analyzing the dataset in Section \ref{sec_exp_HYWIND} (multimodal photovoltaic energy), for each method considered. The same notation as in Fig. \ref{fig_res_heat_RS} is used here.}
	\label{fig_res_heat_HYWIND}
\end{figure}

\textbf{L1 - adaptivity:} 
Figures \ref{fig_res_heat_RS} to \ref{fig_res_heat_HYWIND} report the results we achieved by assessing the ability of extracting functional information by applying clustering methods to the outcomes of graph learning frameworks.
For each column we report the value of $mP$, $mARI$, and $mNMI$ computed according to the setting mentioned in the previous bullet point list. 
Independently by the configuration we used to assess the results, the proposed method is able to outperform the other schemes in all datasets and set-ups. 
Indeed, the approach we introduce appears to be more suitable to adapt to the properties of the considered datasets. 

In particular, it is worth noting that the proposed diffusion model provides a solid platform that can be used by several graph analysis approaches to effectively explore the data properties (stronger fluctuations of the proposed metrics are registered for heat diffusion-based schemes across different clustering algorithms).  
This  is crucial to extract functional characteristics of the records taken into account, so that an analysis at semantic level can be accurately performed. 
Hence, the flexibility produced by the fluid diffusion model to analyze the graph representation provides a great advantage with respect to the heat diffusion-based approaches.

\begin{figure*}[htb]
	\centering
	\includegraphics[width=2\columnwidth]{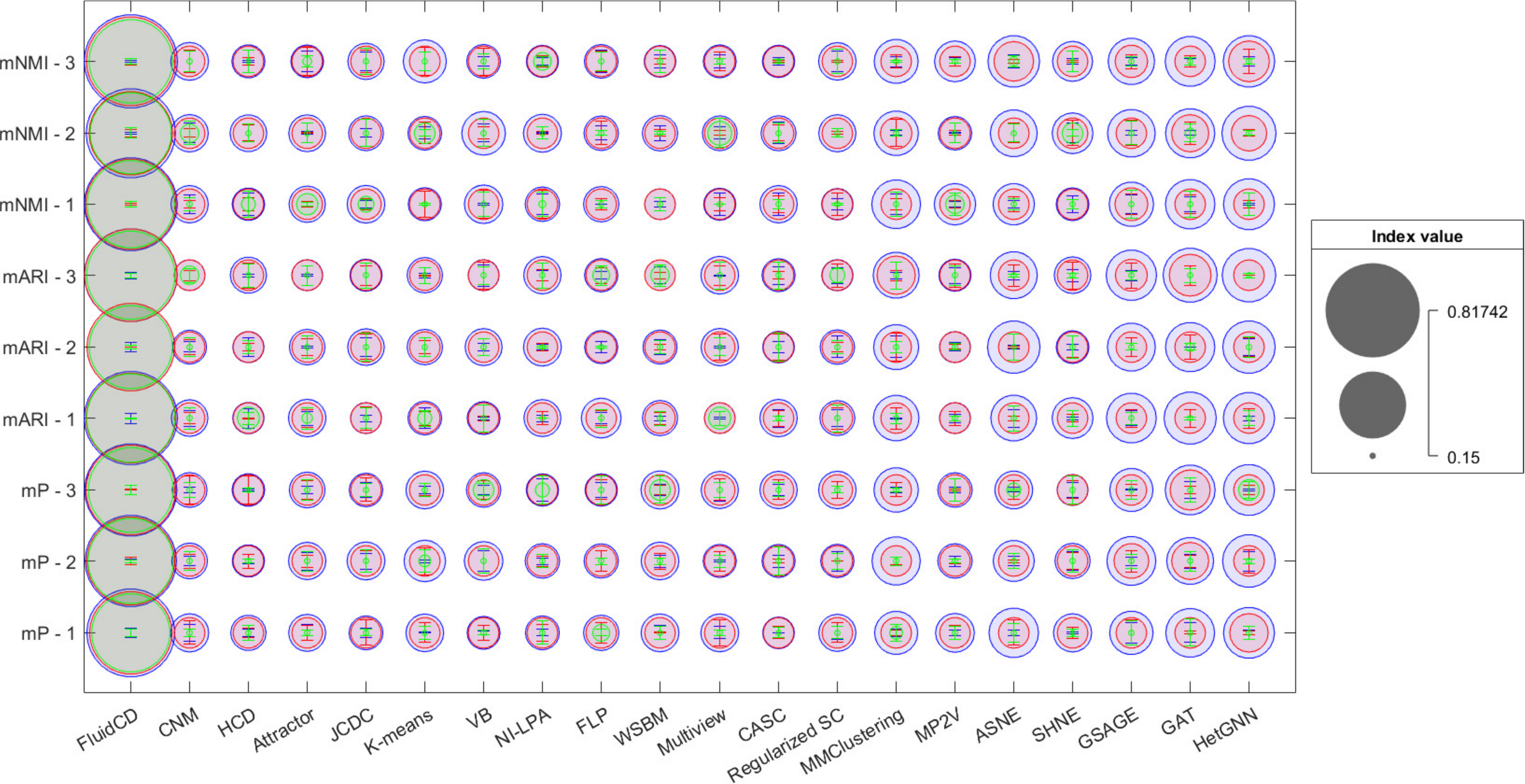} 
	\caption{Community detection performance for each considered method when $M_P \cdot N$ features are missing for each sample in the multimodal remote sensing dataset. The average value of each community detection index is represented by the size of each bubble in the diagram. The results for $M_P$ set to $\{0.08, 0.1, 0.15\}$ are displayed in blue, red, and green color, respectively. The confidence interval for each set-up is reported as error bar within each bubble.}
	\label{fig_bubble_RS}
\end{figure*}

\begin{figure*}[htb]
	\centering
	\includegraphics[width=2\columnwidth]{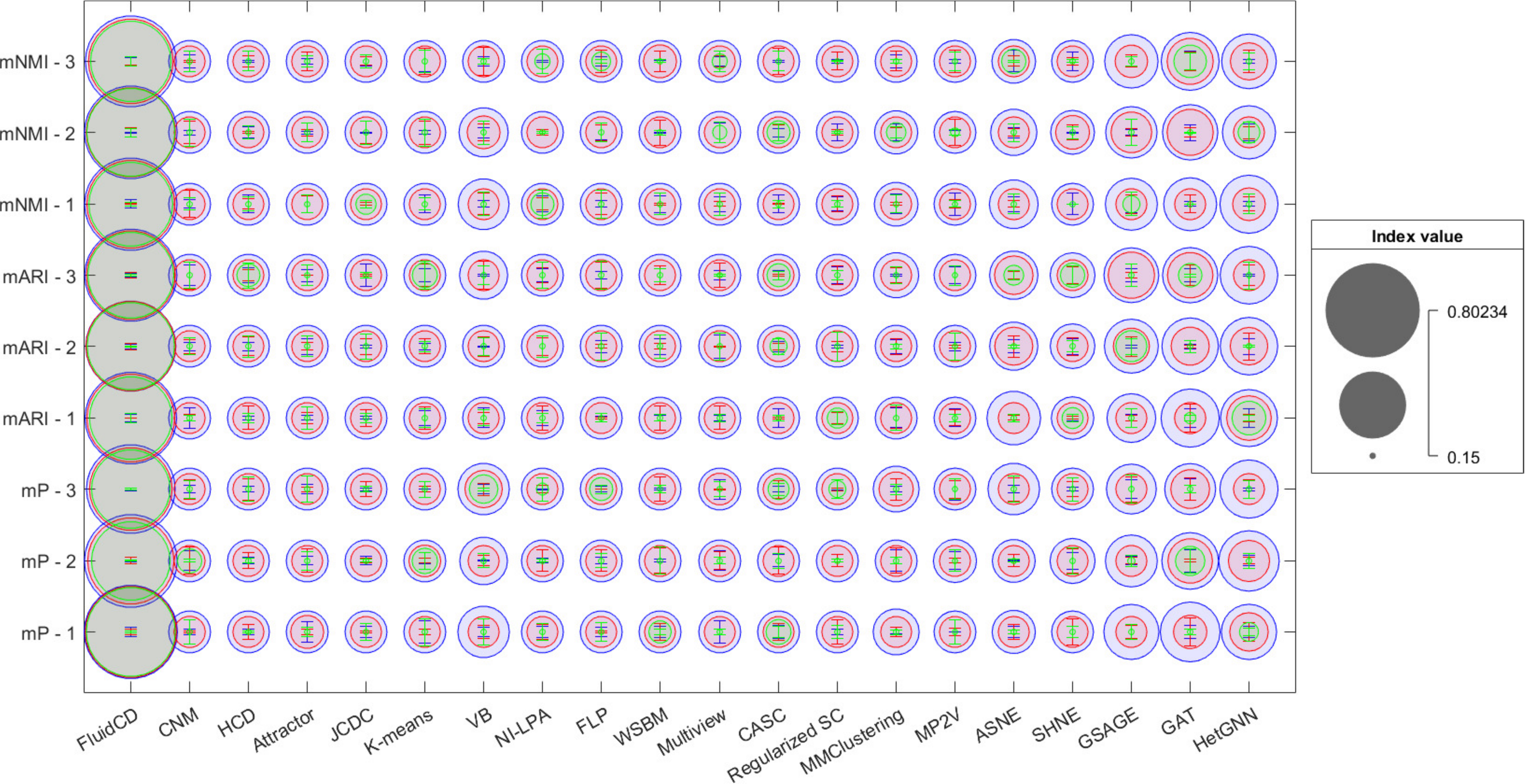} 
	\caption{Community detection performance for each considered method when $M_P \cdot N$ features are missing for each sample in the multimodal brain-computer interface dataset. The same notation as in Fig. \ref{fig_bubble_RS} applies here.}
	\label{fig_bubble_BCI}
\end{figure*}

\begin{figure*}[htb]
	\centering
	\includegraphics[width=2\columnwidth]{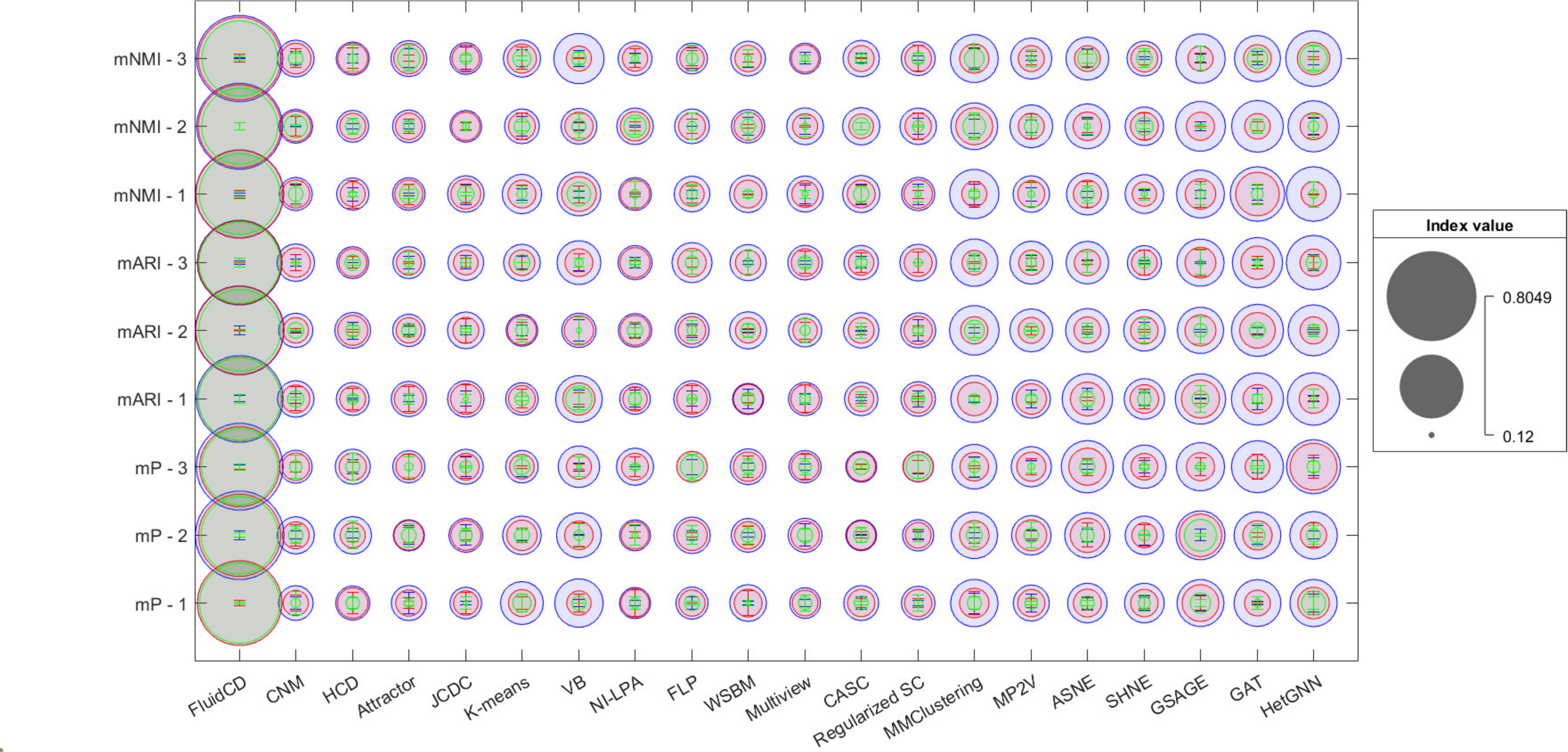} 
	\caption{Community detection performance for each considered method when $M_P \cdot N$ features are missing for each sample in the multimodal photovoltaic energy dataset. The same notation as in Fig. \ref{fig_bubble_RS} applies here.}
	\label{fig_bubble_PV}
\end{figure*}

\begin{figure*}[htb]
	\centering
	\includegraphics[width=2\columnwidth]{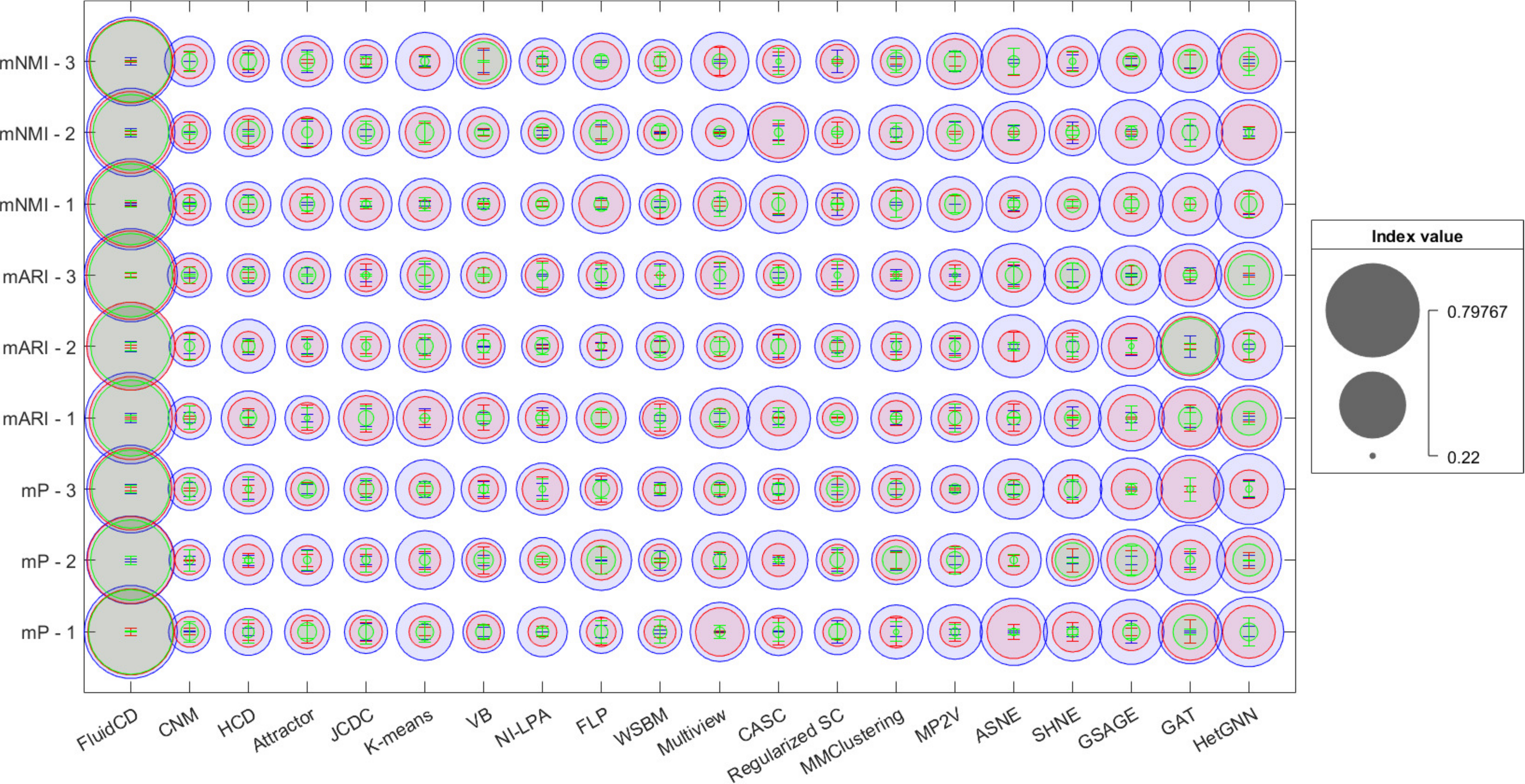} 
	\caption{Community detection performance for each considered method when $M_P \cdot N$ features are missing for each sample in the unimodal temporal OSWF monitoring dataset. The same notation as in Fig. \ref{fig_bubble_RS} applies here.}
	\label{fig_bubble_HYWIND}
\end{figure*}

\textbf{L2 - missing data:} We used the datasets in Section \ref{sec_exp_res} to test the robustness of the proposed method in case of datasets affected by missing data. 
Specifically, we randomly picked a 
set of $M_P \cdot n$ features to be set to a null value for every sample: this operation is meant to simulate the absence of $M_P \cdot n$ features out of the original $n$ features each sample is characterized of. 
We performed community detection on the resulting dataset. 
Then, we iterated these steps 100 times. 
Also, the values of $M_P$ were set to $\{0.08,0.1, 0.15\}$. 

We reported the community detection results in Figures \ref{fig_bubble_RS} to \ref{fig_bubble_HYWIND}. 
These figures show how the increase of missing records in the dataset to be analyzed affects the community detection performance of the methods introduced in technical literature. 
In particular, the progressive reduction of the ability to detect communities in the dataset as the number of missing features increase is strong, as well as the variance of the results for all these methods. 

The strategy that we propose in this work instead is very robust to missing data. 
In fact, the FluidCD method achieves community detection performance that is substantially equal to those displayed in Figures \ref{fig_res_heat_RS}-\ref{fig_res_heat_HYWIND} when $M_P=0.08$. 
Moreover, this performance does not strongly decrease when $M_P$ increases, and the variance of the outcomes for each index is smaller than those of the other methods. 

This behaviour is caused by the ability of the fluid graph representation to incorporate (in particular in the ${\cal K}$ tensor) the relevance of the features to be used to compute the similarities among samples. 
In this way, when some features are missing in the representation of a sample, the corresponding elements of the ${\cal K}$ tensor are set to 0, i.e., those features are automatically excluded from the computation of the distance between two samples. 
This is in contrast to the state-of-the-art methods in technical literature, where the missing features are still used (by setting their values to 0) to estimate the samples' similarities. This means that the distance computation (and hence the graph representation) is prone to artificial bias, hence leading to inaccuracy in understanding the true links among samples.


\begin{figure}[htb]
	\centering
	\includegraphics[width=1\columnwidth]{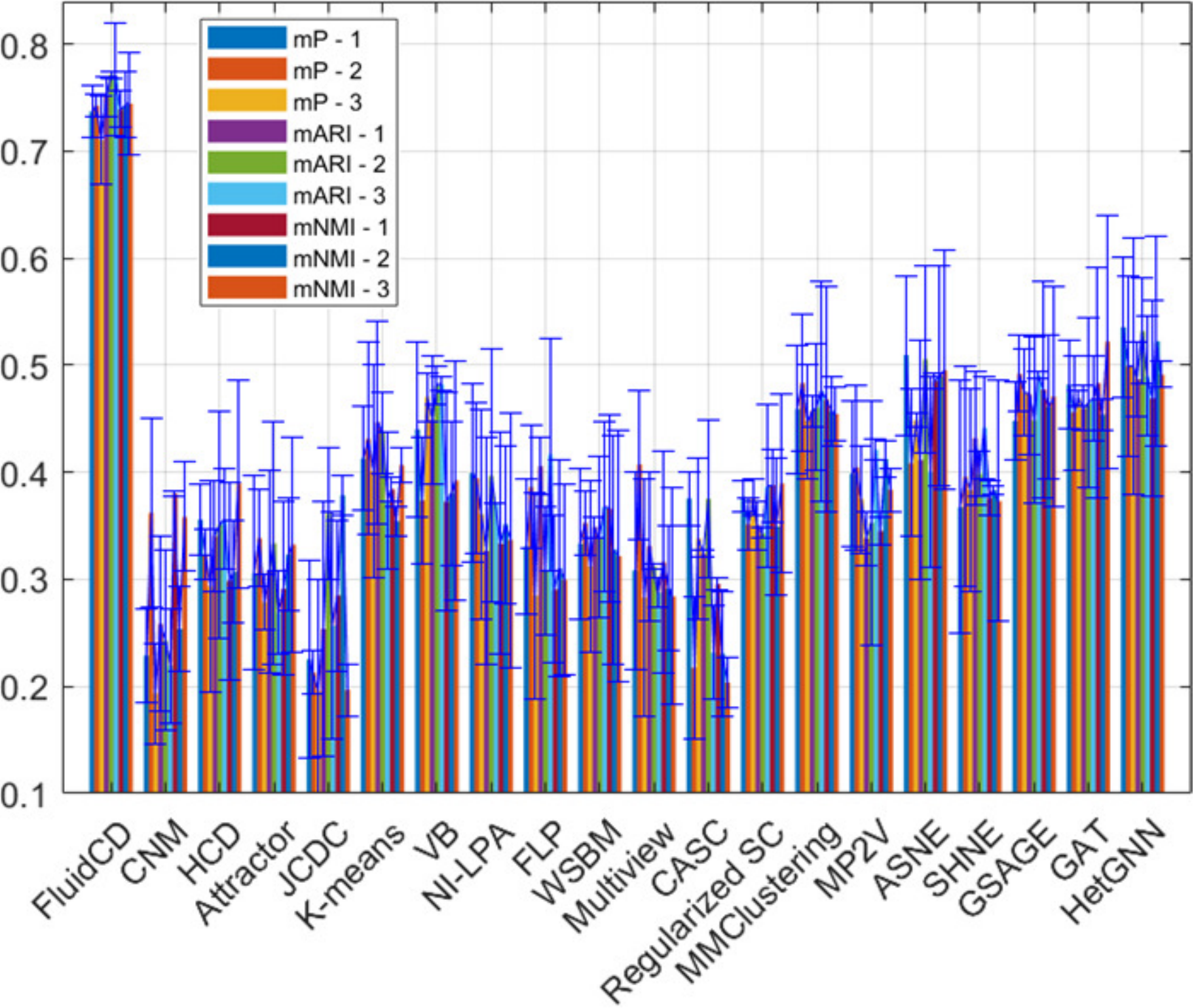} 
	\caption{Community detection results achieved when analyzing the unbalanced datasets obtained from the multimodal remote sensing dataset. The index values are shown as bars. For each index, the confidence interval obtained after 100 runs is shown as blue errorbars.}
	\label{fig_bar_RS}
\end{figure}

\begin{figure}[htb]
	\centering
	\includegraphics[width=1\columnwidth]{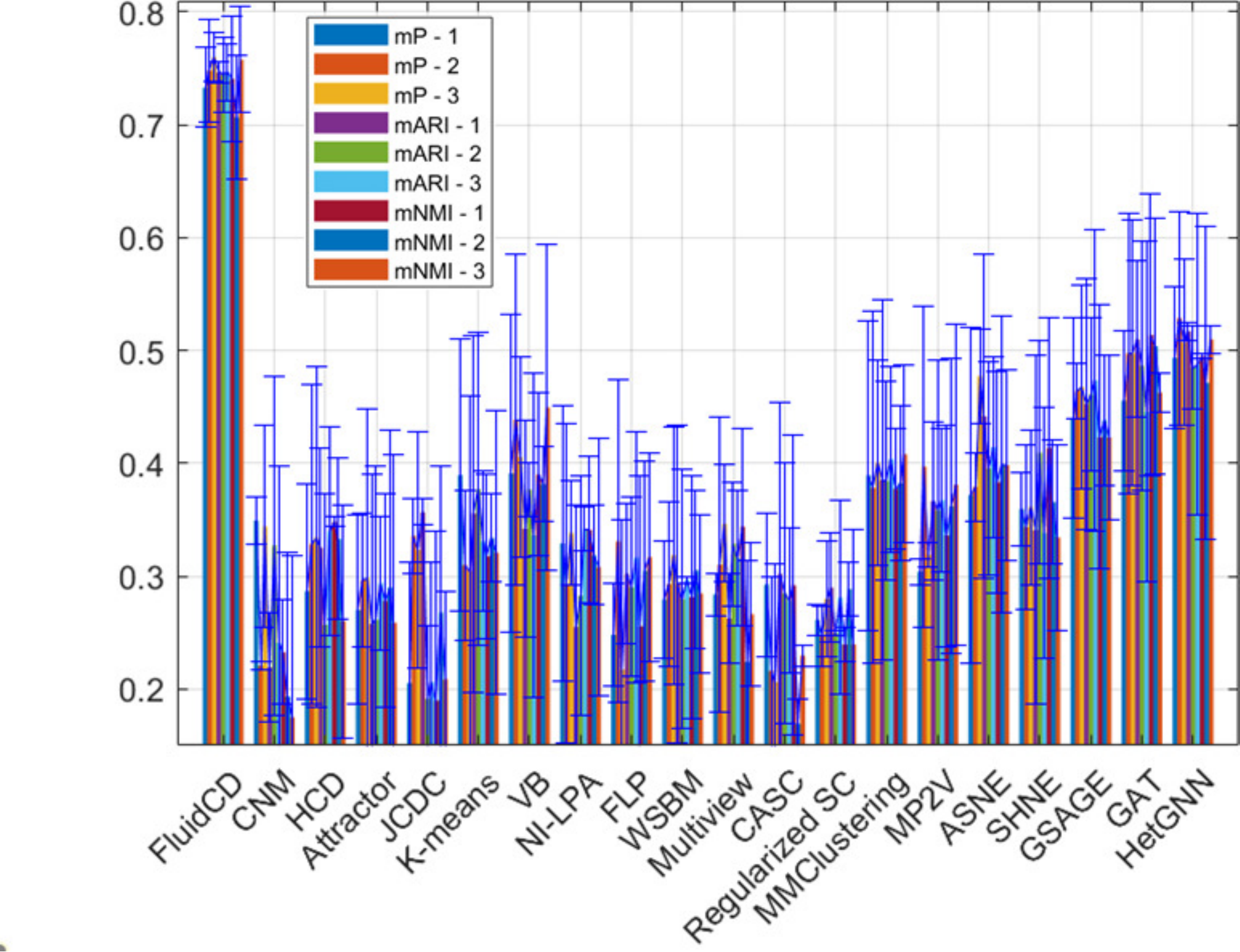} 
	\caption{Community detection results achieved when analyzing the unbalanced datasets obtained from the multimodal brain-computer interface dataset. The same notation as in Fig. \ref{fig_bar_RS} applies here.}
	\label{fig_bar_BCI}
\end{figure}

\begin{figure}[htb]
	\centering
	\includegraphics[width=1\columnwidth]{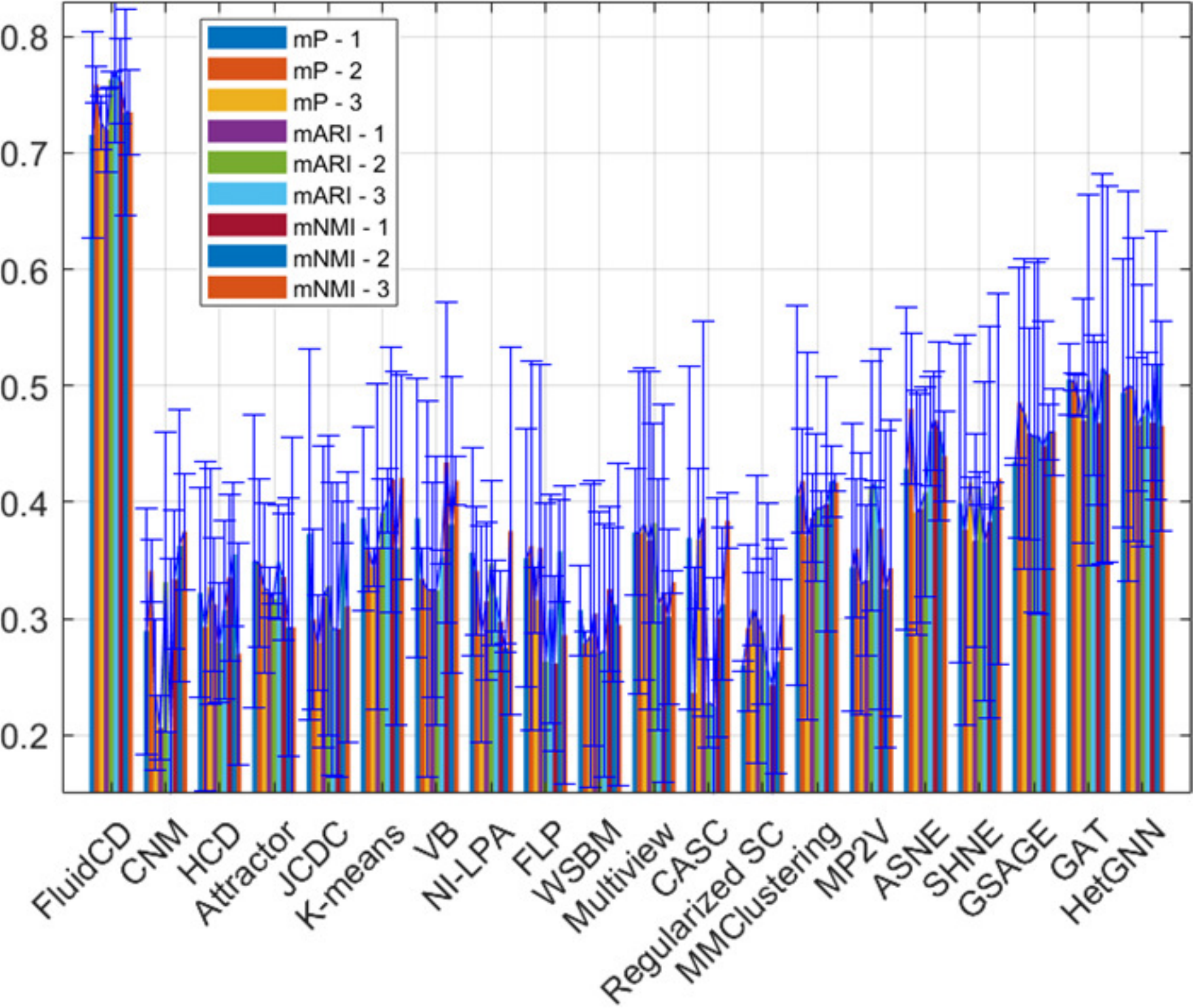} 
	\caption{Community detection results achieved when analyzing the unbalanced datasets obtained from the multimodal photovoltaic energy dataset. The same notation as in Fig. \ref{fig_bar_RS} applies here.}
	\label{fig_bar_PV}
\end{figure}

\begin{figure}[htb]
	\centering
	\includegraphics[width=1\columnwidth]{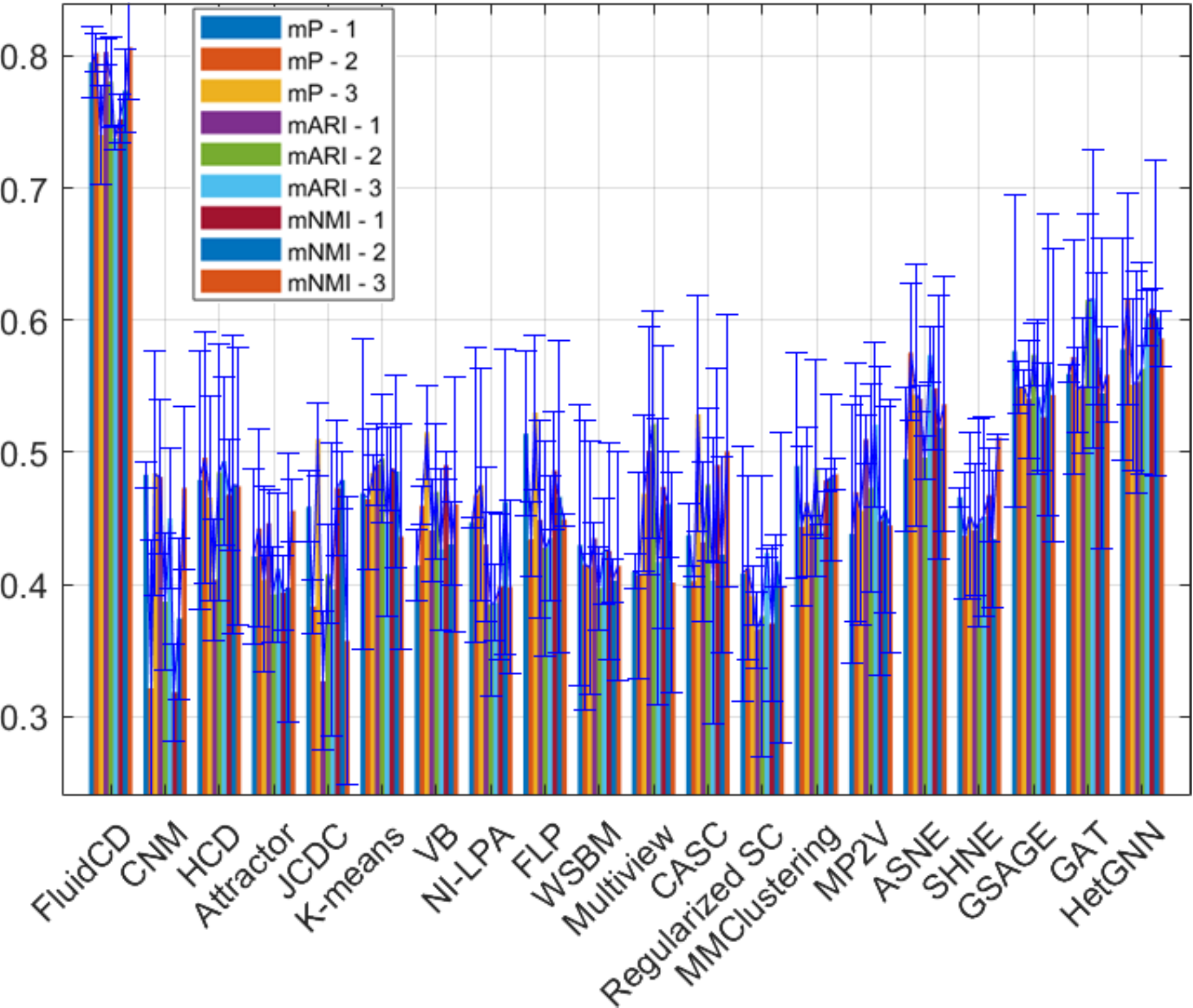} 
	\caption{Community detection results achieved when analyzing the unbalanced datasets obtained from the unimodal temporal OSWF monitoring dataset. The same notation as in Fig. \ref{fig_bar_RS} applies here.}
	\label{fig_bar_HYWIND}
\end{figure}

\begin{figure}[htb]
	\centering
	\includegraphics[width=1\columnwidth]{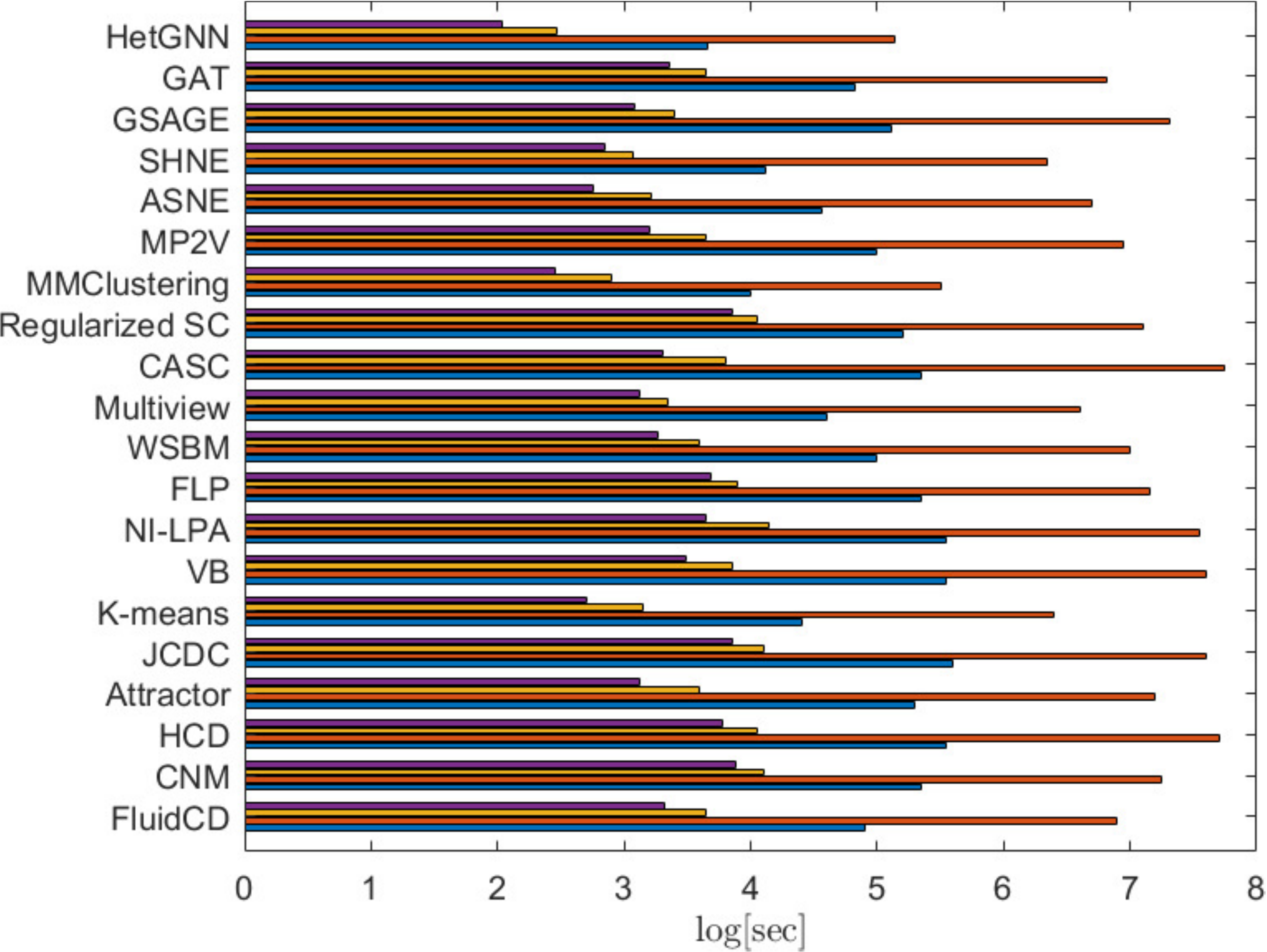} 
	\caption{Execution time (in log(seconds)) required by the community detection algorithms to achieve the results in Fig. \ref{fig_res_heat_RS}-\ref{fig_res_heat_HYWIND} for the multimodal remote sensing, brain computer interface, photovoltaic energy, and off-shore wind farm datasets (blue, red, orange, and purple bars, respectively).}
	\label{fig_compcompl}
\end{figure}

\textbf{L3 - unbalanced data:} We used the datasets in Section \ref{sec_exp_res} to test the ability of the proposed fluid community detection strategy to extract reliable information when unbalanced datasets are taken into account. 
To this aim, we modified the given datasets by changing the distribution of the samples to be analyzed. 
Specifically, we made the samples belonging to one chosen class for each dataset (i.e., 'building' in the multimodal remote sensing dataset; 'move left arm up' in the multimodal brain-computer interface dataset; '50$\%$ production' for the multimodal photovoltaic energy; 'sinusoidal movement' for the unimodal temporal OSWF monitoring dataset) become the 60$\%$ of the samples to be investigated. 
The remaining 40$\%$ of the datasets to be considered are randomly picked from the samples associated with the other $K_C - 1$ classes, to make them uniformly distributed in this second part of the dataset to be analyzed. 
Then, we performed community detection by means of the methods that have been previously discussed in this Section. 
We repeated this operation 100 times. 

Figures \ref{fig_bar_RS} to \ref{fig_bar_HYWIND} report the results we obtained. 
For each community detection methods, the average values achieved for each index in Appendix \ref{app_metrics} over 100 runs are displayed as bars. 
Moreover, the confidence interval for each index is displayed as blue errorbar. 
The observations we drew for the previous tests still apply here. 
In fact, we can notice that the proposed fluid community detection scheme clearly outperforms the other architectures, and show smaller confidence intervals than those of the other methods across all datasets. 
On one hand, these results show how unbalanced datasets can actually affect the performance of automatic learning even when performed in unsupervised fashion. 
On the other hand, these results are consistent with those that we reported throughout the paper. 
Hence, they show that the proposed strategy is robust to unbalanced datasets.


\textbf{Execution time:} 
Finally, Fig. \ref{fig_compcompl} displays the execution time (in $\log$[sec]) for the aforesaid algorithms to achieve community detection (shown in Figure \ref{fig_res_heat_RS}-\ref{fig_res_heat_HYWIND}) on the multimodal remote sensing, brain-computer interface, photovoltaic energy, and off-shore wind farm monitoring datasets. 
For each scheme, these results are shown in blue, red, orange, and purple bars, respectively. 
The proposed community detection algorithm based on fluid graph representation delivers a performance that is comparable with the other methods in this respect. 
Its computational complexity can be expressed as $O(N\theta)$, where $N$ identifies the number of samples of the given dataset, and $\theta$ is the number of steps necessary for the optimization in (\ref{eq_specclust4}) to converge \cite{normalizedCut}. 
It is possible to appreciate how the size of the datasets is typically the driving force behind these outcomes. 
Nevertheless, these results show how taking advantage of the modularity property of deep learning-based approach (such as that in \cite{CommunityDetection_multi3}) could reduce the computational load of the architecture.
Hence, to improve the scalability of the approach we presented in this work, a deep learning analysis relying on the proposed fluid graph representation will be considered in future works.

%
%

%
%
 \section{Conclusion}
 \label{secconcl}
 
 In this paper, we introduce a novel approach for graph representation with special focus of multimodal data analysis. 
 The proposed scheme is based on the use of a fluid diffusion model to characterize the interactions among samples, and hence the mechanism for information propagation in graphs. 
 This approach is meant to address several issues in modern multimodal data analysis, when large scale datasets collected by heterogeneous sources of information are investigated. 
 In particular, the proposed framework aims to provide an accurate and versatile automatic characterization of the relationships among samples, so that a robust community detection can be derived for complex datasets where multiple statistical, geometrical, and semantic distributions are collected. 
 In this respect, the main contributions of this work are:
 \begin{itemize}
 	\item the introduction of a novel model for graph information propagation based on fluid diffusion; 
 	\item the development of a compact description of the interactions among data that takes advantage of the eigenanalysis of flow velocity matrix, so to guarantee a data driven set-up for multimodal data characterization;
 	\item the development of an architecture for community detection based on fluid dynamics, which allows to obtain a solid characterization of the connections among samples in complex datasets (e.g., where samples show different levels of reliability and where the relevance of the feature might vary across the data).
 \end{itemize} 

We tested our approach on three diverse real multimodal datasets in terms of functional information retrieval. 
The experimental results we achieved show the solidity of our approach, as the proposed framework is able to outperform the state-of-the-art methods in community detection, which are all based on heat diffusion model. 
Thus, it is possible to state that the fluid diffusion model could be a valid option to improve the characterization of multimodal data analysis and to enhance the understanding of the functions and phenomena underlying the multimodal records, so to fully exploit the potential provided by the diversity of modalities collected in the datasets under exam. 
This approach can thus represent the platform on which multimodal data analysis could be based to address the main issues in modern multimodal data analysis. 
Future works will be then devoted to explore the development of fluid diffusion-based data analysis schemes for specific tasks, from semisupervised learning to explainable data analysis, to prediction and inference in complex operational scenarios. 
Moreover, the results we achieved might signal the need to investigate the geometry of multimodal spaces by means of new computational strategies, which theoretical and methodological basis will be studied in the next steps of this work. 

\section{Acknowledgements}

This work is funded in part by Centre for Integrated Remote
Sensing and Forecasting for Arctic Operations (CIRFA) and
the Research Council of Norway (RCN Grant no. 237906), the Visual Intelligence Centre for
Research-based Innovation funded by the Research Council of
Norway (RCN Grant no. 309439), 
the Automatic Multisensor remote sensing for Sea Ice
Characterization (AMUSIC) Framsenteret ”Polhavet” flagship project 2020, and the IMPETUS project funded by the European Union Horizon 2020 research and innovation program under grant agreement nr. 101037084.

\appendix
\subsection{Motivations of a new graph representation: a multimodal example}
\label{app_motiv}

In this Section, we provide an experimental example to support the statements we made in Section \ref{sec_meth_motnew} to support the need for a new graph representation for multimodal datasets.  

To this aim, let us consider a multimodal dataset that is considered a benchmark in the remote sensing community \cite{Trentodata}. 
This dataset consists of hyperspectral and Lidar observations acquired over a
rural area in the south of the city of Trento, Italy. 
Hence, by means of the hyperspectral sensor it is possible to obtain a characterization of the physical composition of the scene. 
As such, this source of information can help in discriminating between different elements in the considered region of interest, e.g., vegetation vs. asphalt. 
On the other hand, LiDAR observations lead to the generation of digital surface models (DSMs), which provide details on the height of the objects showing up in the considered area. 
Therefore, combining these records can help in obtaining an accurate characterization of specific elements (e.g., tree species) on the Earth's surface.  
The main properties of this dataset are summarized as follows:
$i)$	the size of the dataset is 600 $\times$ 166 pixels (each pixel is a sample of the dataset);
$ii)$ the sensor used to generate the LiDAR DSM and the associated features (summing up to 14 features in total) was Optech ALTM 3100EA sensor, having spatial resolution of 1m;
$iii)$the hyperspectral data acquired by AISA Eagle sensor consist of
63 bands ranging from 402.89 to 989.09nm, where the spectral
resolution is 9.2nm, with spatial resolution of 1m; 
$iv)$ six classes of interests were extracted,
including Building, Woods, Apple trees, Roads, Vineyard, and
Ground.

\begin{figure}[htb]
	\centering
	\includegraphics[width=1\columnwidth]{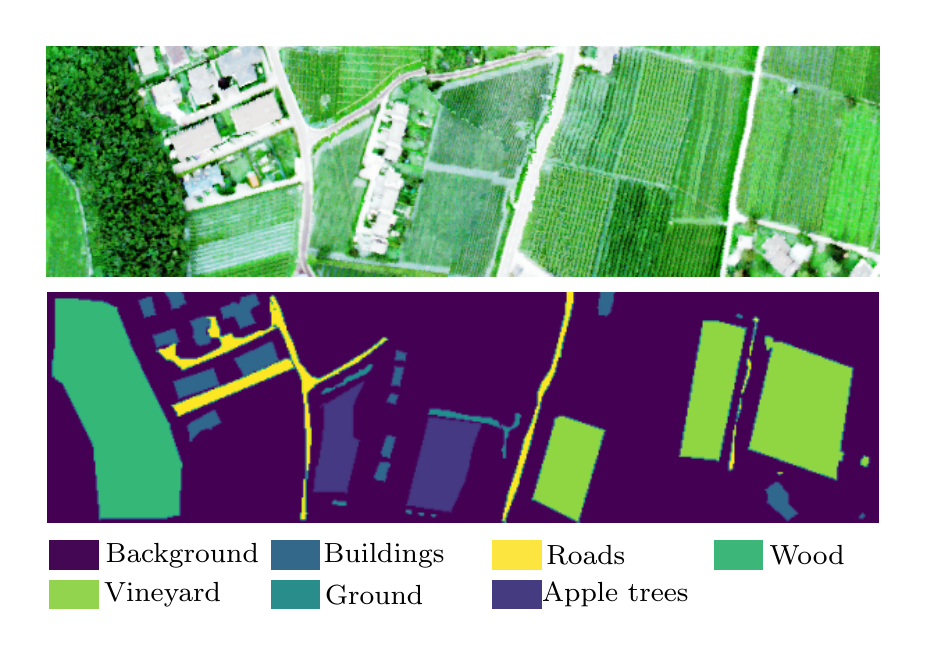} 
	\caption{Multimodal remote sensing dataset acquired over the region of Trento, Italy \cite{Trentodata}: RGB composite (top) and groundtruth map (bottom).}
	\label{fig_Trento_dataset}
\end{figure}

Fig. \ref{fig_Trento_dataset} displays the RGB composite of this dataset, as well as its groundtruth. 
It is worth noting that pixels in which geographical unit more materials associated with the aforementioned six thematic classes in the dataset 
are grouped in the "Background", which is not used for learning and performance comparison purposes.
%
%
%
Furthermore, the geographical extent of the dataset avoids the occurrence of heterogeneous atmospheric effects throughout the data. Hence, it is possible to state that the atmosphere composition could show uniform properties across the whole region, hence avoiding the need for focused correction processing across the records. 
Moreover, the classes showing up in the dataset are pretty distinct from each other, both in the spectral and spatial domains. 
However, it is possible to prove that analyzing this dataset by means of classic graph representation might not lead to accurate and reliable understanding of the considered scene.
Indeed, although acquired in ideal conditions, the diversity of the records make the classic graph representation based on heat diffusion model not adequate for the characterization of this dataset, as it can be proven by considering the description in Section \ref{sec_methmot} and especially computing the trend of the quantities in (\ref{eq_isoleigen_cond}). 

Specifically, let us start to consider the relevance of the features for each class in the dataset. 
In fact, investigating and selecting the most relevant features in the samples associated with each class we can obtain an initial estimate of how the different records could contribute to the characterization of the dataset. 
To this aim, considering an algorithm for feature selection that can take into account the nonlinearities and sparsity of the information across the samples can help in understanding the actual interactions among features, hence their statistical relationships. 
Therefore, without losing generality, we can run over the samples associated with each class  the algorithm for adaptive dimensionality reduction proposed in \cite{JSTARS_GKMI21} to select the most relevant features for each subregion showing homogeneous properties in the dataset in Fig. \ref{fig_Trento_dataset}. 

\begin{figure}[htb]
	\centering
	\includegraphics[width=1\columnwidth]{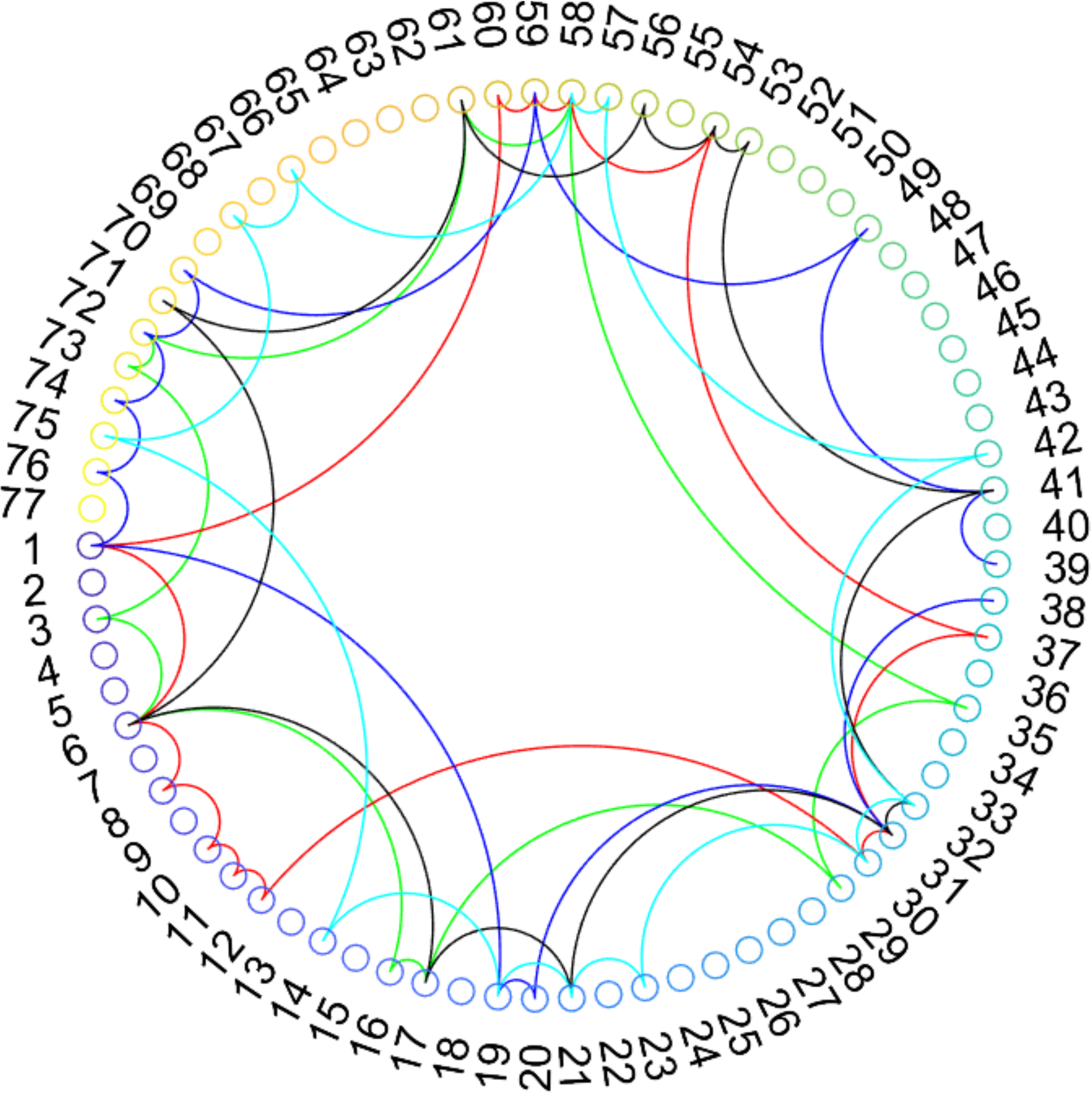} 
	\caption{Circular graph plot showing the selected features for five pixels associated with the class "Apple trees" in the dataset in Fig. \ref{fig_Trento_dataset}. The features selected for each pixel by means of the algorithm in \cite{ensembleADR} are displayed by lines of different colors connecting circles associated with the corresponding indices, according to Table \ref{tab_legend_Trento}.}
	\label{fig_fluid_circle_Trento}
\end{figure}

In order to visualize the features (i.e., spectral channels in case of the hyperspectral sensor, properties of the digital surface model for the LiDAR sensor) that are selected for different pixels belonging to the same class, we can use a circular graph plot as the one in Fig. \ref{fig_fluid_circle_Trento}. 
In this diagram, the features are displayed as circles distributed around a circular disk marked by a unique index: the correspondence between these indices and the features that have been previously mentioned is reported in Table \ref{tab_legend_Trento}
. 
At this point, the features selected for each pixel by means of the aforesaid dimensionality reduction method are displayed by lines connecting the nodes associated with these selected records. 
It is worth recalling that we are interested in exploring the interactions among features within the samples associated with each class in the dataset. 
Thus, we can draw a circular graph plot for each of the six classes occurring in the dataset in Fig. \ref{fig_Trento_dataset} and display the features that are selected for every pixels in the given class by lines with different colors. 
For sake of clarity, in Fig. \ref{fig_fluid_circle_Trento} we display the features selected for five pixels in the class "Apple trees": nevertheless, we achieved similar outputs carrying the same analysis when considering the other classes in the dataset. 

\begin{table}[!th]
	\renewcommand{\arraystretch}{1.3}
	\caption{Legend of the features shown in Fig. \ref{fig_fluid_circle_Trento}}
	\label{tab_legend_Trento}
	\centering
	\begin{tabular}{|c|c|}
		\hline
		\bfseries Index & \bfseries Meaning  \\
		\hline
		1 - 14 & LiDAR  \\
		\hline
		15 - 19 &  402.89 nm - 440.7 nm [Violet]   \\
		\hline
		20 - 24 & 450.16 nm - 487.98 [Blue]     \\
		\hline
		25 - 32 &  497.43 nm - 563.62 nm [Green]  \\
		\hline
		33 - 35 &  573.07 nm - 591.91 nm [Yellow]   \\
		\hline
		36 - 39 & 601.44 nm - 629.8 nm [Orange]   \\
		\hline
		40 - 45 & 639.2 nm - 686.53 nm [Light red]   \\
		\hline
		46 - 52 & 695.99 nm -  752.71 nm [Dark red] \\
		\hline
		53 - 77 & 762.17 nm - 989.09 [Near Infrared] \\
		\hline
	\end{tabular}
\end{table}

Let us take a closer look to Fig. \ref{fig_fluid_circle_Trento}.
In particular, it is possible to observe that each considered pixel is characterized by a different subset of features, as the lines of different colors connect distinct circles (i.e., features) in the disk. 
This is a very interesting result, since one might expect that pixels belonging to one single class would be characterized by one very homogeneous set of features. 
Fig. \ref{fig_fluid_circle_Trento} clearly contradicts this intuition. 
Indeed, the graph in Fig. \ref{fig_fluid_circle_Trento} results from the complex set of nonlinearities induced by different conditions of illumination, scattering, morphology that are intrinsically affecting the observations of the multimodal remote sensing dataset, and that are taken into account by the feature selection algorithm in \cite{JSTARS_GKMI21}. 
As such, this method is able to highlight how different subsets of features are actually channeling the relevant information for different pixels. 
In particular, it is worth noting that the selected features are not only different from pixel to pixel, but also associated with very distinct observations in the dataset.
For instance, to characterize the pixel associated with the red line in Fig. \ref{fig_fluid_circle_Trento} the features provided by the LiDAR sensor are apparently playing a key role, whilst the records acquired in the near infrared portion of the electromagnetic spectrum are not so significant. 
On the other hand, the pixel associated with the blue line in Fig. \ref{fig_fluid_circle_Trento} shows a dual behavior, i.e., the features in the near infrared seem very relevant for it characterization, whilst the LiDAR records could be discarded. 

\begin{figure}[htb]
	\centering
	\includegraphics[width=1\columnwidth]{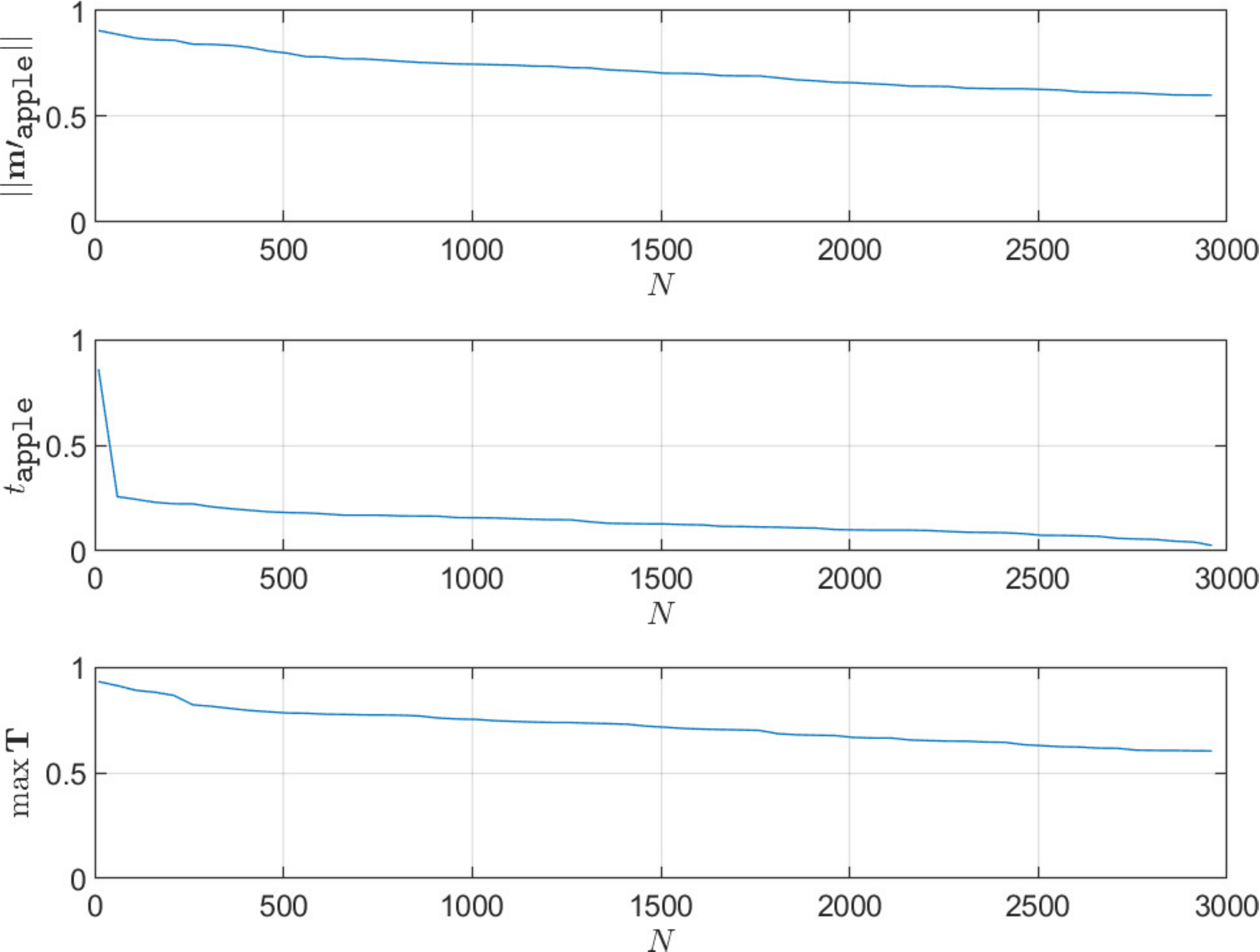} 
	\caption{Trends of the quantities in (\ref{eq_isoleigen_cond}) for the dataset in Fig. \ref{fig_Trento_dataset} as the number of pixels used in their computation increases. The values of $||\textbf{m}_l'||$ and $t_l$ have been computed for the class "Apple trees" (top and middle plots, respectively). The bottom plot displays the trend of the maximum value of the matrix $\textbf{T}$. }
	\label{fig_fluid_trend_Trento}
\end{figure}

The results in Fig. \ref{fig_fluid_circle_Trento} provide material to discuss the composition of the matrices and vectors in (\ref{eq_isoleigen_cond}). 
%
%
In fact, since the relevant features for each pixel belonging to one class might vary substantially, it is possible to expect that the off-diagonal elements of the covariance matrices $\textbf{C}_l'$ might show not negligible values. 
As a result, the energy will not be concentrated in the diagonal elements of these matrices, which is implicitly a major factor for the conditions in (\ref{eq_isoleigen_cond})
to hold \cite{COUILLET16}. 
In fact, when progressively computing the quantities 
in (\ref{eq_isoleigen_cond})
as the number of samples increases, it is possible to obtain their trends: 
Fig. \ref{fig_fluid_trend_Trento} reports the results we achieved as the number of pixels used in their computation increases. 
Specifically, the values of $||\textbf{m}_l'||$ and $t_l$ have been computed for the class "Apple trees" (top and middle plots, respectively). 
On the other hand, the bottom plot in Fig. \ref{fig_fluid_trend_Trento} displays the trend of the maximum value of the matrix $\textbf{T}$ that is constructed according to the definitions in Section \ref{sec_meth_heat}. 

As these trends are valid also for the other classes in the dataset, it is therefore apparent that the quantities 
in (\ref{eq_isoleigen_cond}) 
converge to a finite non-zero value. 
More importantly, this proves that the values of the parameters 
in (\ref{eq_isoleigen_cond})
would not diverge to infinity, 
hence jeopardizing (according to the discussion we previously reported) the validity and appropriateness of a classic graph representation based on heat diffusion model for the considered dataset in Fig. \ref{fig_Trento_dataset}. 
As a result, carrying out a characterization of the dataset by means of classic graph-based data analysis (i.e., processing and investigating the data over a classic graph representation structure) might not be the best choice, and indeed might lead to a strong degradation of the overall analysis. 

To further support this statement, we analyzed the dataset in Fig. \ref{fig_Trento_dataset} by means of state-of-the-art supervised and semisupervised methods based on the classic graph representation introduced in Section 
\ref{sec_meth_heat} 
\cite{GCN_Cheby,GCN_Classic,GCN_S,SIGN}. 
Moreover, for comparison, we used a method based on ensemble learning approach \cite{ensembleADR}. 
It is worth noting that deeply investigating the capacity and limitations of all these methods is out of the scope of this paper. 
Nonetheless, taking a look to the results that these algorithms provide when analyzing the dataset in Fig. \ref{fig_Trento_dataset} with comparable hyperparameter set-ups can help us in consolidating the statements and observations we previously made. 

In particular, focusing once again on the "apple trees" class, it is possible to compute the average misclassification error between the "apple trees" and "wood" classes obtained over all the experiments we ran: Table \ref{tab_GCN_Trento} reports these results. 
It is therefore possible to appreciate that the algorithms based on the analysis of the graph structure defined according to the guidelines in Section \ref{sec_meth_heat} 
show a substantial increase of the error with respect to the ensemble learning-based architecture in \cite{ensembleADR}. 
Hence, the classic graph representation based on heat diffusion model apparently leads to a substantial degradation of the ability of the architectures to characterize and interpret the samples in the dataset. 
Recalling the considerations we previously made on the quality of the Trento dataset, these error distributions emphasize that the classic graph representation might not be adequate for a lot of situations encompassed by modern data analysis, leading to unacceptable degradation in the characterization performance. 

\begin{table}[!th]
	\renewcommand{\arraystretch}{1.3}
	\caption{Average misclassification between "apple trees" and "wood" classes in the dataset in Fig. \ref{fig_Trento_dataset}.}
	\label{tab_GCN_Trento}
	\centering
	\begin{tabular}{|c|c|}
		\hline
		\bfseries Method & \bfseries Error  \\
		\hline
		Graph convolutional network (GCN) \cite{GCN_Classic} & 14 $\pm$ 4.3  \\
		\hline
		Simple GCN (S-GCN) \cite{GCN_S} &  12.3 $\pm$ 2.5   \\
		\hline
		Deep neural network with & 11.8 $\pm$ 1.9 \\
		Chebyshev polynomial approximation (ChebNet) \cite{GCN_Cheby} &      \\
		\hline
		Scalable inception graph neural network (SIGN) \cite{SIGN} &  9.9 $\pm$ 3.3  \\
		\hline
		Ensemble learning with & 2.1 $\pm$  1.2 \\
		adaptive dimensionality reduction (ADR-EL) \cite{ensembleADR}  &  \\
		\hline
	\end{tabular}
\end{table}

\subsection{Derivation of the Fokker-Planck equation for fluid diffusion in porous media}
\label{app_deriv1}

In this Section, we report the steps that are taken to derive the Fokker-Planck equation for fluid diffusion in porous media starting from the analysis of the diffusion process in the form of (\ref{eq_dynsyst_fluid})
by means of the It\^{o}'s lemma \cite{fluid_FokkerPlanck7}. 
In particular, let us consider (\ref{eq_Ito4_fluid}).  
Since the function $g(\textbf{x})$ is arbitrary by construction, this equation is satisfied when the following condition applies:

\begin{equation}
\frac{\partial p(\textbf{x},t)}{\partial t} + \nabla \cdot \left[\textbf{a}(\textbf{x})p(\textbf{x},t) \right] = \beta'(\textbf{x},t), 
\label{eq_Ito_beta_1}
\end{equation} 

\noindent where $\beta'(\textbf{x},t) =  \sum_{i,j=1}^{n} \partial x_i \partial x_j (\tilde{B}_{ij} p(\textbf{x},t))$. 
Let us focus on each term of this sum: we can then write as follows:

\begin{eqnarray}
\partial x_i \partial x_j (\tilde{B}_{ij} p(\textbf{x},t)) &=& \left[\partial x_i \partial x_j  \tilde{B}_{ij} \right] p(\textbf{x},t)) \nonumber \\
&+& \left[\partial x_j \tilde{B}_{ij}\right] \partial x_i p(\textbf{x},t)) \nonumber \\ 
& + & \partial x_i \left[\tilde{B}_{ij} \partial x_j p(\textbf{x},t)) \right].
\label{eq_Ito_beta_2} 
\end{eqnarray}

At this point, $\beta'(\textbf{x},t)$ becomes: 

\begin{eqnarray}
\sum_{i,j=1}^{n} \partial x_i \partial x_j (\tilde{B}_{ij} p(\textbf{x},t)) &=& \left[ \nabla \cdot \left(  \nabla \tilde{\textbf{B}}(\textbf{x})\right) \right] p(\textbf{x},t)) \nonumber \\
& + & \left[ \nabla \tilde{\textbf{B}}(\textbf{x}) \right] \cdot \nabla p(\textbf{x},t) \nonumber \\
& + & \nabla \cdot \left[ \tilde{\textbf{B}}(\textbf{x}) \nabla p(\textbf{x},t) \right], 
\label{eq_Ito_beta_3}
\end{eqnarray}

\noindent where $\nabla = [\frac{\partial}{\partial x_i}]_{i=1, \ldots, n}$. 
Substituting this term in (\ref{eq_Ito_beta_1}), then we obtain:

\begin{eqnarray}
\frac{\partial p(\textbf{x},t)}{\partial t}  + &  \nabla \cdot \left[ \textbf{a}(\textbf{x})p(\textbf{x},t) \right] - \left[ \nabla \cdot (\nabla \tilde{\textbf{B}}(\textbf{x})) \right] p(\textbf{x},t) \nonumber \\
= & \left[ \nabla \tilde{\textbf{B}}(\textbf{x}) \right] \cdot \nabla p(\textbf{x},t) - \nabla \cdot \left[ \tilde{\textbf{B}}(\textbf{x}) \nabla p(\textbf{x},t)\right],
\nonumber 
\end{eqnarray}

\noindent i.e., eq. (\ref{eq_Ito5_fluid}). 
This hence leads to the formulation of the diffusion process represented by (\ref{eq_dynsyst_fluid}) 
in terms of the Fokker-Planck equation for fluid diffusion in porous media.


\subsection{Derivation of transition probability density function in the fluid diffusion model}
\label{app_deriv2}

In this Section, we report the details of the derivation of the transition probability density function $p(\textbf{x}(t+\epsilon) = \textbf{x}_j|\textbf{x}(t)=\textbf{x}_i)$ characterizing the fluid diffusion model as described in Section \ref{sec_fluid_repr}. 

\begin{figure}[htb]
	\centering
	\includegraphics[width=1\columnwidth]{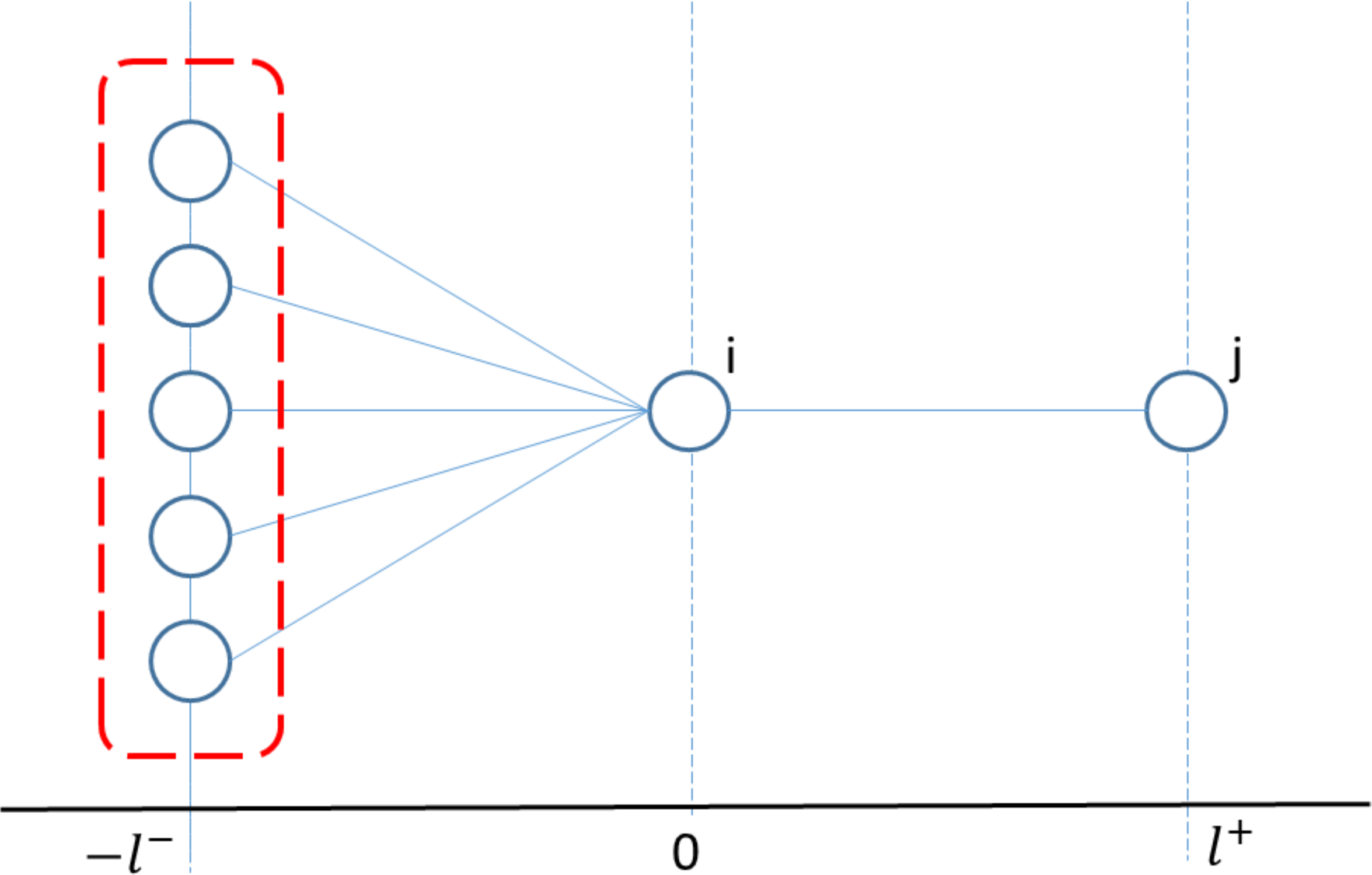} 
	\caption{Diffusion system centered on node $i$ to analyze the transition probability to node $j$ as set by the time domain random walk approach \cite{Fluid_TDRW1}. The red box identify the set ${\cal N}(i)\setminus j$.}
	\label{fig_TDRW}
\end{figure}

The time domain random walk strategy \cite{Fluid_TDRW1,Fluid_TDRW2} solves the Green function problem derived from 
(\ref{eq_FokkerPlanck_fluid})
by imposing initial conditions and absorbing boundary conditions to the diffusion system centered on the node $i$. 
In other terms, this strategy considers boundary conditions at the outer edges of the $i$-th node and a pulse initial condition of unit mass at $\textbf{x}=0$, where the $i$-th node is located. 
Furthermore, we assume to locate the $j$-th node at the right boundary $l^+$, while the remaining nodes in the neighborhood of node $i$ ${\cal N}(i)$ (i.e., the nodes adjacent to node $i$) are located at the left boundary at $-l^-$, where $l^+, l^- >0$. 
Fig. \ref{fig_TDRW} displays the aforesaid set-up. 
Hence, the boundary and initial conditions of the Green function $G(\textbf{x},t)$ (solution of the system in 
(\ref{eq_FokkerPlanck_fluid})) are defined as follows: 

\begin{eqnarray}
G(l^-,t) = G(l^+,t) & = & 0  \nonumber \\
G(\textbf{x},0) &=& \delta(\textbf{x}),
\label{eq_Green1}
\end{eqnarray}

\noindent where $\delta(\cdot)$ is the Dirac delta. 
Moreover, solutions for the Green function problem can be written as follows: 

\begin{equation}
G(\textbf{x},t) = G_-(\textbf{x},t)\Theta(-\textbf{x}) + G_+(\textbf{x},t)\Theta(\textbf{x}),
\label{eq_Green2}
\end{equation}

\noindent where $\Theta(\textbf{x})$ identifies the Heaviside step function \cite{Fluid_TDRW1}. 
At this point, we can impose the continuity condition of $G(\textbf{x},t)$ at $\textbf{x}=0$, which implies that $G_-(0,t) = G_+(0,t)$, according to the aforesaid notation. 

With this in mind, it is possible to derive the definition of $G(\textbf{x},t)$ by considering the projection in the Laplace space of the fluid diffusion system. 
In fact, the 
Laplace transform of 
(\ref{eq_FokkerPlanck_fluid}) over the time variable (denoted by $G^{\cal L}$)
with the condition $G(\textbf{x},0) = \delta(\textbf{x})$ leads to the following
\cite{Fluid_TDRW1,Fluid_TDRW2}: 

\begin{equation}
\xi G^{\cal L}(\textbf{x},\xi) + \nabla \cdot \left[\textbf{v}(\textbf{x})G^{\cal L}(\textbf{x},\xi)\right] - \nabla \cdot \tilde{\textbf{B}}(\textbf{x})\nabla G^{\cal L}(\textbf{x},\xi) = \delta(\textbf{x}).
\label{eq_Green3}
\end{equation}

Moreover, it worth recalling that the $G(\textbf{x},t)$ function can be expressed as in (\ref{eq_Green2}). 
As such, it is possible to write as follows:

\begin{eqnarray}
G^{\cal L}_\pm(\textbf{x},\xi) & = & A_\pm^{(1)}(\xi) \exp \left[ \frac{v_\pm}{2\tilde{B}_\pm} \textbf{x} (1-\alpha_\pm(\xi)) \right]  \nonumber \\
& + & A_\pm^{(2)}(\xi) \exp \left[ \frac{v_\pm}{2\tilde{B}_\pm} \textbf{x} (1+\alpha_\pm(\xi)) \right], 
\label{eq_Green4}
\end{eqnarray}

\noindent where $G^{\cal L}_-(\textbf{x},\xi)$ and $G^{\cal L}_+(\textbf{x},\xi)$ represent the Laplace transform of $G_-(\textbf{x},t)$ and $G_+(\textbf{x},t)$, respectively. Moreover, taking into account the quantities that have been defined in Section 
\ref{sec_fluid_repr}, we can define $v_\pm$ and $\tilde{B}_\pm$ as follows: 

\begin{eqnarray}
v_+ & = & v_{ij}, \\
v_- & = & \sum_{m \in {\cal N}(i) \setminus j} \frac{v_{im}}{|{\cal N}(i)|- 1}, \nonumber \\
\tilde{B}_+ & = & \tilde{B}_{ij}, \nonumber \\
\tilde{B}_- & = & \sum_{m \in {\cal N}(i) \setminus j} \frac{\tilde{B}_{im}}{|{\cal N}(i)|- 1}, \nonumber
\end{eqnarray}

\noindent where ${\cal N}(i)$ identifies the neighborhood of node $i$, i.e., the set of nodes adjacent to node $i$. 
Finally, we define $\alpha_\pm(\xi)= \sign(v_\pm) \sqrt{1+4\tilde{B}_\pm\xi/v_\pm^2}$. 

Recalling that $G(i^-,t) = G(i^+,t) =  0$, this equation can be rewritten as follows: 

\begin{eqnarray}
G^{\cal L}_\pm(\textbf{x},\xi) & = & A_\pm^{(1)}(\xi) \exp\left[\frac{v_\pm}{2\tilde{B}_\pm} \textbf{x}\right] \left \{ \exp\left[-\frac{v_\pm}{2\tilde{B}_\pm} \textbf{x} \alpha_\pm(\xi)\right] \right. \nonumber\\
& - & \left. \exp\left[-\frac{v_\pm}{2\tilde{B}_\pm} (\textbf{x}\mp 2l^\pm) \alpha_\pm(\xi)\right] \right \}
\label{eq_Green5}
\end{eqnarray}

\noindent Recalling the condition for which $G_-(0,t) = G_+(0,t)$, it is possible to write $G^{\cal L}_-$ and $G^{\cal L}_+$ as follows: 

\begin{eqnarray}
G^{\cal L}_-(\textbf{x},\xi) & = & A_-^{(1)}(\xi) \exp\left[\frac{v_-}{2\tilde{B}_-} \textbf{x} \right] 
\left \{ \exp\left[-\frac{v_-}{2\tilde{B}_-} \textbf{x} \alpha_-(\xi) \right] \right. \nonumber\\
& - & \left. \exp \left[-\frac{v_-}{2\tilde{B}_-} (\textbf{x}+ 2l^-) \alpha_-(\xi) \right] \right \},
\label{eq_Green6}
\end{eqnarray}

\begin{eqnarray}
G^{\cal L}_+(\textbf{x},\xi) &= & A_-^{(1)}(\xi) \frac{1-\exp\left[\frac{l^- v_-}{\tilde{B}_-} \right] \alpha_-(\xi)}{1-\exp\left[-\frac{l^+ v_+}{\tilde{B}_+} \right] \alpha_+(\xi)} \exp \left[\frac{v_+}{2 \tilde{B}_+} \textbf{x} \right] \nonumber \\
& \cdot& \left \{ \exp \left[- \frac{v_+}{2 \tilde{B}_+} \textbf{x} \alpha_+(\xi) \right] \right. \\
&-& \left. \exp \left[ \frac{v_+}{2 \tilde{B}_+} (\textbf{x} - 2l^+) \alpha_+(\xi) \right] \right \}. \nonumber
\label{eq_Green7}
\end{eqnarray}

Thus, it is possible to summarize these equations as follows (where for convenience the variable $A_-^{(1)}$ is renamed $A(\xi)$)): 

\begin{equation}
G^{\cal L}_\pm(\textbf{x},\xi) = A(\xi) \sinh \left[ (l^\pm \mp \textbf{x}) v_\pm^* \alpha_\pm(\xi) \right] \csch \left[v_\pm^* \alpha_\pm(\xi) \right], 
\label{eq_Green_Gpm}
\end{equation}

\noindent where $v_\pm^* = l^\pm v_\pm / 2\tilde{B}_\pm$ and $\csch[z] = 1/\sinh[z] = 2/(\exp[z] - \exp[-z])$. 
At this point, it is worth to recall that the definition of the Green function $G(\textbf{x},t)$ is regulating the diffusion process through the nodes. 
Furthermore, the functions $G_-$ and $G_+$ determine the fluxes through the boundaries that the time domain random walk approach imposes. 
Hence, they can be used to determine the first
arrival time densities at the boundaries, which can be written as follows: 

\begin{equation}
\phi_\pm(t) = \tilde{B}_\pm \left. \nabla G_\pm(\textbf{x},t) \right|_{\textbf{x} = \pm l^\pm}, 
\label{eq_Green_Gpm2}
\end{equation}

\noindent where $\nabla = [\frac{\partial}{\partial x_i}]_{i=1, \ldots, n}$. 

According to the notation that we used so far, $\phi_+(t) dt$
would then denote the joint probability of the transition 
to occur towards the right boundary (i.e., from node $i$ to node $j$) with an
arrival time in $[t,t+\epsilon[$ \cite{Fluid_TDRW1,Fluid_TDRW2}. 
As such, we can then define the  transition probability density
$p(\textbf{x}(t+\epsilon) = \textbf{x}_j|\textbf{x}(t)=\textbf{x}_i)$ 
as follows \cite{Fluid_TDRW1}: 

\begin{equation}
p(\textbf{x}(t+\epsilon) = \textbf{x}_j|\textbf{x}(t)=\textbf{x}_i) = p_+ = \int_0^{+\infty} \phi_+ dt
\label{eq_fluid_trans_prob}
\end{equation}

Therefore, it is crucial to derive the analytical expression of the $A(\xi)$ parameter in (\ref{eq_Green_Gpm}) so that the transition probability in (\ref{eq_fluid_trans_prob}) can be retrieved. 
To this aim, let us integrate
(\ref{eq_Green3}) over \textbf{x}: this would lead to the following equation: 

\begin{eqnarray}
\xi \int G^{\cal L}(\textbf{x},\xi) d\textbf{x} & = & 1 + A(\xi) \nonumber \\
& \cdot & \left \{ v_- \alpha_-(\xi) \exp \left[- \frac{l^- v_-}{2 \tilde{B}_-}(1-\alpha_-(\xi)) \right] \right. \nonumber \\
& - & v_+ \alpha_+(\xi) \exp \left[ \frac{l^+ v_+}{2 \tilde{B}_+}(1-\alpha_+(\xi)) \right] \\
& \cdot & \left. \frac{1 - \exp \left[\frac{l^- v_-}{\tilde{B}_-} \alpha_-(\xi) \right]}{1 - \exp \left[-\frac{l^+ v_+}{\tilde{B}_+} \alpha_+(\xi) \right]} \right \}. \nonumber
\label{eq_Green8}
\end{eqnarray}

On the other hand, considering the definition of $G^{\cal L}_-(\textbf{x},\xi)$ and $G^{\cal L}_+(\textbf{x},\xi)$ in (\ref{eq_Green6}) and (\ref{eq_Green7}), integrating over \textbf{x} and multiplying by $\xi$ the Laplace transform of (\ref{eq_Green2}), we can write as follows: 

\begin{eqnarray}
\xi \int G^{\cal L}(\textbf{x},\xi) d\textbf{x} & = & \frac{1}{2} A(\xi) \left \{ v_+ \frac{1 - \exp [\frac{l^- v_-}{\tilde{B}_-} \alpha_-(\xi)]}{1 - \exp [-\frac{l^+ v_+}{\tilde{B}_+} \alpha_+(\xi) ]} \right. \nonumber \\
&\cdot & \left[ 1- \exp [-\frac{l^+ v_+}{\tilde{B}_+} \alpha_+(\xi) ] + \alpha_+(\xi) \right. \nonumber \\
& \cdot & \left( 1-2\exp [ \frac{l^+ v_+}{2\tilde{B}_+} (1-\alpha_+(\xi)) ] \right. \\
& +& \left. \left. \exp [-\frac{l^+ v_+}{2\tilde{B}_+} ] \right) \right] \nonumber \\
& - & v_- \left[ 1- \exp[\frac{l^- v_-}{\tilde{B}_-} \alpha_-(\xi)] + \alpha_-(\xi) \right. \nonumber \\
& \cdot & \left( 1-2\exp[-\frac{l^- v_-}{2\tilde{B}_-} (1-\alpha_-(\xi))] \right. \nonumber \\
&+&\left. \left.  \left. \exp[-\frac{l^- v_-}{2\tilde{B}_-}] \right) \right] \right \}. \nonumber
\label{eq_Green9}
\end{eqnarray}  

By equating (\ref{eq_Green8}) and (\ref{eq_Green9}), it is possible to obtain the definition of $A(\xi)$ as follows: 

\begin{equation}
A(\xi) = \left \{ \sum_{u \in \{+,-\}} v_u \left[\alpha_u(\xi) \coth \left(v_u^* \alpha_u(\xi) \right) + \tilde{u} \right] \right \}^{-1},
\label{eq_Green10}
\end{equation}

\noindent where $\tilde{u}=-1$ if $u$ is -, and $\tilde{u}=+1$ if $u$ is +.  
Consequently, we can derive the Laplace transform of the $\phi$ functions in (\ref{eq_Green_Gpm2}). In particular, $\phi_+^{\cal L}(t)$ can be written as follows: 

\begin{equation}
\phi_+^{\cal L}(\xi) = v_+^* \alpha_+(\xi) A(\xi) \csch \left[ v_+^* \alpha_+(\xi) \right].
\label{eq_Green11}
\end{equation}

Finally, we can derive the value of $p_+$ in (\ref{eq_fluid_trans_prob}) that can be written as follows:

\begin{equation}
p_+ = \frac{|v_+^*| \exp[v_+^*] \csch[|v_+^*|]}{\sum_{u \in \{+,-\}} |v_u^*| \exp[\tilde{u}\cdot v_u^*] \csch[|v_u^*|]},
\label{eq_fluid_trans_prob2_nonorm}
\end{equation}

\noindent where $\tilde{u}$ assumes the same value as in (\ref{eq_Green10}). 
It is worth noting that for a fully connected graph, and especially when unsupervised graph analysis is considered, the values of $l^\pm$ should be normalized. 
In fact, in this case, the random walk would model a unitary jump from node to node \cite{Fluid_TDRW1,Fluid_TDRW2}. 
Therefore, we can set $l^+=l^-=1$, $v_\pm^\dagger = v_\pm/2 \tilde{B}_\pm$, and write $p_+$ as follows:

\begin{equation}
p_+ = \frac{|v_+^\dagger| \exp\left[v_+^\dagger\right] \csch \left[|v_+^\dagger| \right]}{\sum_{u \in \{+,-\}} |v_u^\dagger| \exp \left[\tilde{u}\cdot v_u^\dagger \right] \csch \left[|v_u^\dagger| \right]}.
\label{eq_fluid_trans_prob2}
\end{equation}

Hence, we can use this equation to define the transition probability density
$p(\textbf{x}(t+\epsilon) = \textbf{x}_j|\textbf{x}(t)=\textbf{x}_i)$ $\forall (i,j) \in \{1, \ldots, N\}^2$, $i \neq j$, and then the elements of the \textbf{Q} matrix that has been mentioned in Section \ref{sec_fluid_repr}. 


\subsection{Effect of the Q matrix: a visual example}
\label{app_Qmat}

\begin{figure}[htb]
	\centering
	\includegraphics[width=1\columnwidth]{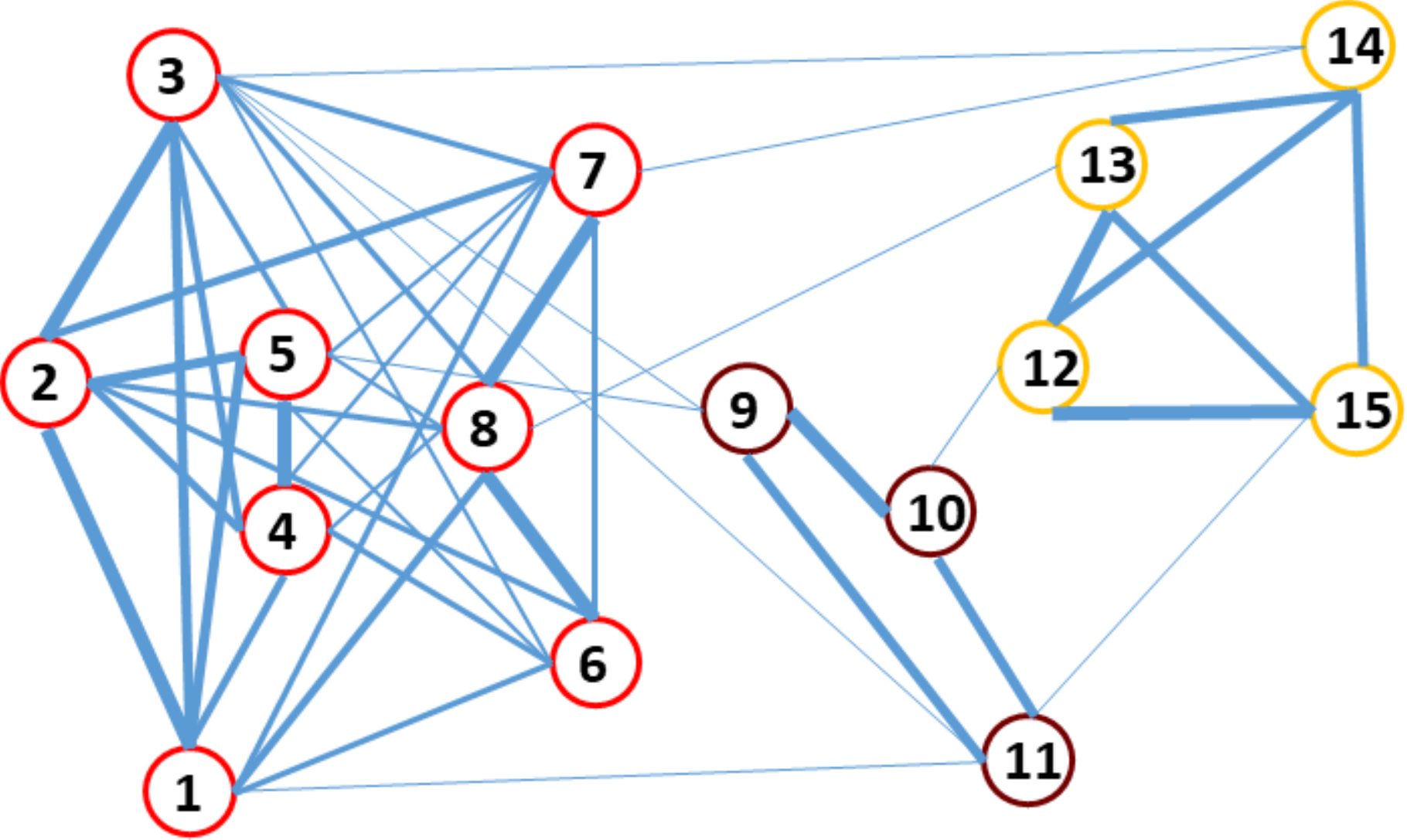} 
	\caption{Graph of the synthetic dataset used to illustrate the difference between $\textbf{Q}$ matrix and the weight matrix computed according to the classic graph representation based on heat diffusion mechanism. The different colors of the nodes identify the classes they are associated with, i.e., "Vineyard", "Road", and "Wood" in the Trento dataset that has been previously mentioned (see Fig. \ref{fig_Trento_dataset}).}
	\label{fig_fluid_example_graph}
\end{figure}

To visualize how the proposed definition of the $\textbf{Q}$   matrix (introduced in Section \ref{sec_fluid_repr})
could help in obtaining more accurate and reliable characterization of the data interactions in the considered dataset, let us consider a toy example. 
In particular, we synthetically generated a small dataset consisting of 15 samples. 
Each sample was characterized by 77 features. 
Specifically, the dataset was set up as follows.
Initially, 8 samples, 3 samples and 4 samples were 
associated with three 
sets of features (or signatures) corresponding to three pixels labeled as "Vineyard", "Road", and "Wood" in the Trento dataset that has been previously mentioned (see Fig. \ref{fig_Trento_dataset}), respectively. 
Then, we applied noise to the samples in a not uniform fashion across the features: the graph induced by the resulting dataset can be summarized in Fig. \ref{fig_fluid_example_graph}, where the nodes represent the samples and the width of the edges is proportional to the similarity between the nodes they connect. 
It is worth noting that in principle the graph would be fully connected. 
However, for sake of visualization, we avoided to display in Fig. \ref{fig_fluid_example_graph} the connections having weights close to 0. 

\begin{figure}[htb]
	\centering
	\includegraphics[width=1\columnwidth]{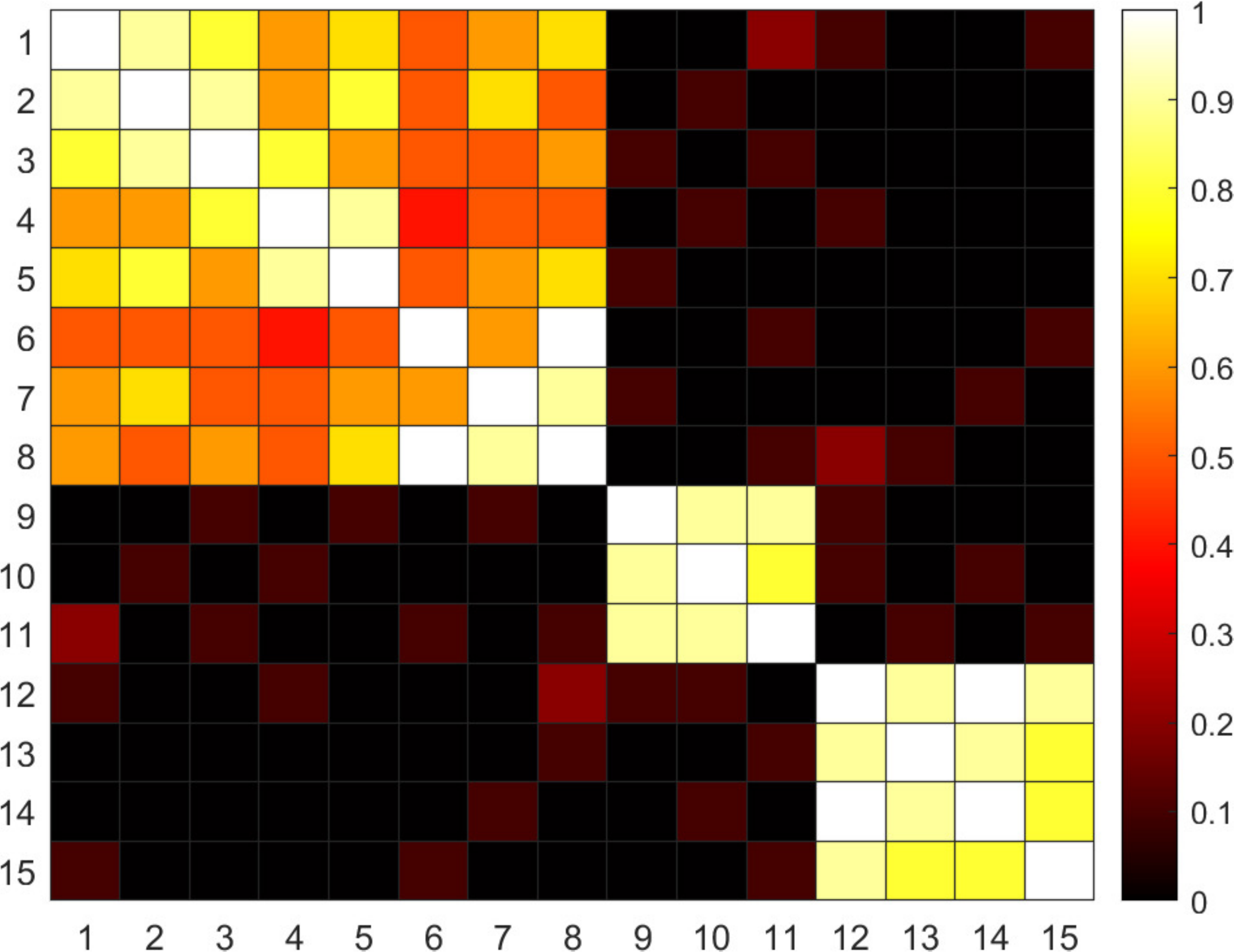} 
	\caption{Heatmap of the $\textbf{Q}$ matrix computed according to the proposed graph representation based on fluid diffusion mechanism on the synthetic dataset in Fig. \ref{fig_fluid_example_graph}. The values are quantized over 10 levels.}
	\label{fig_fluid_example_Qmat}
\end{figure}

\begin{figure}[htb]
	\centering
	\includegraphics[width=1\columnwidth]{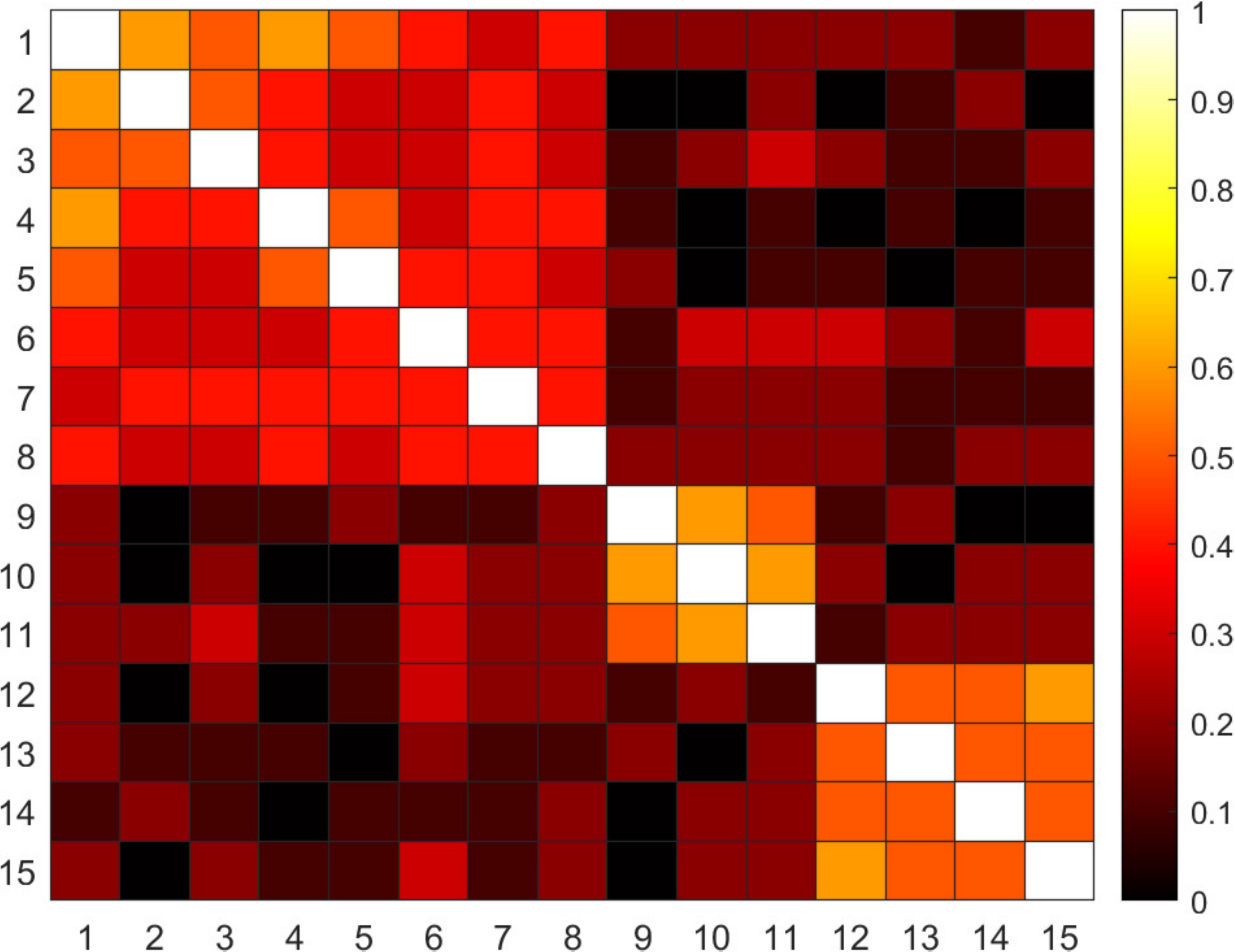} 
	\caption{Heatmap of the weight matrix \textbf{W} computed according to the classic graph representation based on heat diffusion mechanism on the synthetic dataset in Fig. \ref{fig_fluid_example_graph}. The values are quantized over 10 levels.}
	\label{fig_fluid_example_Wmat}
\end{figure}

At this point, we computed the $\textbf{Q}$ matrix and the weight matrix \textbf{W} according to the classic graph representation based on heat diffusion mechanism. For the \textbf{Q} matrix, we considered the definition as in (\ref{eq_fluid_trans_prob_pij}),
we used the method in \cite{ensembleADR} to determine the values of the ${\cal K}$ tensor, and assumed that the distribution of the $\tilde{\textbf{B}}$ matrix would be uniform. 
For the classic definition of the weight matrix $\textbf{W}$ based on heat diffusion mechanism, we used a Gaussian kernel to define the function $\eta$ mentioned in Section \ref{sec_meth_heat}
 \cite{Luxburg,CommunityDetection_Fortunato}. 
We displayed the heatmaps associated with the \textbf{Q} and \textbf{W} matrices in Fig. \ref{fig_fluid_example_Qmat} and \ref{fig_fluid_example_Wmat}, respectively: for sake of visualization, the values of the elements were quantized in 10 levels. 
Moreover, we can recall that high values in these matrices (i.e., lighter colors in the heatmaps) mean high degree of similarity between the considered samples. 

It is worth noting that by construction the ideal configuration of the matrix summarizing the similarities among samples should show as a symmetric diagonal block matrix, since the samples are originally associated with three distinct classes. 
The \textbf{Q} matrix clearly shows this property. 
In fact, the \textbf{Q} matrix is apparently able to retrieve the main characteristics of the informative groups of samples although the noise we added to the samples' signatures we used. 
On the other hand, it is possible to appreciate from Fig. \ref{fig_fluid_example_Wmat} that the classic approach to derive the similarities among samples would lead to a more confused distribution of the weights for graph characterization, leading to a less accurate understanding of the samples' interactions. 
In particular, the higher precision of the $\textbf{Q}$ matrix directly translates in a more accurate eigenanalysis of the associated Laplacian matrix, which is crucial to identify the relevant features and the informative patterns within the data, i.e., to retrieve a reliable information extraction from the considered datasets \cite{COUILLET16,Luxburg,CommunityDetection_Fortunato}. 
As such, the $\textbf{Q}$ matrix seems to represent a valid candidate to achieve a more accurate and reliable understanding of the considered datasets. 


\subsection{Community detection algorithms: a short review}
\label{app_review}

As mentioned in Section \ref{sec_methback},
community detection algorithms can be grouped in categories defined according to the strategy they employ \cite{CommunityDetection_Fortunato}. 
In particular, it is possible to categorize these methods in seven main groups: 1) graph partitioning; 2) hierarchical clustering; 3) partitional clustering; 4) spectral clustering; 5) dynamic community detection; 6) statistical inference-based community detection; and 7) hybrid methods. It is worth noting that these methods have been originally developed considering heat diffusion to derive the graph representation. Future works could be dedicated to explore the opportunity to develop these algorithms while considering the graph representation based on fluid dynamics proposed in this paper. 
The following subsections report the main characteristics of these classes.

\subsubsection{Graph partitioning} 
Graph partitioning methods aim to minimize the number of edges lying between the groups, so to discriminate  communities of predefined size in the dataset \cite{CommunityDetection_Fortunato}. 
%
Graph partitioning problem has been historically addressed as an optimization problem where a function (e.g., modularity, connectivity, conductance, ratio cut, normalized cut) of the difference between the number of edges inside
the modules and the number of edges lying between
them must be maximized \cite{normalizedCut,CommunityDetection_Fortunato,CommunityDetection_GP1,CommunityDetection_GP2, CommunityDetection_GP3,CommunityDetection_GP5}. 
%
It is worth to recall that graph partitioning algorithms require to provide as input the number of clusters (and in some cases also some additional information on the clusters, such as size or degree of compactness) \cite{CommunityDetection_Fortunato,CommunityDetection_GP3,CommunityDetection_GP1,CommunityDetection_GP2,CommunityDetection_GP7}. 
This is a major drawback in several application scenarios, as in most of the cases no information on the characteristics of the data to be explored is available: this is particularly true when addressing operational scenarios.  
Also, these algorithms are typically prone to high sensitivity with respect to the initial state of the optimization process (e.g., bisectioning the graph) \cite{CommunityDetection_GP4,CommunityDetection_GP5,CommunityDetection_GP6,CommunityDetection_GP7}.
Hence, to overcome these issues graph partitioning algorithms are often combined with hierarchical, spectral and/or dynamic community detection principles. 

\subsubsection{Hierarchical clustering}

Several times the graph might show a hierarchical structure, where it is possible to identify a tree-like interplay among the communities in the dataset. Hierarchical clustering aim to exploit this property to detect the communities in the dataset \cite{CommunityDetection_Fortunato}. 
Two main categories of hierarchical clustering techniques can be mentioned: $i)$ agglomerative algorithms, i.e., clusters are iteratively
merged if their similarity is sufficiently
high;
$ii)$ divisive algorithms, i.e.,  clusters are iteratively
split by removing edges connecting vertices with
low similarity.
These approaches are intrinsically opposite strategies, since agglomerative algorithms are 
bottom-up strategies (i.e., nodes are initially considered to be all separate clusters and are iteratively merged according to their similarity), while divisive algorithms are top-down 
(i.e., nodes are initially considered to be all part of a single cluster and are iteratively separated according to their similarity). 
%
Although these schemes have shown good performance in several application scenarios, it is true that they are very sensitive to design choices (e.g., choice of similarity metric). 
Moreover, to reduce the latency of the algorithms, nonlinear processing solutions have been considered, although this leads to higher sensitivity of the initial conditions 
\cite{CommunityDetection_HC4,CommunityDetection_HC5,CommunityDetection_HC6,CommunityDetection_HC7,CommunityDetection_Fortunato,CommunityDetection_HC2,CommunityDetection_HC1}.

\subsubsection{Partitional clustering}

Another popular approach for community detection is represented by so-called partitional clustering. 
In this case, the nodes of the graph are mapped into a multidimensional space, and the edges connecting nodes show weights proportional to the distance between the nodes in this space \cite{CommunityDetection_Fortunato,partitionalClustering}. 
This distance is thus used as a metric to assess the dissimilarity among nodes, so that it can be used to design objective functions to be optimized in order to  identify the group of nodes that would maximize a similarity condition. 
The methods belonging to this category differentiate according to the function to be optimized. 
The most popular algorithm in this class is k-means scheme, and its fuzzy variant \cite{CommunityDetection_Fortunato,CommunityDetection_PC1,partitionalClustering}. 
These schemes are typically brilliant in terms of efficiency, as the mapping and the optimization tasks can be performed very rapidly and with low computational complexity. 
On the other hand, these algorithms require to specify the number of clusters to be detected, which might not be always possible in practical scenarios, as previously mentioned. 
Additionally, the mapping onto the multidimensional space used for the optimization step must be properly designed according to the given dataset, in order to avoid the introduction of artificial bias for the community detection task \cite{CommunityDetection_Fortunato,CommunityDetection_PC1}.

\subsubsection{Spectral clustering}
\label{sec_back_spec}

Spectral clustering methods rely on the definition of a similarity matrix \textbf{W} that aims to summarize the affinity of the samples \cite{Luxburg}. 
Typically, the similarity matrix can be defined in terms of distance between samples computed according to a specific criterion (traditionally, Euclidean distance in the multidimensional feature space), although several metrics have been proposed in technical literature to measure the similarity between samples \cite{Luxburg,COUILLET20,COUILLET18,CommunityDetection_SC1,CommunityDetection_SC2, CommunityDetection_SC3, CommunityDetection_SC4}. 
Spectral clustering methods investigate the space induced by the eigenvectors of the Laplacian matrix (introduced in Section \ref{sec_meth_heat} 
) associated with \textbf{W}. 
The most common definition of the Laplacian matrix is   
$\textbf{D} - \textbf{W}$, where \textbf{D} is the diagonal matrix whose element $D_{ii}$ equals
the degree of the $i$-th node in the graph, i.e., the sum of the weights in the $i$-th row of \textbf{W}. 

It is possible to appreciate that the sum of the
elements of each row of the Laplacian 
is zero by construction. 
This implies that the Laplacian matrix would always show  at
least one zero eigenvalue, corresponding to the eigenvector
with all equal components.  
Moreover, eigenvectors corresponding to different eigenvalues are all orthogonal
to each other \cite{Luxburg,COUILLET18,specclust1,specclust2,specclust4}. 
it is possible to state that the ideal case of $K$ non-overlapping communities showing up neatly separated in the graph would lead to $K$ eigenvalues of the Laplacian matrix equal to zero. 
On the other hand, in a realistic scenario, $K$ subgraphs
would be weakly linked to each other, so that the spectrum of
the unnormalized Laplacian will have one zero eigenvalue,
while the others would be positive \cite{specclust4,specclust1}. 
Nevertheless, it is possible to prove that 
the lowest $K-1$ eigenvalues would still be close to
zero. 
Hence, identifying the $K+1$-th eigenvalue that is clearly different from zero would lead to a robust community detection. 
Furthermore, the detection of significant gaps between eigenvalues is crucial for the effectiveness of spectral clustering algorithms, as well as for their efficiency \cite{COUILLET18,COUILLET20,specclust4,specclust2,specclust1,Luxburg}. 
To address this problem several spectral clustering methods have been introduced in technical literature 
\cite{CommunityDetection_SC3,CommunityDetection_SC2,CommunityDetection_SC1,COUILLET20,COUILLET18,specclust1,CommunityDetection_SC4}. 
However, it is also true that these schemes might show some limitations in addressing community detection in large scale and sparse graphs, where the separability of the aforesaid eigenvalues
may be cumbersome to guarantee and achieve \cite{COUILLET18,COUILLET20}.

\subsubsection{Dynamic community detection}
Another approach for community detection relies on the representation of graphs as results of dynamic processes occurring from node to node, so that the nodes that minimize the energy of the given process would be assumed to belong to the same community 
\cite{CommunityDetection_DA1,CommunityDetection_DA3,CommunityDetection_DA4,CommunityDetection_DA2}.
These methods are typically based on a Markov random field description of the state dynamics, so that higher order moment of informativity can be incorporated in iterative detection schemes in a very straightforward way \cite{CommunityDetection_DA1,CommunityDetection_DA3}. 
These algorithms are typically pretty easy to implement, which represent the main properties for their success in scientific community. 
On the other hand, it is also true that their complexity might become very high when sparse graphs are taken into account. 
Moreover, these algorithms are typically prone to high sensitivity to initial conditions and might not be reliable when communities' distribution is very skewed (i.e., high degree of imbalanced data to be analyzed), such that the convergence of the methods might be jeopardized \cite{CommunityDetection_DA2,CommunityDetection_DA3,CommunityDetection_Fortunato}.

\subsubsection{Statistical inference-based community detection}

In this category, the methods aim to fit the graph topology to a (set of) statistical hypotheses that are drawn according to the diverse set-ups that could be considered. 
These methods employ different strategies in achieving this goal. 
Nonetheless, it is also true that these schemes rely on a common trait, i.e., the definition of the transition probability (used to define the inference mechanisms) as a function of the similarity between nodes \cite{mackaybook}. 
These functions might be linear or nonlinear, or eventually based on higher order moments of the features to be considered \cite{CommunityDetection_SI1,CommunityDetection_SI3,CommunityDetection_SI4}. 
The main idea that drive these methods is that the nodes 
of the same group are linked with a probability $p_x$,
while nodes of different groups are linked with a probability $p_y$. 
Indeed, $p_x$ and $p_y$, as well as the number of the communities, should be input to the system \textit{a priori}: this might limit the impact of the statistical inference-based architectures, although in recent times methods for the automatic optimized selection of these probabilities have been proposed \cite{CommunityDetection_SI2,CommunityDetection_SI1,CommunityDetection_SI5}. 
The complex characteristics that might show up in the dataset (especially in operational scenario) might provide a major issue of these algorithms. 
In fact, the probabilities used in these schemes are typically assumed to be independent. 
As previously mentioned, this might not be matched by the considered dataset: this is particularly true when multimodal community detection is addressed \cite{CommunityDetection_SI1,CommunityDetection_Fortunato}. 

\subsubsection{Hybrid methods}

Finally, it is worth to recall that several methods for community detection that have been used in diverse research fields rely on the combination of multiple strategies that have been previously summarized in this Section (e.g., hierarchical clustering and graph partitioning, spectral clustering and partitional clustering), with the ultimate goal to combine the major benefits of the considered techniques whilst overcoming their possible drawbacks.

The local optimization based on objective functions retrieved from combining modularity and compactness is used to design community detection methods \cite{CommunityDetection_Hyb1,CommunityDetection_Hyb2,CommunityDetection_Hyb4,CommunityDetection_LP1, CommunityDetection_LP2, CommunityDetection_LP3, CommunityDetection_LP4}. 
The local nature of the process enables high efficiency and low computational complexity, although increasing the sensitivity to initial conditions 
\cite{CommunityDetection_Hyb3}. 
Leveraging on edge betweenness in the regions close to boundaries might reduce the impact of this issue. 
This idea has been extended to manifold analysis, where similarities among nodes are computed in terms of geodesic distances. 
Communities
appear as portions of the graph with a large curvature \cite{CommunityDetection_curv}. 
The manifold approach has been proven to be successfully applied to massive datasets obtained form one single source of information (e.g., social media networks) \cite{CommunityDetection_Hyb1,CommunityDetection_Hyb4}. 
However, they also requires the communities to show up uniformly across the dataset, as well as the features to be affected by linear perturbations across the samples \cite{CommunityDetection_Fortunato}. 
This is typically a major limiting factor for this approach. 


Deep learning-based community detection has then triggered the attention of a large number of scientists worldwide, either in terms of autoencoder-based structures or deep investigation of latent features by means of convolutional neural networks  \cite{CommunityDetection_DL,CommunityDetection_DL1,CommunityDetection_DL3, CommunityDetection_DL4, CommunityDetection_DL5, CommunityDetection_DL6, CommunityDetection_DL2}. 
In fact, embedding communities in deep learning fashion has proven to be an efficient approach, since the network aims to learn node distributions of communities in a low-dimensional
space \cite{CommunityDetection_DL6,CommunityDetection_DL4}
Nevertheless, there ares still several open issues that can be identified for this community detection strategy \cite{CommunityDetection_DL}. 
In fact, in order to perform accurate community detection, deep learning techniques require the number of communities to be identified as an input parameter of the processing, so that the analysis of the latent space they rely on can be carried out accurately \cite{CommunityDetection_DL,Fluid_multimodML}. 
Moreover, the different probability distributions
associated with each data type to be processed need to be addressed at the design step of models and algorithms. 
This is a relevant limiting factor for the use of deep learning-based community detection, especially  when complex and/or multimodal datasets are considered \cite{CommunityDetection_DL,CommunityDetection_DL1,CommunityDetection_DL3, CommunityDetection_DL4, CommunityDetection_DL5, CommunityDetection_DL6, CommunityDetection_DL2}. 
Finally, the community detection performance of these methods can be jeopardized when unbalanced datasets (eventually showing non-uniform statistical properties across the records) are analyzed. 

\subsection{Study of methods for the definition of the permeability tensor}
\label{app_K_ablation}
We report in this Section the results that we obtained when studying the ability of the method in Section \ref{sec_meth_ADR} to reliably identify relevant features across complex datasets, so that the construction of the permeability ${\cal K}$ tensor in (\ref{eq_flowrate}) could be carried out.  

We considered the three datasets in Section \ref{sec_exp_res}, and added  records generated by noise-like process characterized by a Gaussian distribution and a signal-to-noise ratio (SNR) set to 20dB to each sample. 
Hence, we obtained for each dataset an additional subset of attributes (approximately 30$\%$ of the original amount, i.e., we added 50, 23, and 4 features for the datasets in Section \ref{sec_exp_RS}-\ref{sec_exp_PV}, respectively) that were clearly irrelevant for the characterization of the phenomena under exam and therefore simulating a set of corrupted, noisy, and non-informative records. 

\begin{figure}[htb]
	\centering
	\includegraphics[width=1\columnwidth]{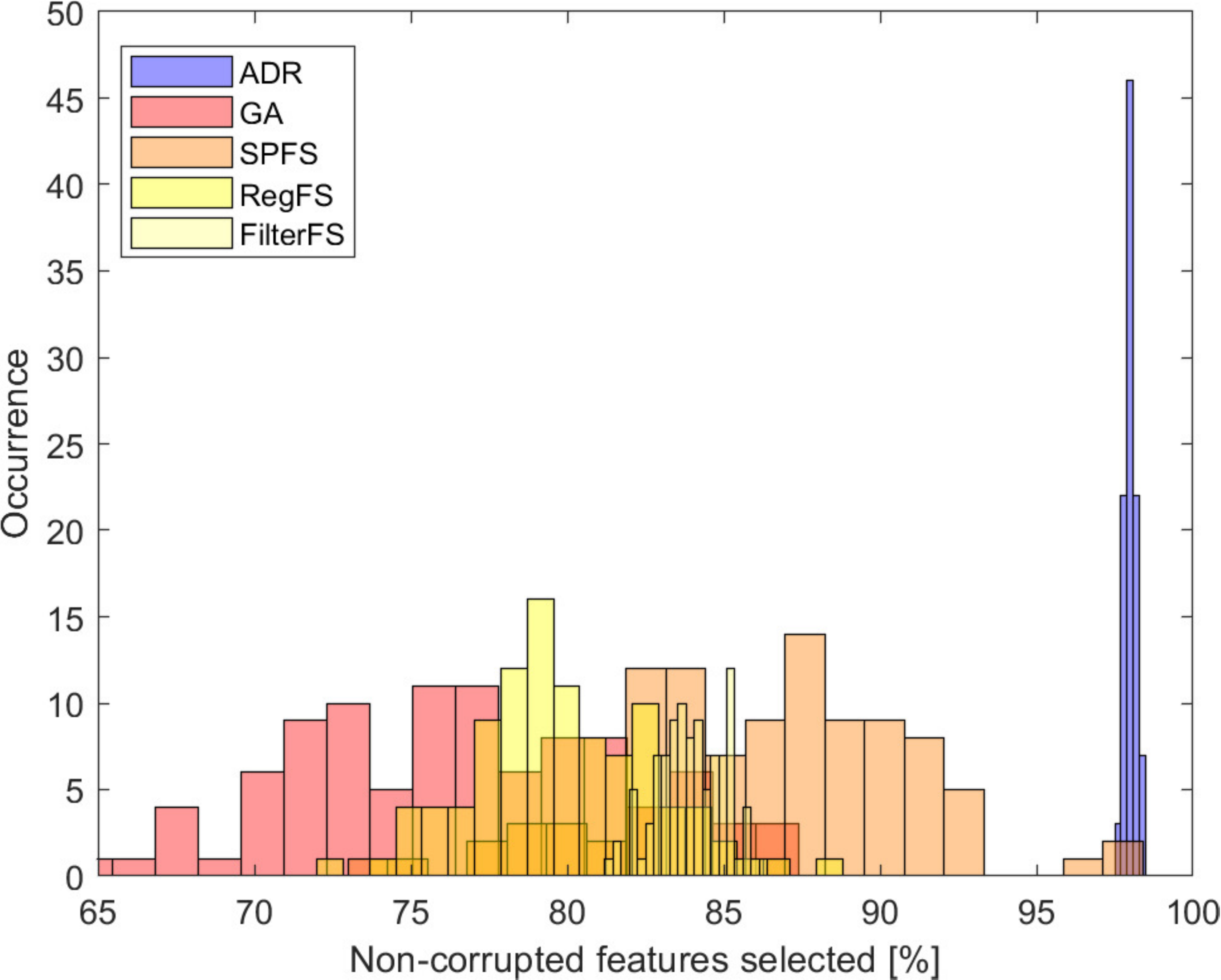} 
	\caption{Fraction of non-corrupted features selected for the definition of the ${\cal K}$ in (\ref{eq_fluid_trans_prob_pij}) when analyzing the extended dataset in Section \ref{sec_exp_RS} (multimodal remote sensing) achieved by using the method described in Section \ref{secmeth} (ADR), genetic algorithm (GA) \cite{CommunityDetection_FS4}, structure preserving feature selection (SPFS) \cite{CommunityDetection_FS1}, regularization-based feature selection (regFS) \cite{CommunityDetection_FS2}, and filter-based feature selection (FilterFS) \cite{CommunityDetection_FS3}. The ADR method is the approach we propose to establish the fluid graph representation in this paper.}
	\label{fig_res_FS_RS}
\end{figure}

\begin{figure}[htb]
	\centering
	\includegraphics[width=1\columnwidth]{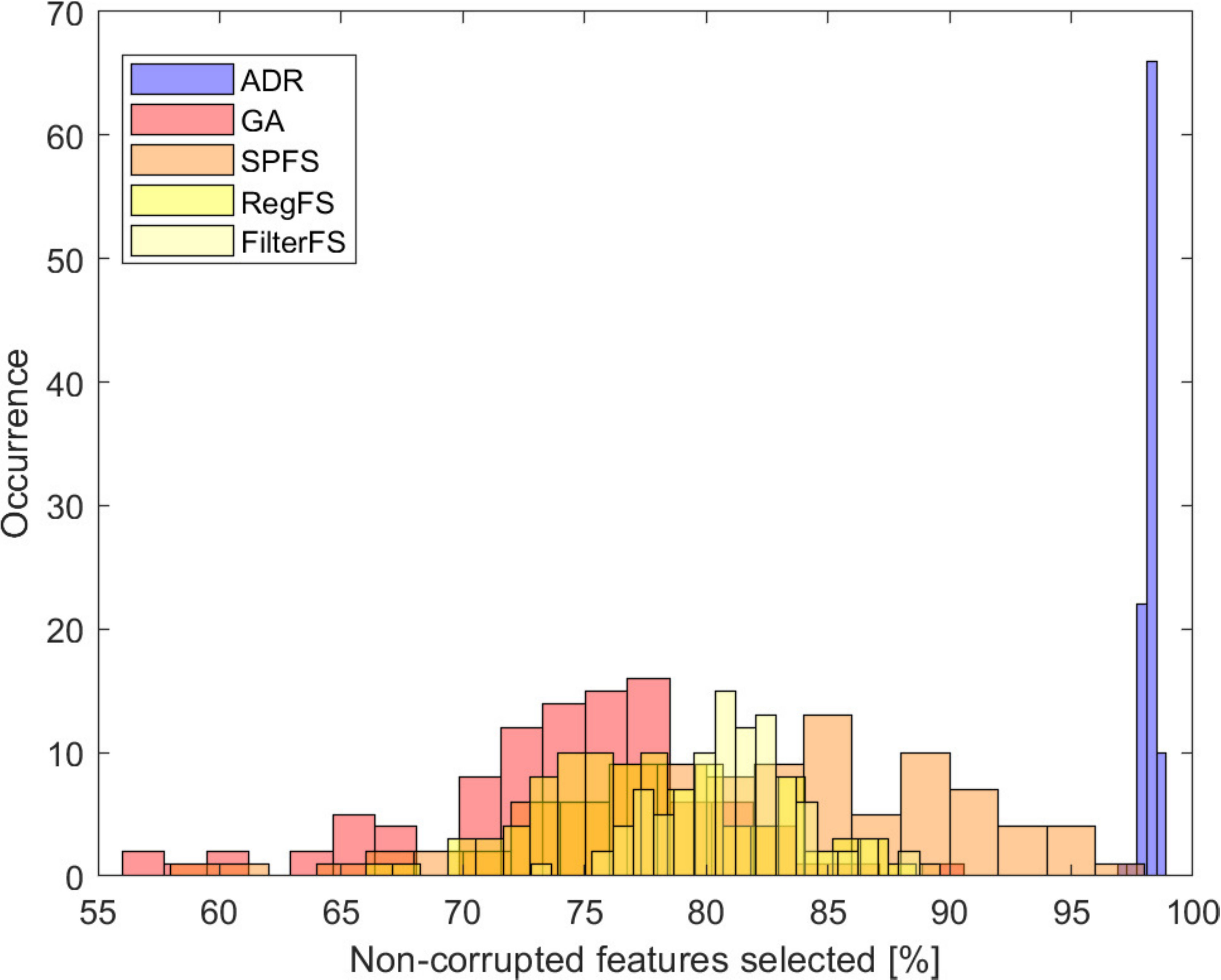} 
	\caption{Fraction of non-corrupted features selected for the definition of the ${\cal K}$ in (\ref{eq_fluid_trans_prob_pij}) when analyzing the extended dataset in Section \ref{sec_exp_BCI} (multimodal brain-computer interface). The same notation as in Fig. \ref{fig_res_FS_RS} applies here.}
	\label{fig_res_FS_BCI}
\end{figure}

\begin{figure}[htb]
	\centering
	\includegraphics[width=1\columnwidth]{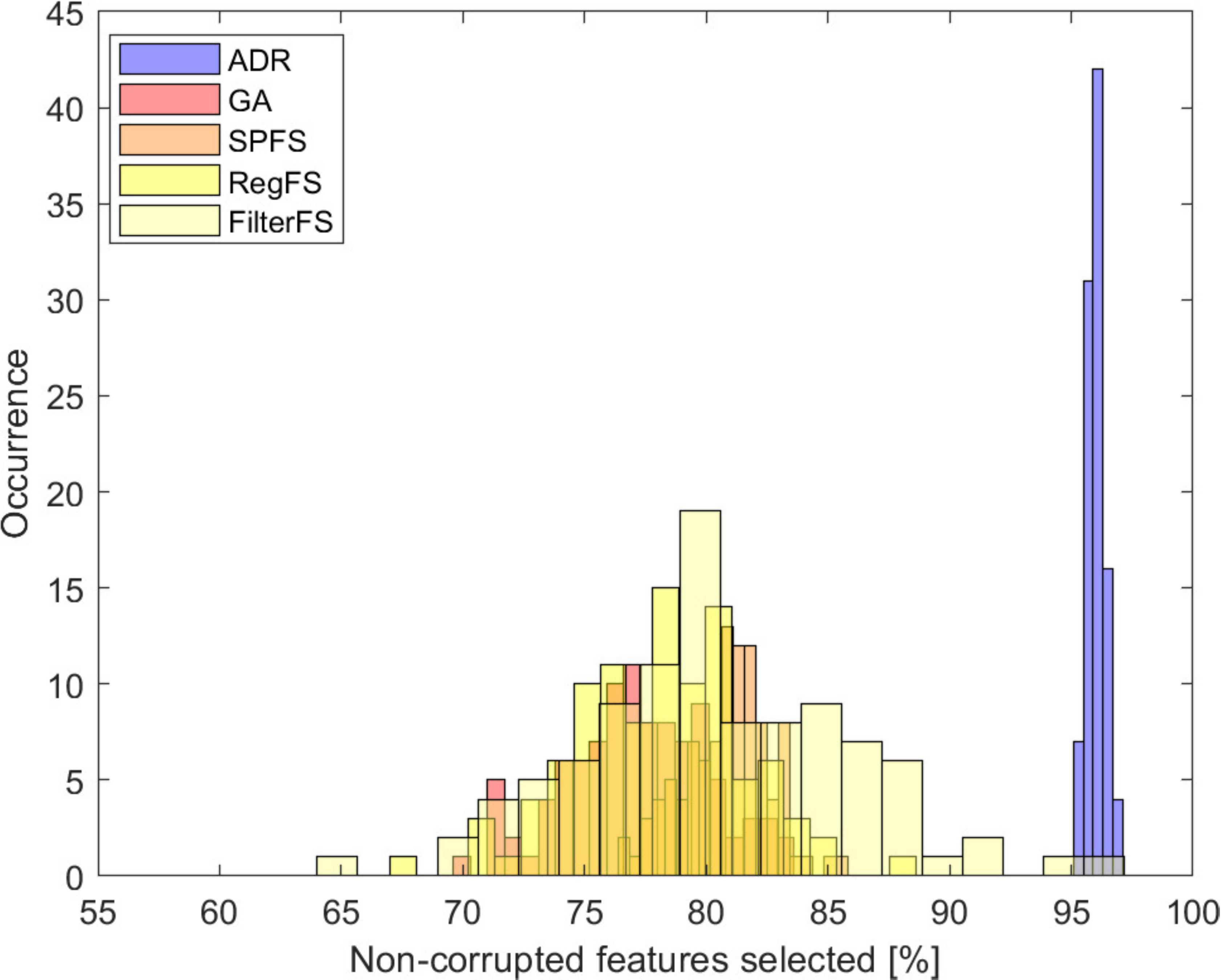} 
	\caption{Fraction of non-corrupted features selected for the definition of the ${\cal K}$ in (\ref{eq_fluid_trans_prob_pij}) when analyzing the extended dataset in Section \ref{sec_exp_PV} (multimodal photovoltaic energy). The same notation as in Fig. \ref{fig_res_FS_RS} applies here.}
	\label{fig_res_FS_PV}
\end{figure}

Then, we applied the adaptive dimensionality reduction approach for the determination of ${\cal K}$, as well as other four state-of-the-art feature selection methods on the new total datasets: the algorithms we considered were based on diverse strategies, i.e., genetic algorithm (GA) \cite{CommunityDetection_FS4}, structure preserving feature selection (SPFS) \cite{CommunityDetection_FS1}, regularization-based feature selection (RegFS) \cite{CommunityDetection_FS2}, and filter-based feature selection (FilterFS) \cite{CommunityDetection_FS3}. 
Our goal is to check whether these methods are able to discriminate the real records from the added (irrelevant) ones in the obtained extended datasets.  
We carried out 100 runs, and summarized the results in Fig. \ref{fig_res_FS_RS}-\ref{fig_res_FS_PV}. 

Taking a look to these histograms, it is possible to appreciate how the approach we proposed in this work is actually able to outperform the state-of-the-art methods by selecting almost exclusively the original (relevant) features for dimensionality reduction. 
Moreover, the high kurtosis of the histograms achieved by means of the adaptive dimensionality reduction method emphasizes the robustness of this algorithm to select relevant features across all the tests we have performed. 
On the other hand, the variability of the performance of the other methods appear very high, so that their outcomes do not appear solid. 
Thes observations are valid for all the datasets we presented in Section \ref{sec_exp_res}, hence supporting our choice of using the adaptive dimensionality reduction method presented in Section \ref{sec_meth_ADR} for the construction of the fluid Laplacian matrix.

\begin{table}[!th]
	\renewcommand{\arraystretch}{1.3}
	\caption{Minimum and maximum value [$\%$] of non-corrupted features selected  for ablation study on the kernel design choice in (\ref{eq_GKMI1})}
	\label{tab_ablation1}
	\centering
	\begin{tabular}{|c|c|c|c|}
		\hline
		\bfseries Method & \bfseries RS & \bfseries BCI & \bfseries PV \\
		& (multimodal & (multimodal & (multimodal \\
		& remote sensing) & brain-computer & photovoltaic \\
		&  & interface) & energy) \\
		\hline
		\textbf{Gaussian} & 97.4-98.1 & 97-98.8 & 95-97.6 \\
		\hline
		\textbf{Euclidean} &  81.3-94.3 & 81.8-93.2 & 78.2-91.9\\
		\hline
		\textbf{Linear} & 82.4-96.1  & 83.3-96.2  & 79.4-92.5  \\
		\hline
		\textbf{Polynomial} & 84.6-94.4  & 85.3-97.1 &  84.1-93.2 \\
		\hline
	\end{tabular}
\end{table}

The method in \cite{ensembleADR} employs a Gaussian kernel to quantify the difference between two features for each sample. 
To assess the validity of this choice, we conducted an ablation study on the definition of this metric in equation (\ref{eq_GKMI1}). 
Therefore, we analyzed the support of the histograms for the selection of non-corrupted features in the extended datasets we previously described in this section. 
%
when using kernel functions that are most commonly
employed, i.e.:
\begin{itemize}
	\item Euclidean distance: $|| \underline{x}_{m n_1}- \underline{x}_{m n_2} ||$;
	\item linear kernel:  $\underline{x}_{m n_1}^T \underline{x}_{m n_2} + l$;
	\item polynomial kernel:  $(a\underline{x}_{m n_1}^T \underline{x}_{m n_2} + l)^b$.
\end{itemize}

\noindent These kernels allow a fair comparison with the definition in (\ref{eq_GKMI1}) because they fulfill the Mercer’s
theorem (i.e., being positive-definite similarity matrices), which is a basic condition when no prior knowledge on the
interclass and intraclass statistical distributions is available. 

We reported in Table \ref{tab_ablation1} the minimum and maximum values of non-corrupted features that we obtained when using these kernels on the extended datasets we investigated in Fig. \ref{fig_res_FS_RS}-\ref{fig_res_FS_PV}. 
These results lead to observations on the choice of the distance in (\ref{eq_GKMI1}), as well as on the choice of the algorithm for the definition of the ${\cal K}$ tensor. 
In fact, on one side we can observe that all the values in Table \ref{tab_ablation1} are higher than the supports of the histograms displayed in Fig. \ref{fig_res_FS_RS}-\ref{fig_res_FS_PV} obtained when using the methods in \cite{CommunityDetection_FS1,CommunityDetection_FS2,CommunityDetection_FS3,CommunityDetection_FS4}. 
This shows how the overall architecture for the selection of the relevant features in the dataset based on adaptive dimensionality reduction scheme in \cite{ensembleADR} is solid and can adapt to various data characteristics. 
At the same time, it is worth noting how the Gaussian kernel outperforms all the other kernel choices. 
Therefore, 
the Gaussian kernel proves to be more flexible than other kernels in characterizing the data structure in complex systems, especially when
an investigation of large-scale and heterogeneous datasets is
conducted.


\color{black}

\subsection{Metrics used to assess the community detection performance}
\label{app_metrics}

In order to give a compact definition of the metrics to assess community detection performance, let $\Omega =\{\omega_k\}_{k=1, \ldots, C_\Omega}$ be the set of detected clusters, whilst 
$\Psi =\{\psi_l\}_{l=1, \ldots, C_\Psi}$ is the set of ground-truth classes. 
Then, let us consider the modified purity index. 
To define $mP$, it is important to take into account the purity of a node  for a partition $\Omega$ 
relatively to
another partition $\Psi$, i.e., a function that identifies if the class of $\Psi$  containing node $u$ is majority in that of $\Omega$ 
also containing $u$,
and otherwise. This function can be written as: 

\begin{equation}
	P(u,\Omega, \Psi)=\delta(\psi_j \: s.t. \: |\omega_i \cap \psi_j| is \; maximum),
	\label{eq_purity_1}
\end{equation}

\noindent where $u \in \omega_i$, $u \in \psi_j$, and $\delta$ is the Kronecker delta function.
At this point, assuming that $w_u$ is the weight of node $u$ the modified purity ($mP$) can be defined as follows:

\begin{equation}
	mP(\Omega,\Psi) = \sum_{i} \sum_{u \in \omega_i} \frac{w_u}{\sum_j w_j} P(u, \Omega, \Psi )
	\label{eq_mP} 
\end{equation} 

\noindent It is worth noting that $mP$ measures whether each detected community is assigned to the ground-truth label which is most frequent in the community without bias (that could affect the classic purity metric \cite{fluid_metrics,fluid_metrics2}). Moreover, the upper bound of this metric is 1, which corresponds to a perfect match between the partitions $\Omega$ and $\Psi$. On the other hand, its
lower bound is 0 and corresponds to a complete mismatch between partitions $\Omega$ and $\Psi$. 

With the notion of weight of a node in mind, it is possible to modify the classic definition of the adjusted Rand index in order to take into account the topological configuration of the graph. 
In this respect, it is necessary to define for any subset of nodes $\Phi$ the quantity $\kappa(\Phi) = \sum_{t,u \in \Phi} w_t w_u$. 
Hence, the modified adjusted Rand index ($mARI$) can be written as follows \cite{fluid_metrics,fluid_metrics2}: 

\begin{gather} 
	mARI(\Omega, \Psi) =  \nonumber \\ 
	\frac{\sum_{ij} \kappa(\omega_i \cap \psi_j) - \frac{1}{\kappa({\cal V})}\sum_i \kappa(\omega_i) \sum_{j} \kappa(\psi_j)}{\frac{1}{2} (\sum_i \kappa(\omega_i) +  \sum_{j} \kappa(\psi_j)) - \frac{1}{\kappa({\cal V})}\sum_i \kappa(\omega_i) \sum_{j} \kappa(\psi_j)} 
	\label{eq_mARI}
\end{gather}


\noindent This metric interprets the evaluation of community detection performance in terms of the decisions that have been taken for the nodes in the graph. 
In particular, $mARI$ is used to assess the community detection performance against the case for which the detected clusters $\Omega$ would be randomly generated, taking into account the individual effect of each node in the graph and in a chance-corrected manner \cite{fluid_metrics,fluid_metrics2}. 
Specifically, $mARI$ is upper bounded to 1. 
I.e., $mARI$ assumes the value 1 when $\Omega$ and $\Psi$ perfectly match. 
On the other hand, $mARI$ is equal to or less than 0 when the similarity between $\Omega$ and $\Psi$ is equal or less than what is expected from
two random partitions. 

Finally, to define the modified normalized mutual information index ($mNMI$), we should consider a modified joint probability of cluster $\omega_i$ and class $\psi_j$, which can be written as $\tilde{p}_{ij} = \sum_{u \in \omega_i \cap \psi_j} \frac{w_u}{\sum_l w_l}$. 
Accordingly, we can define $mNMI = mNMI(\Omega, \Psi)$ as follows: 

\begin{equation}
	mNMI= \frac{-2 \sum_{ij} \tilde{p}_{ij} \log( \frac{\tilde{p}_{ij}}{\sum_i \tilde{p}_{ij} \sum_j \tilde{p}_{ij} } )}{\sum_i \sum_j \tilde{p}_{ij} \log (\sum_j \tilde{p}_{ij}) + \sum_j \sum_i \tilde{p}_{ij} \log(\sum_i \tilde{p}_{ij})  }
	\label{eq_mNMI}
\end{equation} 

It is possible to notice that the metrics in (\ref{eq_mP}), (\ref{eq_mARI}), and (\ref{eq_mNMI}) all rely on the definition of weight of a node, that is supposed to help in emphasize the role of the graph topology in computing how well the proposed graph representation can help in extracting functional information from the given dataset. 
To this aim, several definitions of $w_u$ can be drawn out \cite{fluid_metrics,fluid_metrics2}. 
In particular, three configurations can be considered in order to cover the main characteristics to be taken into account when assessing the relevance of a node in graph learning frameworks:
\begin{enumerate}
	\item $w_u = d_u / \max_t d_t$, where $d_u$ identifies the degree of node $u$ - in this case, we can talk about a \textit{degree measure};
	\item $w_u = d_u^{\mathtt{INT}} /d_u$, where $d_u^{\mathtt{INT}}$ is the internal degree of node, i.e. the number of connections it has
	in the cluster it belongs to - in this case, we can talk about an \textit{embeddedness measure}; 
	\item $w_u = d_u^{\mathtt{INT}} / \max_t d_t$ - in this case, we can talk about a \textit{weighted embeddedness measure}. 
\end{enumerate}

\subsection{Codes for performance comparison}

The results in Fig. \ref{fig_res_heat_RS}-\ref{fig_res_heat_PV}
have been produced by using codes that can be found at these websites: https://github.com/HPAI-BSC/Fluid-Communities; https://gist.github.com/pszufe;
https://github.com/michaelschaub/michaelschaub.github.io;
https://github.com/chocolates/Community-detection-based-on-distance-dynamics;
https://aaronclauset.github.io/wsbm/;
https://github.com/norbertbin/SpecClustPack;
https://github.com/DTaoo/DMC. 

\bibliographystyle{unsrt}\scriptsize	
\bibliography{BIGDATArefs, BigDATArefs_old}

\end{document}